\documentclass{aastex701}

\usepackage{graphicx}%
\usepackage{amsmath,amssymb,amsfonts}%
\usepackage{amsthm}%
\usepackage{mathrsfs}%
\usepackage[title]{appendix}%
\usepackage{xcolor}%

\usepackage{nicefrac}	

\def\iLFIR{$L_\text{FIR}$}
\def\SigmaIR{$\Sigma_\text{IR}$}
\def\Lsun{L$_\odot$}
\def\Msun{M$_\odot$}
\def\Zsun{Z$_\odot$}

\def\iTdust{$T_\text{dust}$}

\def\iTkin{$T_\text{kin}$}
\def\Hi{H\,{\sc i}}
\def\Hii{H\,{\sc ii}}

\def\Ci{[C\,{\sc i}]}
\def\Cii{[C\,{\sc ii}]}

\def\Oi{[O\,{\sc i}]}

\def\Oiii{[O\,{\sc iii}]}
\def\Oiv{[O\,{\sc iv}]}
\def\Nii{[N\,{\sc ii}]}
\def\Niii{[N\,{\sc iii}]}

\def\Neii{[Ne\,{\sc ii}]}
\def\Neiii{[Ne\,{\sc iii}]}
\def\Nev{[Ne\,{\sc v}]}
\def\Siii{[S\,{\sc iii}]}
\def\Sili{[Si\,{\sc i}]}
\def\Silii{[Si\,{\sc ii}]}
\def\Arii{[Ar\,{\sc ii}]}
\def\Ariii{[Ar\,{\sc iii}]}
\def\Arv{[Ar\,{\sc v}]}

\def\kms{km\,s$^{-1}$}

\def\lsim{\lesssim}
\def\gsim{\gtrsim}

\def\ncrit{$n_\text{crit}$}
\def\ncrite{$n_\text{crit}^{\rm e}$}
\def\ncritH{$n_\text{crit}^{\rm H}$}

\def\ncmmm{cm$^{-3}$}

\usepackage{upgreek}
\renewcommand{\mu}{\upmu}

\makeatletter

\begin{document}

\title[IR FSLs at high $z$]{Infrared fine-structure lines at high redshift}

\author[0000-0003-0699-6083]{Roberto Decarli}\email{roberto.decarli@inaf.it}
\affiliation{INAF Osservatorio di Astrofisica e Scienza dello Spazio di Bologna, via Gobetti 93/3, Bologna, 40129, Italy}

\author[0000-0003-0699-6083]{Tanio D\'{i}az-Santos}\email{tanio@ia.forth.gr}
\affiliation{Foundation for Research and Technology-Hellas, Institute of Astrophysics, Heraklion, 70013, Greece}

\begin{abstract}
Infrared (IR) fine-structure line (FSL) emission arises from the radiative de-excitation of collisionally-excited electrons in atoms and ions. Simple elements such as carbon (C), nitrogen (N), and oxygen (O) are widespread in the interstellar medium (ISM) as a result of metal enrichment. Thanks to their high luminosities and relatively simple physics, IR FSLs have quickly become the workhorse for studying the formation and evolution of galaxies in the nearby and distant Universe. In this review, we introduce the physics of FSL emission and the diagnostics of the ISM that we can derive from them via first principle arguments. We summarize the history of FSL observations with a focus on the far-IR wavelengths and a particular emphasis on the on-going efforts aimed at characterizing galaxies at cosmic noon and beyond. We explore the dependence of emission line trends, such as those observed in `line deficits' or \Cii{}--SFR relations, as a function of redshift and galaxy types. Once selection biases are controlled for, IR FSLs are a powerful tool to constrain the physics of galaxies. The precise redshift information inferred from fine-structure line observations have enabled tracing their ISM properties across cosmic reionization. FSL observations have also led to estimates of the mass of different ISM phases, and of the SFR of distant galaxies. It is thanks to IR FSL observations that we have been able to measure the internal dynamics of high-$z$ galaxies, which in turns has allowed us to test, e.g., the onset of black hole -- host galaxy relations in the first billion years of the Universe and the presence of gas outflows associated with the baryon cycle in galaxies. Finally, FSLs have provided important clues on the physics of the ISM in the most distant galaxies known to date. We demonstrate the strength and limitations of using IR FSLs to advance our understanding of galaxy formation and evolution in the early universe, and we outline future perspective for the field.
\end{abstract}



\tableofcontents

\newpage

The spin-momentum coupling of valence electrons in some atoms and ions results in the fine-structure splitting of the ground energy level into two or three levels. Infrared (IR) fine-structure lines (FSLs) arise from the radiative de-excitation of collisionally-excited electrons in these levels. The modest energies associated with the fine-structure splitting imply that the levels are easily populated at even low temperatures, hence the widespread emission of FSLs in the interstellar medium (ISM) of galaxies. These emission lines are typically optically thin. FSLs beyond rest-frame 50\,$\mu$m suffer from negligible dust extinction in all but the highest column density scenarios. All these arguments imply that the astrophysical interpretation of the observed fluxes is often straightforward. Some FSLs have rest-frame wavelengths in the far-infrared (FIR) regime, hence they are redshifted into the (sub-)mm transparent windows of the atmosphere, where we can observe them using ground-based telescopes and interferometers. Finally, fine-structure lines such as \Cii{} at 158\,$\mu$m and \Oi{} at 63\,$\mu$m are very efficient coolants of the ISM at $10<T {\rm [K]}<1000$, and therefore can account for a few percent of the bolometric luminosity in star-forming galaxies. For all of these reasons, the study of fine-structure line emission arises as a key tool to unveil the astrophysics of distant galaxies.

In this review, we summarize the basics of the physics leading to the fine-structure emission (Sect.~\ref{sec_physics}). We present diagnostics of the ISM properties based on fine-structure lines (Sect.~\ref{sec_diagnostics}). We briefly review the history of fine-structure line observations (Sect.~\ref{sec_history}). Finally, we discuss future perspectives (Sect.~\ref{sec_conclusions}). The interested read will find further information in classical textbooks on the ISM, such as \citet{spitzer78}, \citet{tielens05}, \citet{osterbrock06}, \citet{draine11}. Throughout the review, we adopt a standard flat $\Lambda$ cold dark matter cosmology with $H_0=70$\,\kms{}\,Mpc$^{-1}$, $\Omega_m=0.3$, $\Omega_\Lambda = 0.7$. Following widespread practice in astrophysics, we will generically refer to any element beyond Helium as `metal'.

\section{Physics of the IR FSL emission}\label{sec_physics}

In this section, we discuss the physical ingredients that are at the root of the IR FSL emission. 

\subsection{Electron configurations}

Electrons in atoms and ions are organized in orbitals characterized by the principal quantum number $n$ and the orbital angular momentum $l$ (in units of $\hbar$). Orbitals are designated as $s$, $p$, $d$, $f$ for $l$=0, 1, 2, 3. Because of the Pauli exclusion principle, and the fermionic nature of electrons, each shell $nl$ can host only up to $2l\,(2l+1)$ electrons. The orbital angular momentum and the spin vectors of the electrons can be aligned in different ways, which results in slightly different energies for the corresponding configurations (due to the so-called `L--S coupling'). It is convenient to thus introduce the quantum number of the total angular momentum, {\bf J}, which is the sum of the total orbital angular momentum, {\bf L}, and the total spin momentum, {\bf S}, all in units of $\hbar$; and to adopt the standard notation: $^{2S+1}\mathcal{L}_{J}$, with $\mathcal{L}$\,=\,S, P, D, F for $L$\,=\,0, 1, 2, 3. By construction, J can vary between $|$L-S$|$ and L+S, so that, for instance, a $^3$P configuration (i.e., L=1, S=1) yields J=0, 1, or 2, leading to a fine-structure splitting of the energy level. 

\begin{figure}[ht]
\begin{center}
\includegraphics[width=0.74\columnwidth]{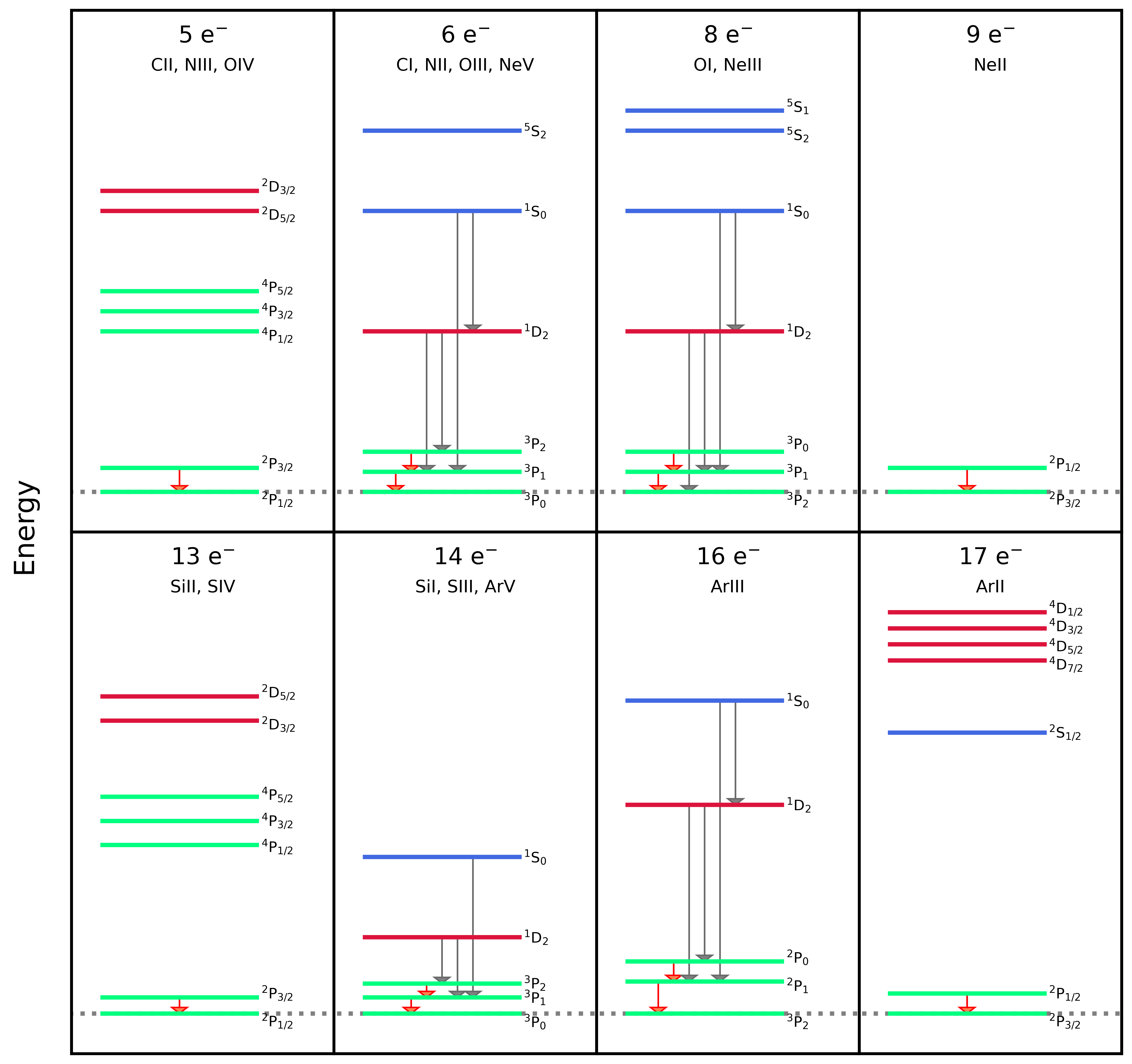}
\end{center}
\caption{Scheme of the energy structure of key atoms and ions, grouped by number of electrons. Configurations with L=0, 1, 2 are color-coded in blue, green, and red, respectively. The ground level is marked with a dotted gray line. The level structure is not in scale, for clarity reasons. Transitions that are typically populated through collisions and depopulated via radiative emission are marked with downward arrows. The IR fine-structure lines are marked with red downward arrows.}
\label{fig_energy_levels}
\end{figure}

Figure~\ref{fig_energy_levels} shows the energy structure of various metals, grouped by number of electrons. The energy level structure depends, to first order, only on the number of electrons, and in particular of valence electrons. Thus, for instance, neutral carbon (C), singly ionized nitrogen (N$^+$), and doubly-ionized oxygen (O$^{++}$) share the same structure, although with different energies.

When the valence electron(s) has(have) orbital angular momentum L=1, the ground energy level splits into two (for an odd number of valence electrons) or three (for an even number of valence electrons) levels, separated by energies $\Delta E/k_{\rm b}=30-1000$\,K. These energy levels can be easily populated via collisions at temperatures that are common in the ISM of galaxies. The radiative de-excitation of such levels results in widespread IR fine-structure line emission tracing a diverse range of gas conditions; the study of these lines and their use as a powerful tool to characterize the ISM physics in galaxies is the topic of this review.

\subsection{Collisional excitation and the energy level population}\label{sec_linepredict}

The population of a given energy level is set via collisions and radiative excitation / de-excitation. If we assume that the gas is in local thermal equilibrium, the kinetic temperature, $T_\text{kin}$, sets the energy distribution of the collision partners. Free electrons are the main collision partners in a fully-ionized gas cloud; electrons, hydrogen atoms (H) and molecules (H$_2$), helium atoms (He), and other species contribute to the collision budget in the neutral and molecular phase. The net rate at which a given energy level $i$ is populated is given by the rate at which electrons in other levels move into this one, minus the rate at which electrons at energy level $i$ move out of it:
\begin{equation}\label{eq_radtransf}
\frac{d n_i}{dt}=\sum_{j\neq i}\,R_{ji}n_j - n_i\,\sum_{j\neq i} R_{ij}.
\end{equation}

At equilibrium, $\frac{d n_i}{dt} = 0$ for any $i$. We label as $R_{ul}$ the rates at which electrons are de-excited from a higher-energy level $u$ to a lower-energy level $l$:
\begin{equation}\label{eq_Rul}
R_{ul} = n_c k_{ul}(T_\text{kin})+(1+n_{\gamma, ul})\,A_{ul},
\end{equation}

and as $R_{lu}$ the rates at which electrons are excited from $l$ to $u$:
\begin{equation}\label{eq_Rlu}
R_{lu}=\frac{g_u}{g_l}\,\left[n_c k_{ul}(T_\text{kin})\,\exp\left(-\frac{E_{ul}}{k_{\rm b}T_\text{kin}}\right)+n_{\gamma,ul}\,A_{ul}\right],
\end{equation}
where $E_{ul}\,=\,E_u-E_l\,=\,h\nu_{ul}$ is the energy difference between two energy levels; $n_c$ is the density of the collision partner; $A_{ul}$ are the Einstein coefficients; $g_{i}$ are the statistical weights of the levels; and $k_{ul}(T_\text{kin})$ are the collision rates, which can be expressed as:
\begin{equation}\label{eq_kul}
k_{ul}(T_\text{kin})=\langle v \sigma \rangle_{u\rightarrow l} = \frac{h^2}{(2 \pi m_e)^{3/2}}\frac{1}{(k_{\rm b} T_\text{kin})^{1/2}}\, \frac{\Omega(l,u)}{g_u}\approx \frac{8.629\times 10^{-6}}{\rm [cm^{3}\,s^{-1}\,K^{1/2}]} \, \frac{\Omega(l,u)}{g_u \,\sqrt{T_\text{kin}}}, 
\end{equation}
where $\Omega(l,u)$ are the collision strengths \citep[see Appendix F of ][for a compilation]{draine11}. Finally, $n_{\gamma,ul}$ is the photon occupation number at the frequency of the $u\rightarrow l$ line: 
\begin{equation}\label{eq_ngamma}
n_{\gamma}= \frac{c^3}{8 \pi \,h\nu^3}\,u_\nu, 
\end{equation}
where $u_\nu$ is the radiation energy density at frequency $\nu$. 

The first term on the right-hand side of Eq.~\eqref{eq_Rul} measures the rate at which level $u$ is de-populated via collisions with a partner of density $n_c$ that leaves the valence electron in the energy level $l$. The second term accounts for de-population of level $u$ via spontaneous ($A_{ul}$) and stimulated ($n_{\gamma,ul} A_{ul}$) emission. 
The FSLs discussed in this review arise from meta-stable states\footnote{Because they are not accounted for by selection rules, FSLs are `forbidden'. Such lines are marked with square brackets (\Cii{}, \Oiii{}, etc).}. Their Einstein coefficients are  of $10^{-6}-10^{-2}$ s$^{-1}$. The right-hand side of Eq.~\eqref{eq_Rlu} sums the contribution of collisions (first term; which is based on the assumption of local thermal equilibrium for the collisional partners) and stimulated excitation (second term).

\begin{figure}[ht]
\begin{center}
\begin{minipage}{0.06\textwidth}
\includegraphics[width=\textwidth]{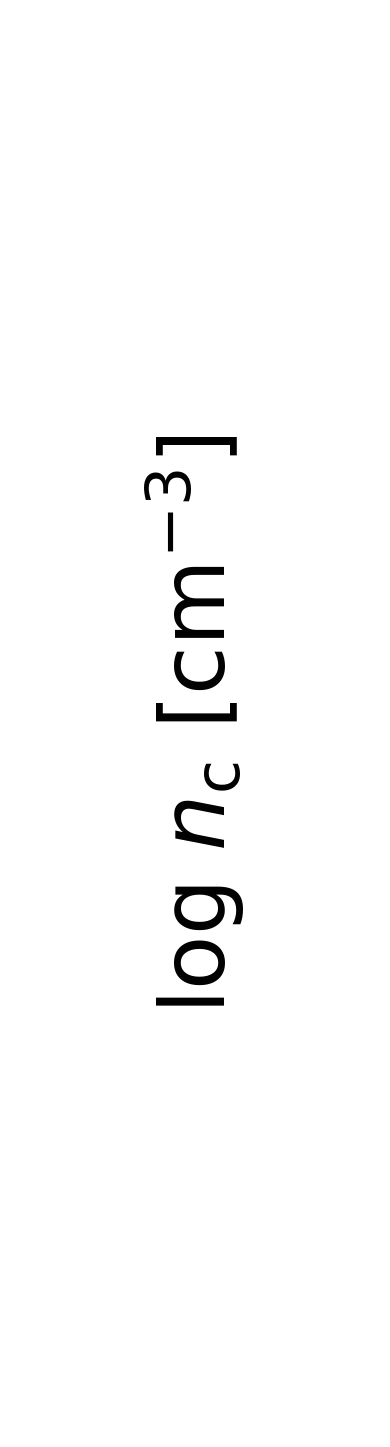}
\end{minipage}
\begin{minipage}{0.93\textwidth}
\includegraphics[width=0.49\columnwidth]{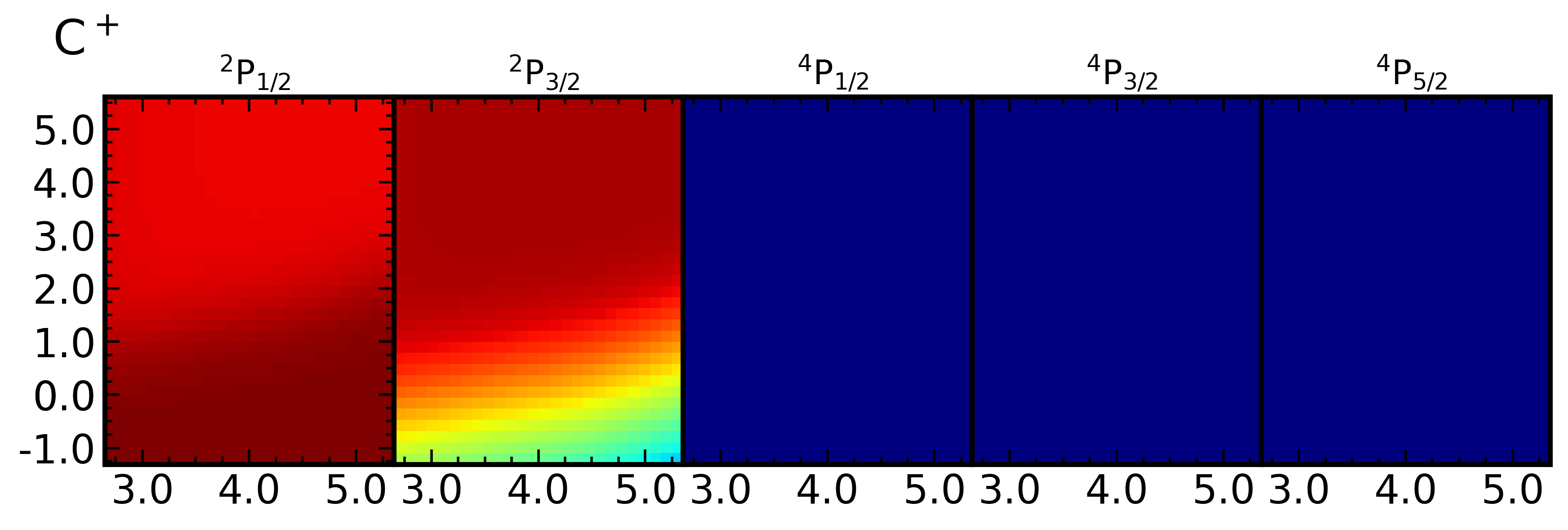}\includegraphics[width=0.49\columnwidth]{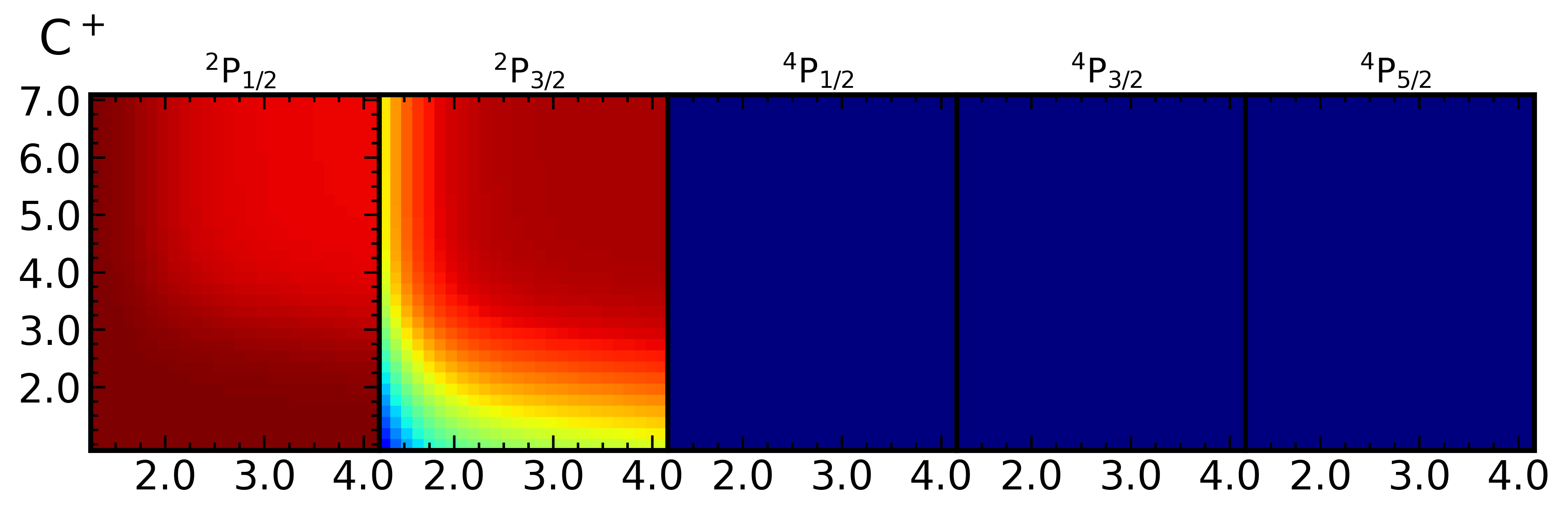}
\includegraphics[width=0.49\columnwidth]{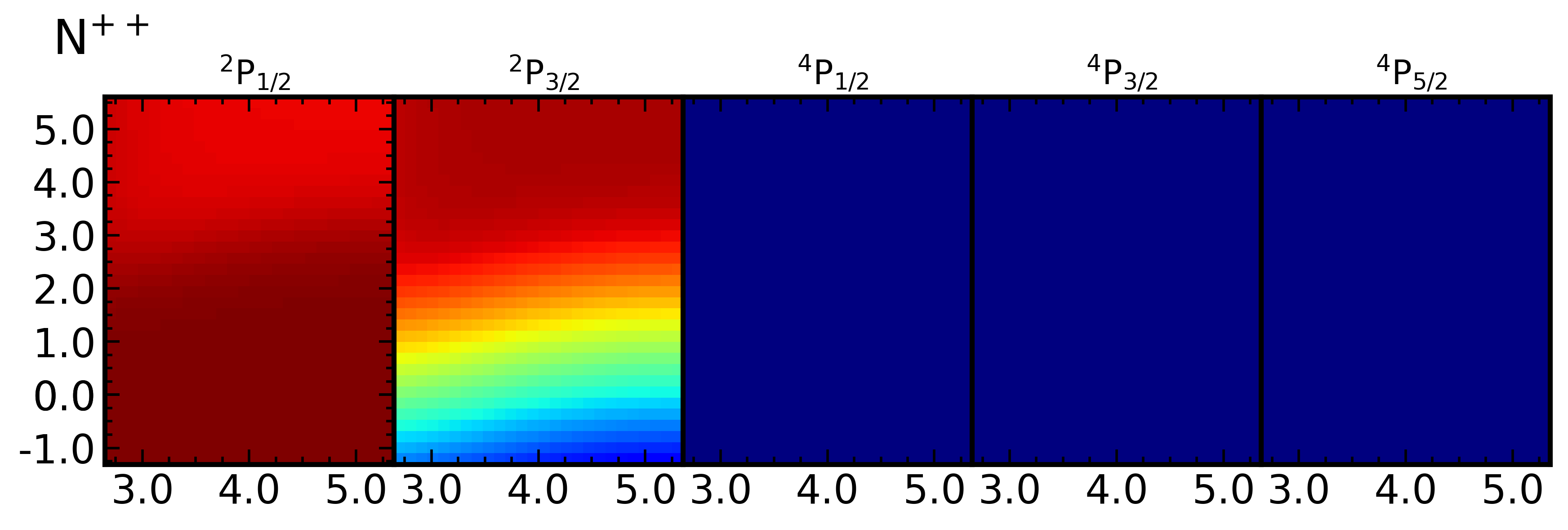}
\includegraphics[width=0.49\columnwidth]{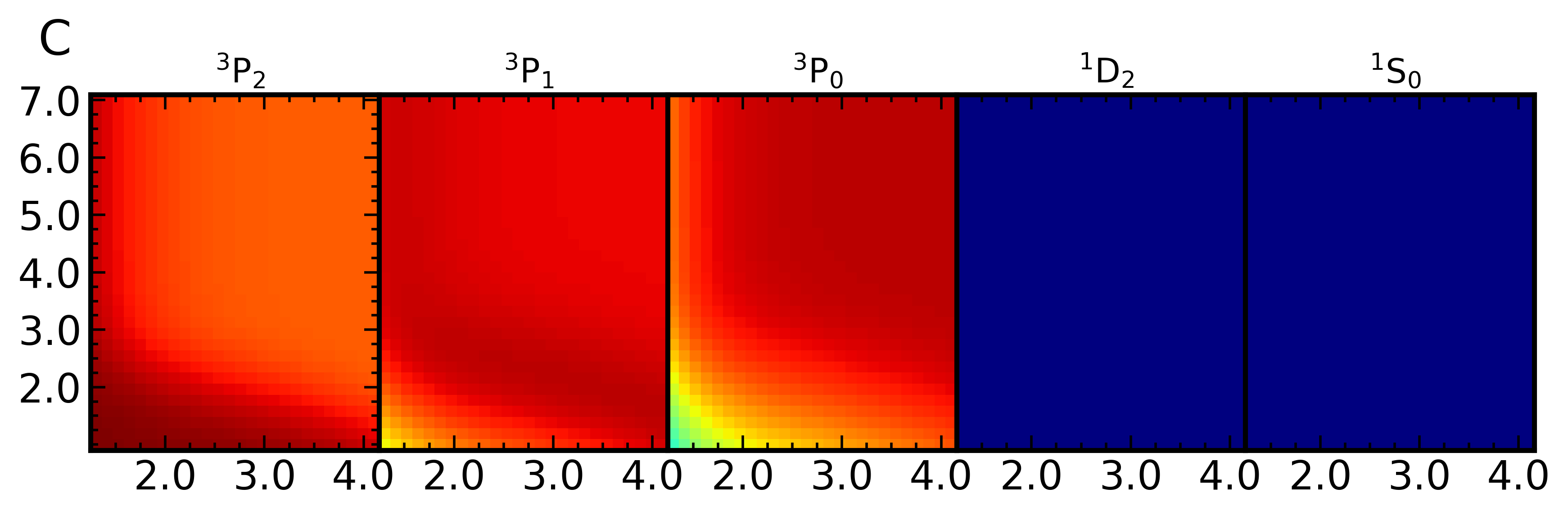}
\includegraphics[width=0.49\columnwidth]{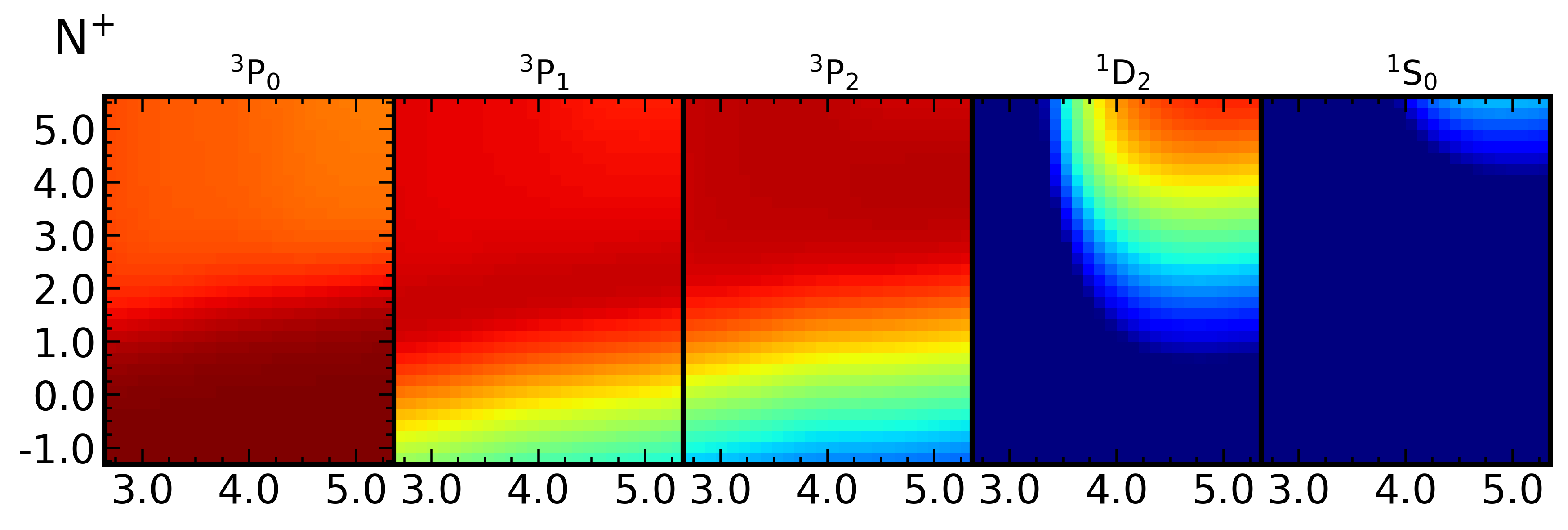}
\includegraphics[width=0.49\columnwidth]{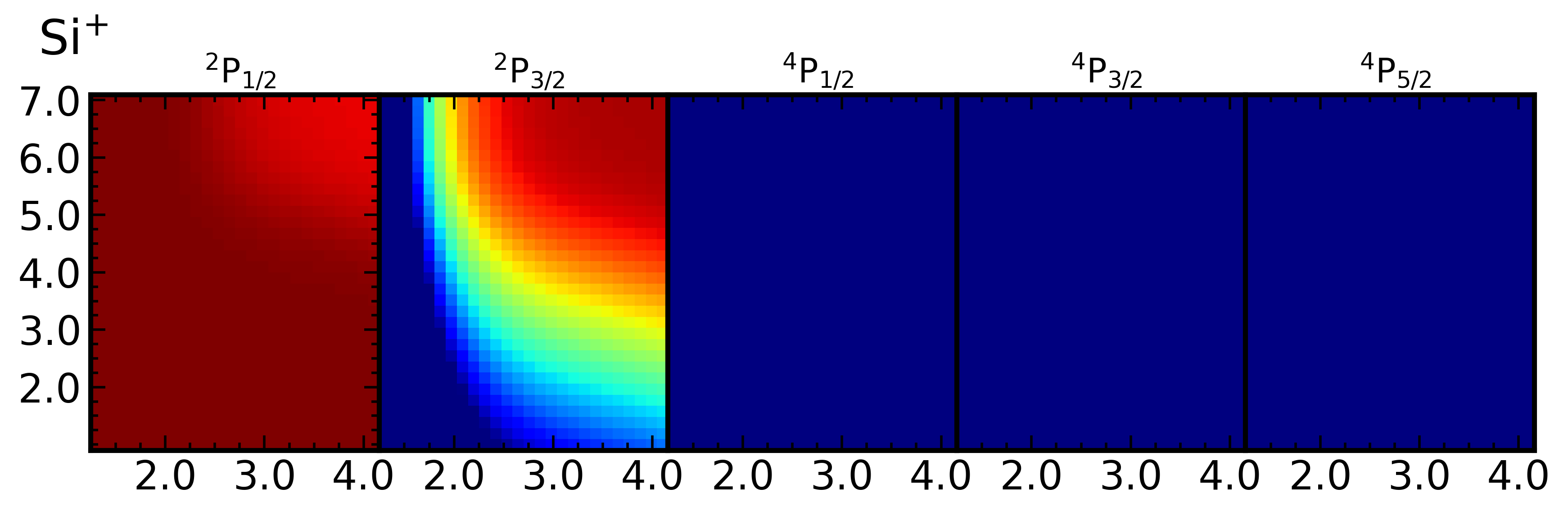}
\includegraphics[width=0.49\columnwidth]{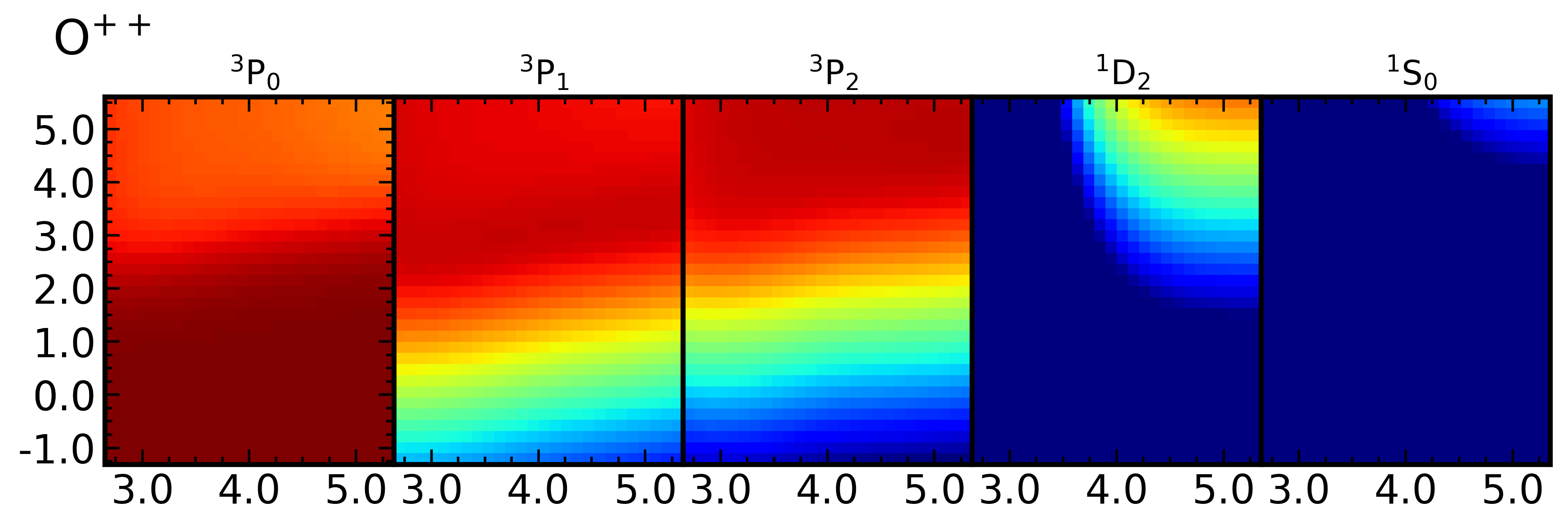}
\includegraphics[width=0.49\columnwidth]{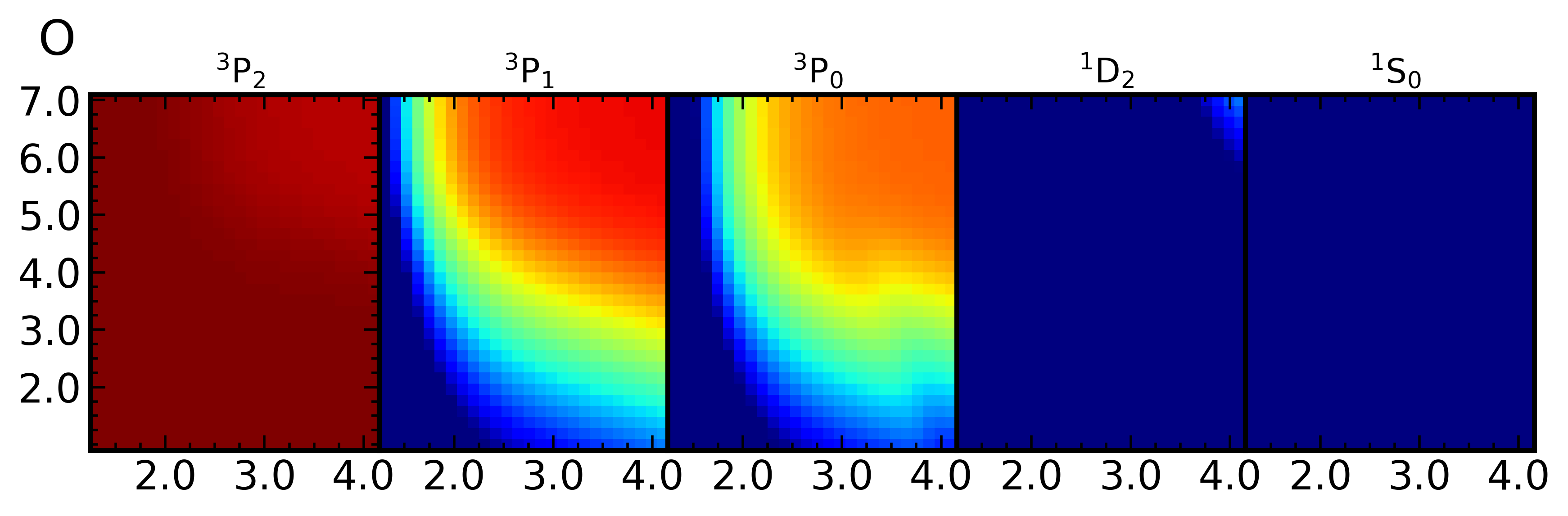}
\includegraphics[width=0.49\columnwidth]{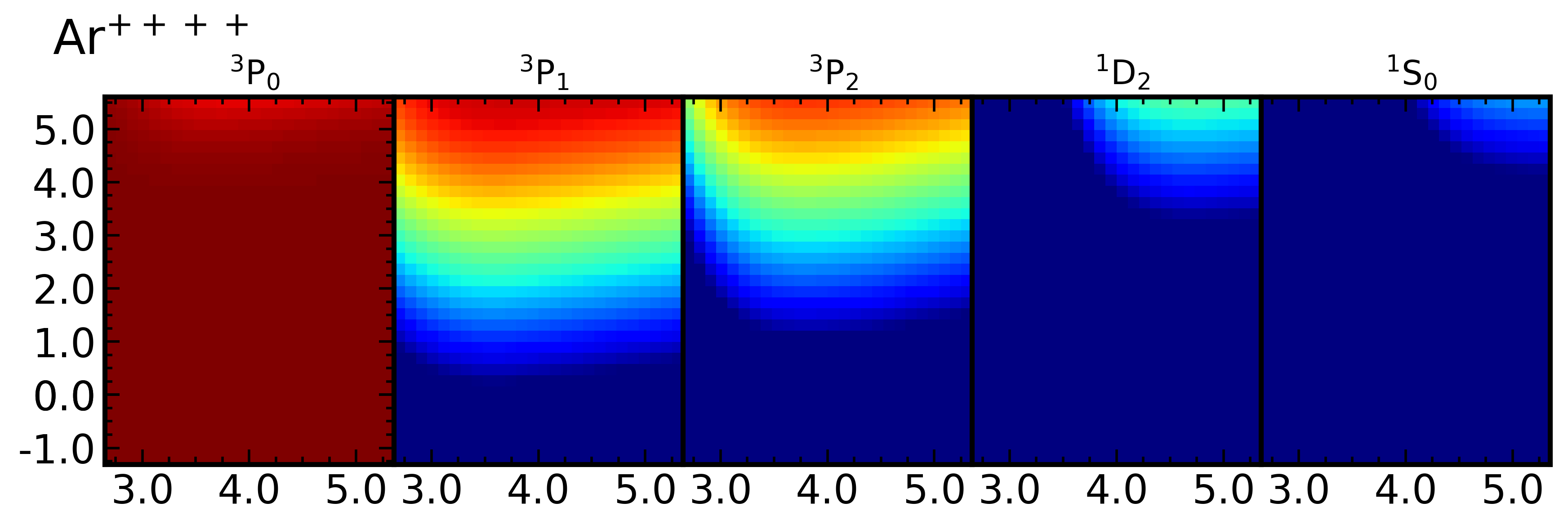}
\includegraphics[width=0.49\columnwidth]{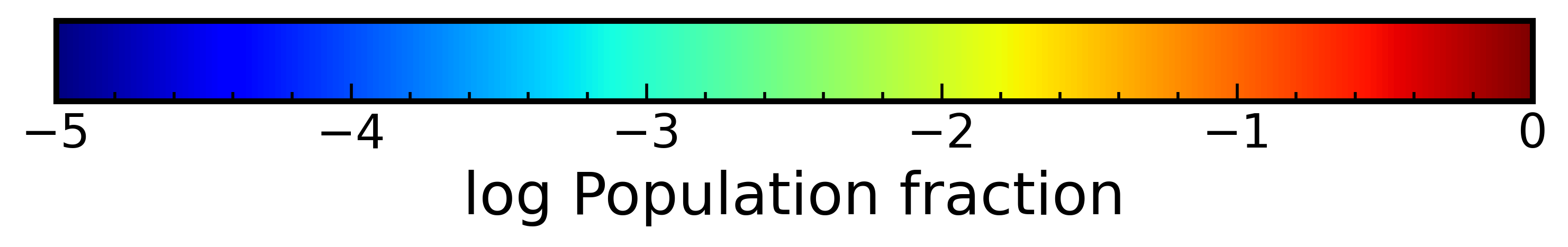}
\includegraphics[width=0.49\columnwidth]{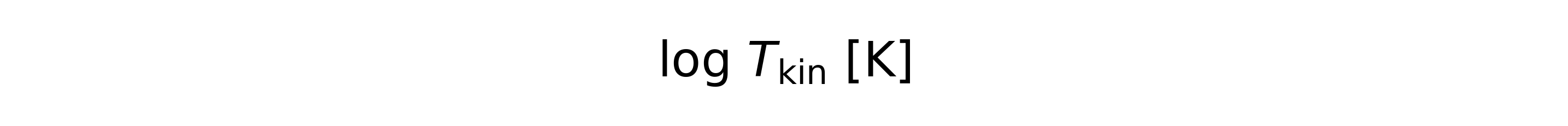}
\includegraphics[width=0.49\columnwidth]{label_lgT.png}
\end{minipage}
\end{center}
\caption{Examples of population fraction $f_i(n_c,T_\text{kin}) = n_i/\sum_i n_i = n_i/n_X$ of the first five energy levels of ions and atoms with fine-structure lines, as a function of density of the colliders, $n_c$, and kinetic temperature $T_\text{kin}$. Left-hand panels show species excited via collisions with electrons; right-hand panels show the case of a neutral medium where collisions are set by a mixture of 50\% H atoms, and 50\% H$_2$ molecules, with temperature-dependent para-to-ortho ratio as in \citet{mcintosh15}. These analytically-derived population fractions can be used to directly infer the expected line luminosities, based on Eq.~\eqref{eq_linelum_theo}. 
}
\label{fig_popfrac}
\end{figure}

Assuming a thermal distribution for the collisional partners, it is possible to analytically infer the balance of excitation and de-excitation of a given species, $X$, by writing a set of Eqs.~\eqref{eq_Rul}--\eqref{eq_Rlu} for each pair of levels $u,l$, and imposing the equilibrium condition in Eq.~\eqref{eq_radtransf}. For two/three level cases, an analytical solution is straightforward. For more levels, it is convenient to solve the system of equations by numerical means. Figure~\ref{fig_popfrac} shows the resulting population fraction in the energy level $i$, $f_i(n_c,T_\text{kin})$, in a 5-energy-levels scenario of elements and ions as a function of the collider density $n_c$ and temperature $T_\text{kin}$, for collisions with electrons (\emph{left} panel), as well as for hydrogen atoms and molecules (\emph{right} panel). Once the system of Eqs.~\eqref{eq_Rul}--\eqref{eq_Rlu} is solved for a given combination of $n_c$ and $T_\text{kin}$, one can infer the associated emissivity (luminosity per unit volume) of a transition $u\rightarrow l$ of the chemical species $X$ as:
\begin{equation}\label{eq_emissivity}
j_{ul}=h \nu_{ul}\,A_{ul}\,n_X\,f_u(n_c,T_\text{kin}),
\end{equation}
where $n_X$ is the number density of the species $X$, and:
\begin{equation}\label{eq_abundances}
n_X = \sum_{i}\,n_i = n_c \frac{n_A}{n_c} \frac{n_X}{n_A}, 
\end{equation}
where $n_A/n_c$ is the relative abundance of the element $A$ with respect to that of the collisional partner, and $n_X/n_A$ is the fraction of atoms of $A$ that are in the species $X$. Under the assumption of optically-thin emission, we can integrate Eq.~\eqref{eq_emissivity} over the source volume, and infer the line luminosity
\begin{equation}\label{eq_linelum_theo}
L_{ul}=h \nu_{ul}\,A_{ul}\,f_u(n_c,T_\text{kin})\,\frac{M_{X}}{m_X},
\end{equation}
where $f_u$ is the population fraction at level $u$, $f_u\,=\,n_u/n_X$, $M_{X}$ is the total mass of $X$, $m_X$ is the mass of a single atom/ion of $X$, and $M_X/m_X$\,=\,$\int n_X\,dV$ is the total number of atoms/ions $X$ in the volume $V$ occupied by the gas.

Because the analytical expression of $f_u$ resulting from solving the set of equations of statistical equilibrium for a system of more than two levels is typically complex, it is useful to define the relative population of two energy levels, $n_u/n_l$, in terms of the expression given by a Boltzmann distribution that depends only on a single variable called excitation temperature, $T_{\rm ex}$, defined so that it satisfies the condition:
\begin{equation}\label{eq_def_Tex}
\frac{n_u}{n_l}=\frac{g_u}{g_l}\exp\left(-\frac{E_{ul}}{k_{\rm b}T_{\rm ex}}\right),
\end{equation}
where $g_i=2\,J+1$ is the statistical weight of level $i$, and:
\begin{equation}\label{eq_def_nu}
f_u=\frac{n_u}{n_X}=\frac{g_u}{Q(T_{\rm ex})}\exp\left(-\frac{E_{u}}{k_{\rm b}T_{\rm ex}}\right).
\end{equation}

The introduction of the excitation temperature via these functional forms, together with the partition function, $Q(T_{\rm ex})$\,=\,$\sum_{i}\,g_i$\,exp($-E_i/k_{\rm b}T_{\rm ex}$), enables to encapsulate the bidimensional $f(n,T)$ dependence in a single parameter, $T_{\rm ex}$. By construction, $T_{\rm ex}\approx T_\text{kin}$ in the high density regime, at $n\gg n_{\rm crit, u}$. At low densities, on the other hand, $T_\text{kin}>T_{\rm ex}$. Generally speaking, unless local thermodynamic equilibrium is achieved, $T_{\rm ex}$ does not have a proper physical meaning. Radiative processes such as pumping or stimulated absorption/emission can alter the level population, and hence $T_{\rm ex}$, even to the point where \emph{negative} excitation temperatures are observed for inverted population levels.

Adapting Eqs.~\eqref{eq_emissivity}--\eqref{eq_linelum_theo} to account for collisions with other partners (e.g. hydrogen atoms and molecules) is trivial, once we adopt the corresponding collision partner densities $n_c$ and collision strength factors in Eq.~\eqref{eq_kul}. In the \emph{right} panel of Fig.~\ref{fig_popfrac} we show examples of level population as a function of density and temperature, assuming that collision partners are 50\% hydrogen atoms, H, and 50\% hydrogen molecules, H$_2$, with a temperature-dependent para-to-ortho ratio set according to \citet{mcintosh15}. However, one should keep in mind that this approach tends to oversimplify the physics of the neutral medium, where multiple gas phases could co-exist and the chemistry of the clouds becomes increasingly more complex. Proper radiative transfer calculations are often necessary in these regimes.

\subsection{Critical densities}\label{sec_ncrit}

We can rearrange Eq.~\eqref{eq_Rul} in order to assess whether a given level $u$ is depopulated mainly via collisions or via radiative de-excitation. The former channel dominates when the density is larger than the critical density, $n_{\rm crit, u}$:
\begin{equation}\label{eq_ncrit}
n_{\rm crit,u}=\frac{\Sigma_{l<u}(1+n_{\gamma})A_{ul}}{\Sigma_{i \neq u}k_{ui}(T_\text{kin})}.
\end{equation}
The numerator in Eq.~\eqref{eq_ncrit} sums over all of the energy levels $l$ with energy below level $u$, since radiative de-excitation would lead to a lower energy level; on the other hand, the denominator sums over all the energy levels except for $u$, as collisions might lead to an even higher excitation of the valence electron. Table~\ref{tab_ncrit} lists the critical densities of various transitions tracing the ionized and the neutral medium. In Sect.~\ref{sec_ne}, we will leverage on the critical densities of different transitions in order to infer diagnostics of the gas density.

\begin{figure}[ht]
\begin{center}
\includegraphics[width=0.49\columnwidth]{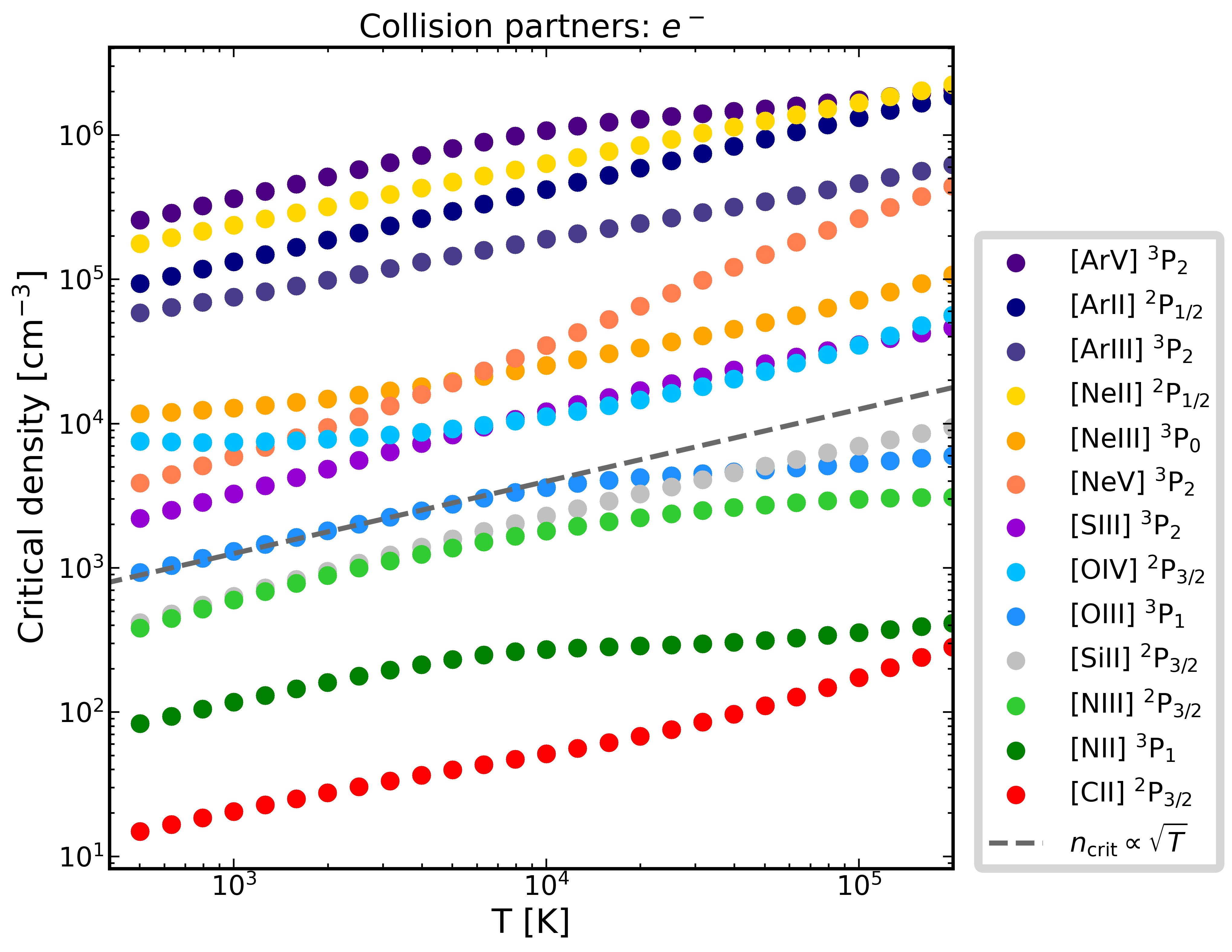}
\includegraphics[width=0.49\columnwidth]{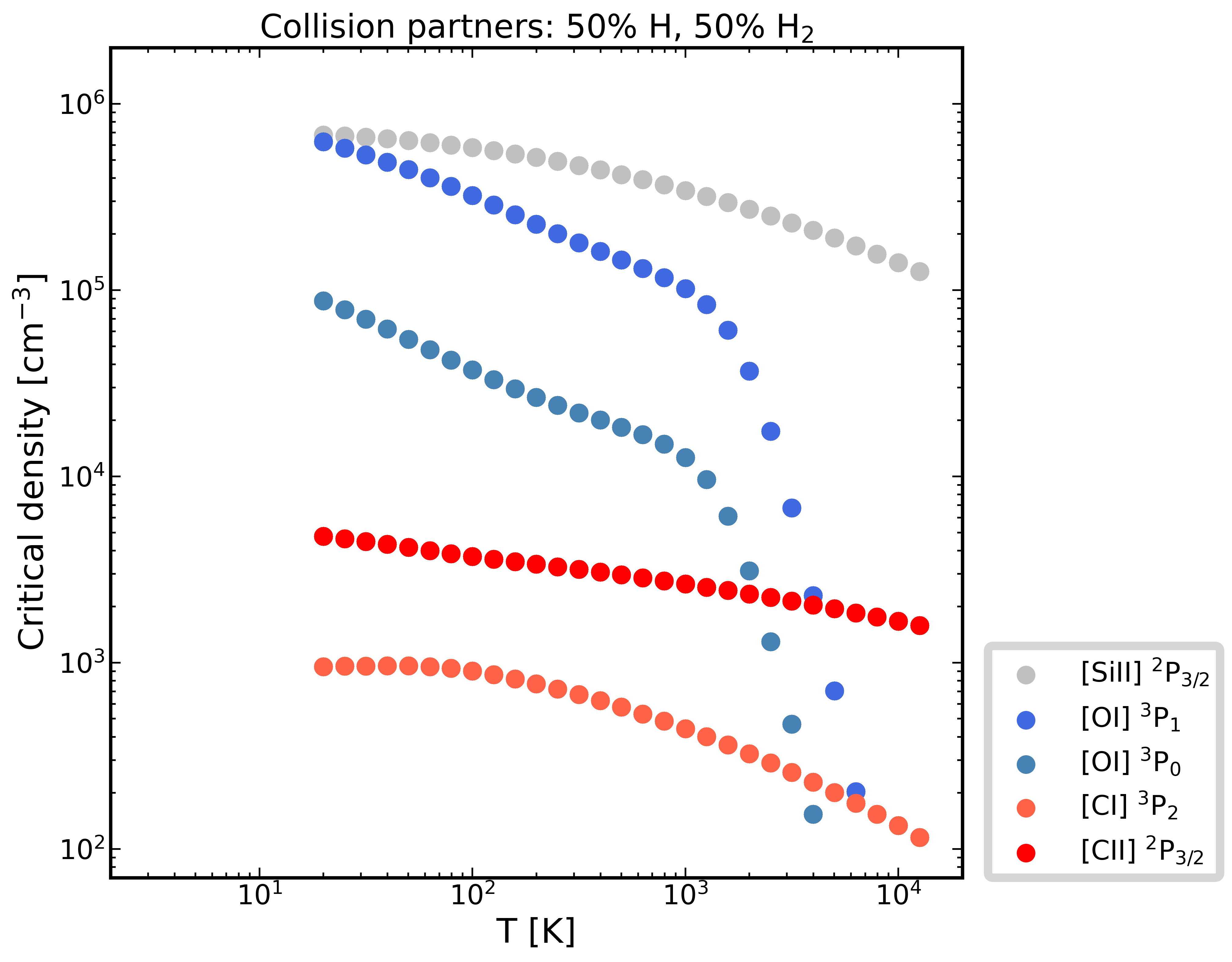}
\end{center}
\caption{Critical density, $n_{\rm crit}$, as a function of kinetic temperature $T_\text{kin}$, for various energy levels of atoms and ions presenting fine-structure lines. The left-hand panel shows the case of electrons as collisional partners for species present in ionized gas; the right-hand panel considers a mixture of 50\% hydrogen atoms, 50\% H$_2$ molecules, and a para-to-ortho ratio set by the temperature $T_\text{kin}$ following \citet{mcintosh15}, for species typical of the neutral and molecular medium. For collisions with electrons, the critical density scales to first order as $n_{\rm crit}\propto \sqrt{T_\text{kin}}$. For collisions with hydrogen atoms and molecules, the critical density typically decreases at increasing $T_\text{kin}$ in a non-linear fashion.
}
\label{fig_ncrit}
\end{figure}

\begin{figure}[ht]
\begin{center}
\includegraphics[width=0.49\columnwidth]{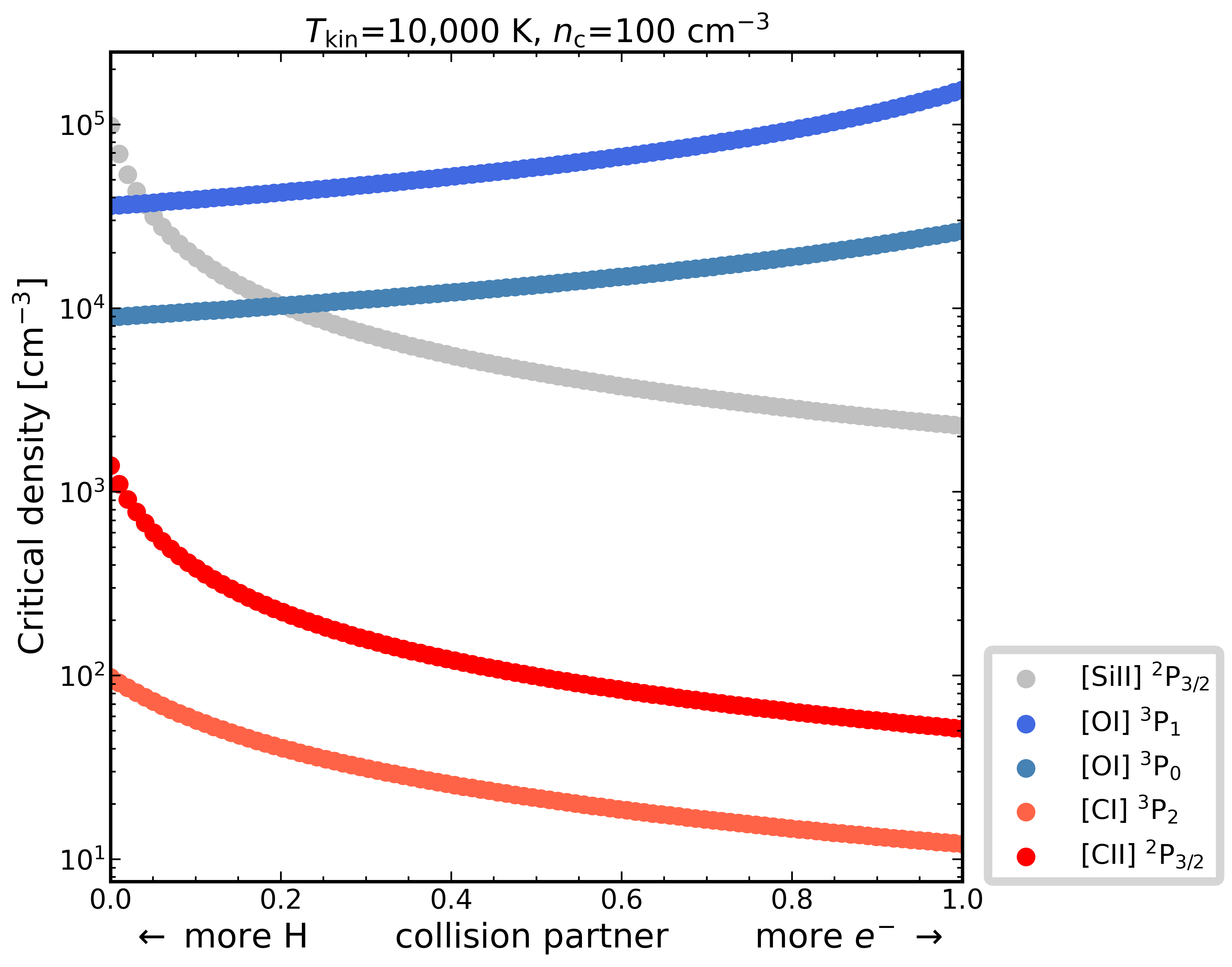}
\includegraphics[width=0.49\columnwidth]{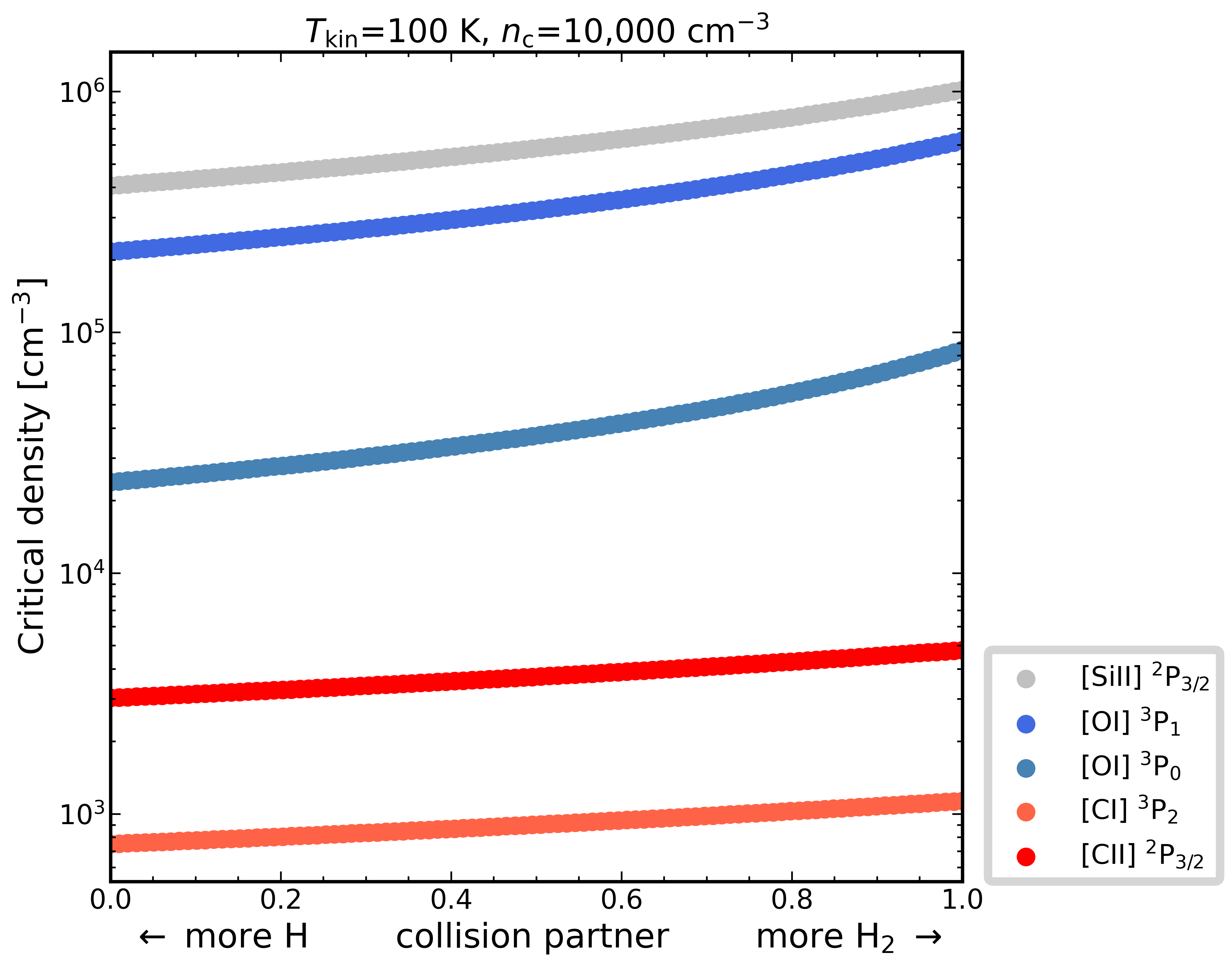}
\end{center}
\caption{Critical density $n_\text{crit}$ as a function of collision partners, computed for a fixed kinetic temperature and density. \emph{Left:} Fraction of collisions due to electrons vs.\ neutral hydrogen atoms. Increasing the contribution of electrons lowers the critical density for some elements and raises it for others.  \emph{Right:} Fraction of collisions of molecular hydrogen vs.\ neutral hydrogen. In this case, increasing the fraction of collisions with H$_2$ molecules with respect to H atoms raises the critical density, due to the larger collisional strength of the H$_2$ molecule.}
\label{fig_ncrit_partners}
\end{figure}

A common misconception is that $n_\text{crit}$ is a fixed number for a given transition. Instead, even ignoring stimulated emission, Eq.~\eqref{eq_ncrit} shows that $n_\text{crit}$ is sensitive to the collision rates, which depend on the collision partners and carry a $T_\text{kin}^{-1/2}$ scaling from Eq.~\eqref{eq_kul}, together with an implicit dependency on $T_\text{kin}$ of the collision strengths $\Omega(l,u)$ (see Appendix F of \citealt{draine11}). Figure~\ref{fig_ncrit} shows the temperature dependence of the critical density of various levels of ions and elements. In the case of collisions with electrons, as mentioned above, the critical density scales roughly as $n_\text{crit}\propto\sqrt{T_\text{kin}}$. In the case of collisions with hydrogen atoms and molecules, the critical density tends to decrease at increasing $T_\text{kin}$ in a non-linear fashion, due to the relative scaling of collision strengths. Figure~\ref{fig_ncrit_partners} shows how the critical density changes as a function of collision partners, assuming fixed $T_\text{kin}$ and $n_c$. 

\begin{figure}[ht]
\begin{center}
\includegraphics[width=0.49\columnwidth]{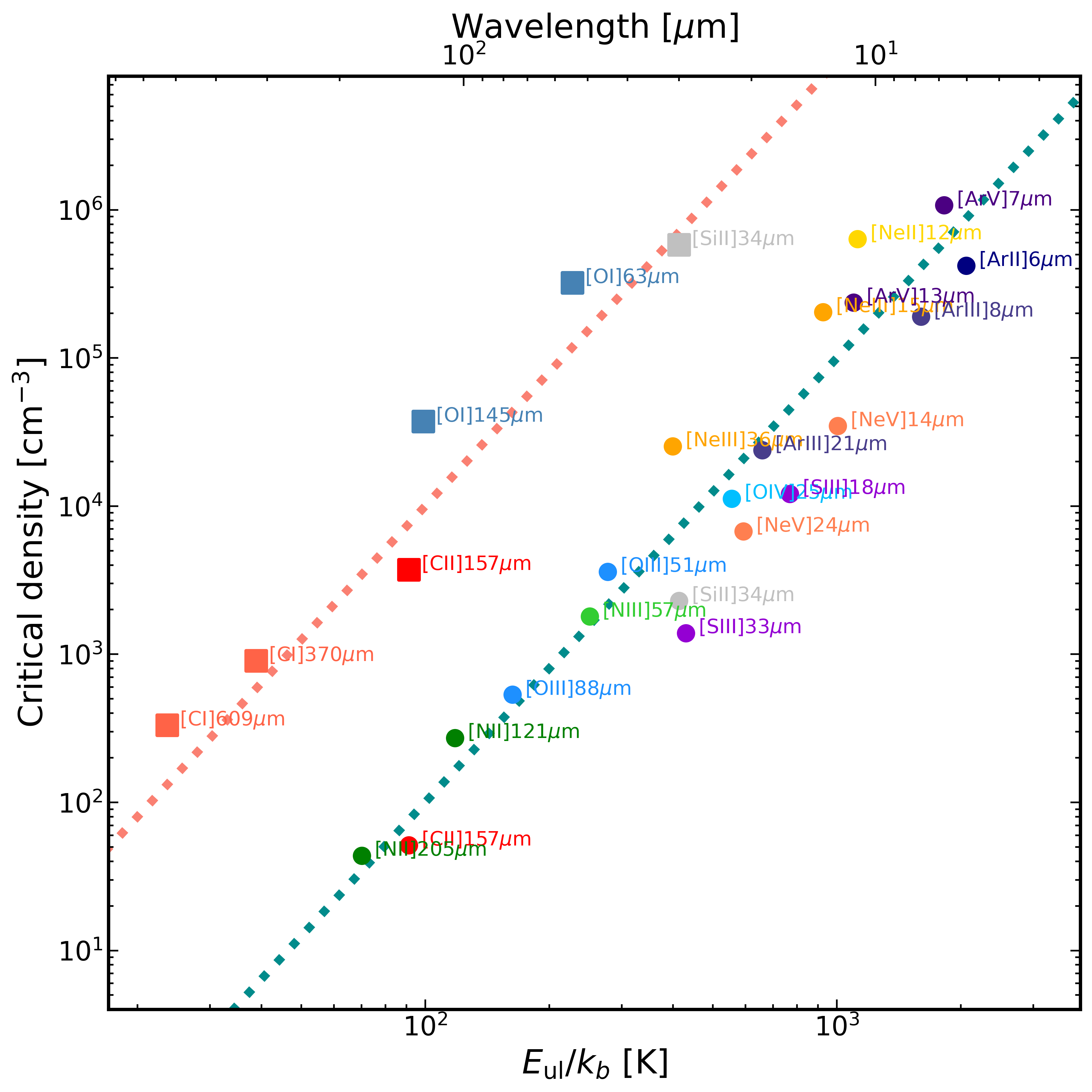}
\end{center}
\caption{Critical density $n_\text{crit}$ as a function of the energy difference of the involved levels, $E_{ul}$ (bottom axis) and of the corresponding wavelength of the emitted line (top axis). Critical densities inferred from heavier collision partners (as in the neutral medium; marked with squares in the plot) tend to be 100$\times$ larger than in the case of electrons as collision partners (circles). Critical densities are computed at $T$=10,000\,K in the case of collisions with electrons, and at $T$=100\,K in the case of collisions with neutral atoms and molecules. The transitions of \Cii{} 158\,$\mu$m and [Si{\sc ii}] 34\,$\mu$m are shown for both cases. Transitions at shorter wavelengths (implying larger energy differences between the levels) show larger critical densities, as highlighted by the dotted lines marking the case $n_\text{crit}\propto E_{ul}^3$. }
\label{fig_ncrit_Eul}
\end{figure}

It is worth noting that the rate of collisional excitation in Eq.~\eqref{eq_Rlu} is a function of the energy \textit{difference} between the two levels, $E_{ul}$, normalized to the thermal energy of the collision partners, $k_\text{b} T_\text{kin}$. As a consequence, in the case of three–level splitting of the ground state, the higher energy level might reach temperature saturation at relatively lower temperatures than the first excited level. Fig.~\ref{fig_ncrit_Eul} shows that the critical density is typically higher for transitions involving the largest energy differences
(shorter wavelengths).

\begin{table}[htbp]
\caption{Quantum numbers, energies, statistical weights, Einstein coefficients, and critical densities of various FSL transitions. Critical densities are computed at \iTkin\,=\,10,000\,K for species tracing the ionized medium, assuming electrons as collision partners; and at \iTkin\,=\,100\,K assuming that collisions partners are 50\% neutral hydrogen atoms, and 50\% molecular hydrogen (H$_2$) for lines arising from the neutral medium, in the case of negligible radiation field $n_\gamma=0$. Column 11 lists the conversion factors $\mathcal{A}$ to convert line luminosities into masses following Eq.~\eqref{eq_mass}.}
\label{tab_ncrit}
\begin{tabular}{cccccccccccc}
\hline
Species  &  Transition & $\lambda$ & $\nu$ & $E_{ul}/k_{\rm b}$ & $E_{u}/k_{\rm b}$ & $E_{l}/k_{\rm b}$ & $g_u$ & $g_l$ & $A_{ul}$ & $n_\text{crit}$  & $\mathcal{A}$ \\
         &             & [$\mu$m]  & [GHz] & [K] & [K] & [K] &  &  & [s$^{-1}$] & [cm$^{-3}$] &  [M$_\odot$/L$_\odot$] \\
(1)      & (2)         & (3)       &  (4)  & (5) & (6) & (7) & (8) & (9) & (10) & (11) & (12) \\
\hline
\multicolumn{12}{l}\textit{Ionized medium: \iTkin\,=\,10,000\,K, collision partners: $e^-$}\\
\Nii{}   & 1$\rightarrow$0     & 205.2  &  1461.132  &   70.1 &    70.1  &	  0    & 3 & 1 & $2.1\times10^{-6}$ &	43.7	&  $1.99\times 10^{-4} $  \\
\Cii{}   & 3/2$\rightarrow$1/2 & 157.7  &  1900.537  &   91.3 &    91.3  &	  0    & 4 & 2 & $2.3\times10^{-6}$ &	51.4	&  $8.97\times 10^{-5} $  \\
\Nii{}   & 2$\rightarrow$1     & 121.9  &  2459.380  &    118 &     188  &	 70.1   & 5 & 3 & $7.5\times10^{-6}$ &	273	&  $1.99\times 10^{-5} $  \\
\Oiii{}  & 1$\rightarrow$0     & 88.36  &  3393.006  &    163 &     163  &	  0    & 3 & 1 & $2.7\times10^{-5}$ &	534	&  $7.61\times 10^{-6} $  \\
\Niii{}  & 3/2$\rightarrow$1/2 & 57.34  &  5230.428  &    251 &     251  &	  0    & 4 & 2 & $4.7\times10^{-5}$ &	1801	&  $1.86\times 10^{-6} $  \\
\Oiii{}  & 2$\rightarrow$1     & 51.81  &  5785.880  &    278 &     441  &	 163   & 5 & 3 & $9.7\times10^{-5}$ &	3612	&  $7.46\times 10^{-7} $  \\
\Neiii{} & 0$\rightarrow$1     & 36.01  &  8324.447  &    399 &    1325  &	 924   & 1 & 3 & $1.0\times10^{-3}$ &	25404	&  $3.14\times 10^{-7} $  \\
\Siii{}  & 1$\rightarrow$0     & 33.48  &  8954.107  &    430 &     430  &	  0    & 3 & 1 & $4.7\times10^{-4}$ &	1391	&  $3.31\times 10^{-7} $  \\
\Oiv{}   & 3/2$\rightarrow$1/2 & 25.89  &  11579.33  &    555 &     555  &	  0    & 4 & 2 & $5.2\times10^{-4}$ &	11242	&  $8.69\times 10^{-8} $  \\
\Nev{}   & 1$\rightarrow$0     & 24.32  &  12328.26  &    593 &     593  &	  0    & 3 & 1 & $1.3\times10^{-3}$ &	6753	&  $5.43\times 10^{-8} $  \\
\Ariii{} & 0$\rightarrow$1     & 21.84  &  13732.92  &    659 &    2260  &	 1600  & 1 & 3 & $5.2\times10^{-3}$ &	23914	&  $6.60\times 10^{-8} $  \\
\Siii{}  & 2$\rightarrow$1     & 18.71  &  16020.55  &    769 &    1199  &	  430  & 5 & 3 & $2.1\times10^{-3}$ &	12110	&  $2.49\times 10^{-8} $  \\
\Neiii{} & 1$\rightarrow$2     & 15.55  &  19272.94  &    925 &     925  &	  0    & 3 & 5 & $6.0\times10^{-3}$ &	204410  &  $7.54\times 10^{-9} $  \\
\Nev{}   & 2$\rightarrow$1     & 14.32  &  20932.74  &   1005 &    1597  &	  592  & 5 & 3 & $4.6\times10^{-3}$ &	34799	&  $5.43\times 10^{-9} $  \\
\Arv{}   & 1$\rightarrow$0     & 13.10  &  22881.08  &   1098 &    1098  &	  0    & 3 & 1 & $8.0\times10^{-3}$ &	237363  &  $8.57\times 10^{-9} $  \\
\Neii{}  & 1/2$\rightarrow$3/2 & 12.81  &  23396.52  &   1123 &    1123  &	  0    & 2 & 4 & $8.6\times10^{-3}$ &	634802  &  $6.50\times 10^{-9} $  \\
\Ariii{} & 1$\rightarrow$2     & 8.991   &  33342.21  &   1600 &    1600  &	  0    & 3 & 5 & $3.1\times10^{-2}$ &	190656  &  $1.52\times 10^{-9} $  \\
\Arv{}   & 2$\rightarrow$1     & 7.902   &  37940.73  &   1837 &    2936  &  1098 &  5 & 3 & $2.7\times10^{-2}$ &	1074695 &  $9.19\times 10^{-10}$  \\
\Arii{}  & 1/2$\rightarrow$3/2 & 6.985   &  42917.79  &   2060 &    2060  &	  0    & 2 & 4 & $5.3\times10^{-2}$ &	419254  &  $1.04\times 10^{-9} $  \\
\hline				  
\multicolumn{12}{l}\textit{Neutral medium: \iTkin\,=\,100\,K, collision partners: 50\% H, 50\% H$_2$}\\ 	   			    
\Ci{}    & 1$\rightarrow$0     & 609.7  &   491.705  &   23.6 &   23.6   &	 0     & 3 & 1 & $7.9\times10^{-8}$ &	331.7	&  $1.35\times 10^{-2} $  \\
\Ci{}    & 2$\rightarrow$1     & 370.4  &   809.375  &   38.8 &    62.4   &	 23.6  & 5 & 3 & $2.7\times10^{-7}$ &	902.9	&  $1.44\times 10^{-3} $  \\
\Cii{}   & 3/2$\rightarrow$1/2 & 157.7  &  1900.537  &   91.3 &   91.3   &	 0     & 4 & 2 & $2.3\times10^{-6}$ &	3716.8  &  $8.97\times 10^{-5} $  \\
\Oi{}    & 0$\rightarrow$1     & 145.5  &  2060.069  &   98.9 &    326   &	 227   & 1 & 3 & $1.8\times10^{-5}$ &	37271	&  $5.64\times 10^{-5} $  \\
\Oi{}    & 1$\rightarrow$2     & 63.18  &  4744.775  &   227  &    227   &	 0     & 3 & 5 & $8.9\times10^{-5}$ &	322002  &  $1.65\times 10^{-6} $  \\
\Silii{} & 3/2$\rightarrow$1/2 & 34.81  &  8610.965  &   413  &    413   &	 0     & 4 & 2 & $2.2\times10^{-4}$ &	582010  &  $4.83\times 10^{-7} $  \\
\hline
\end{tabular}
\end{table}

\subsection{Fractional abundance of atomic species}\label{sec_photoionization}

\begin{figure}[htbp]
\begin{center}
\includegraphics[width=0.49\columnwidth]{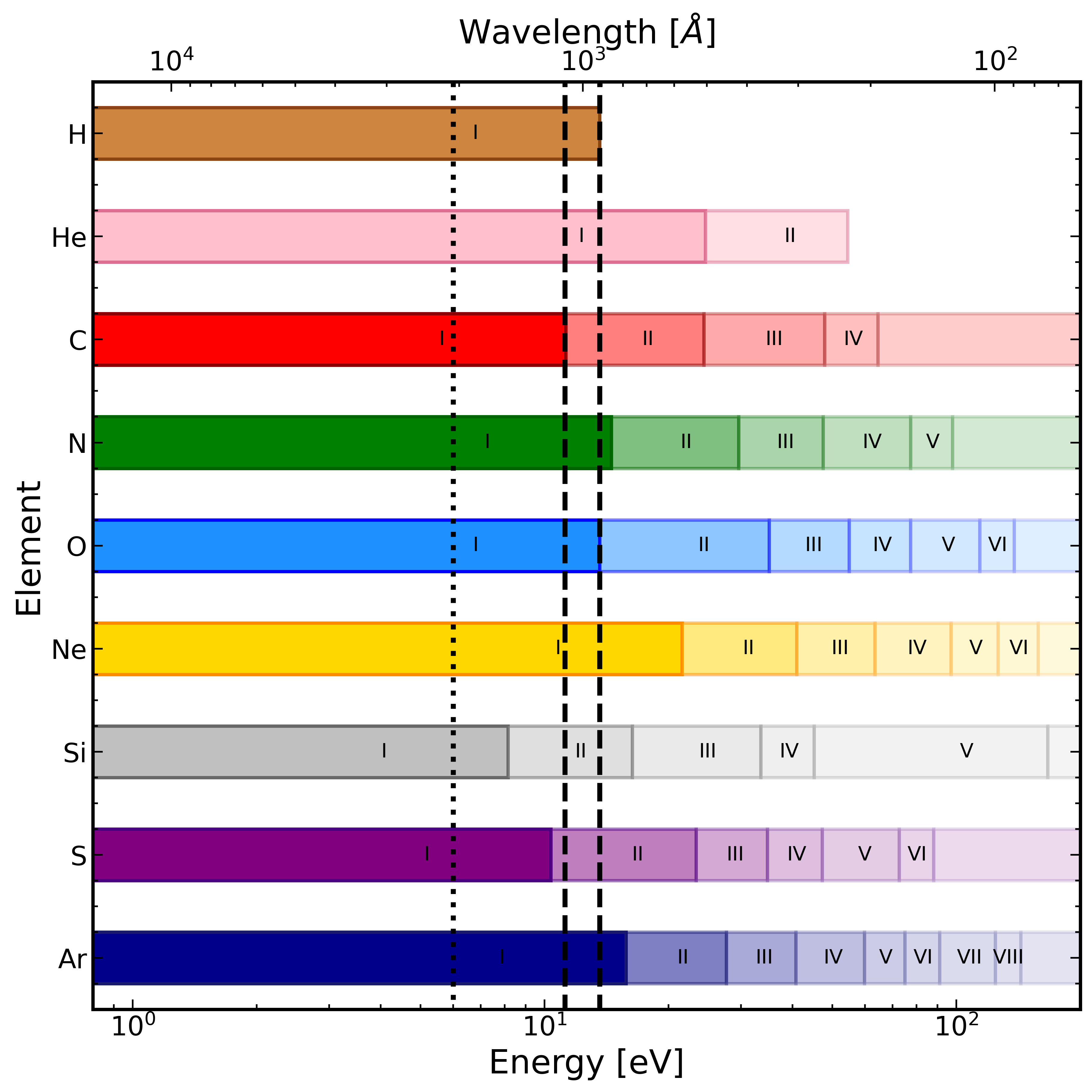}
\includegraphics[width=0.49\columnwidth]{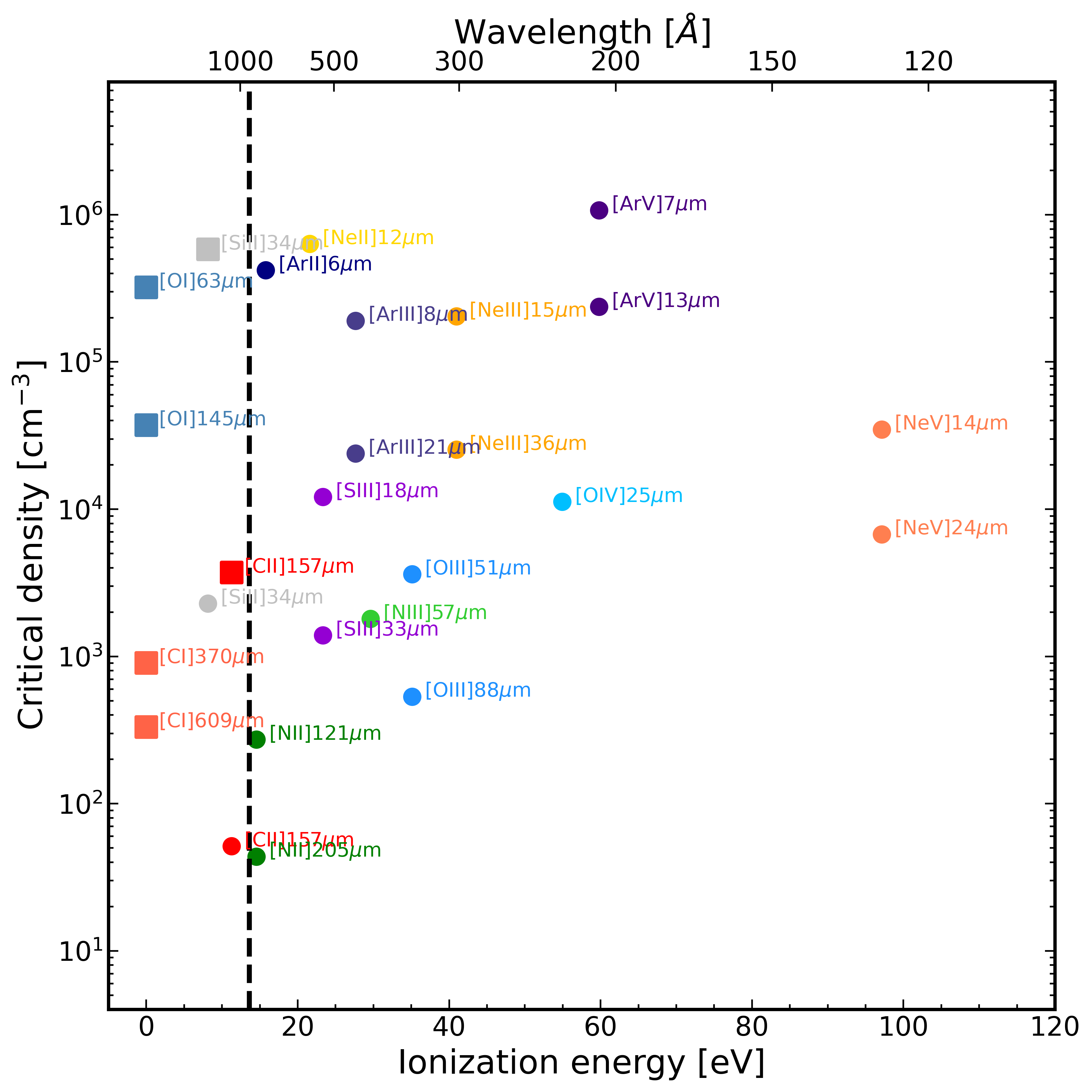}
\end{center}
\caption{\emph{Top:} The ionization energy of various elements. The vertical lines mark the ionization energy of hydrogen and carbon (dashed style; 13.6 and 11.3\,eV respectively), and the 6\,eV limit used in the definition of Habing fluxes \citep[dotted style; see][]{habing68}. \emph{Bottom:} Ionization energy and critical density of various fine-structure line transitions \citep[see][]{spinoglio92}. The vertical line marks the hydrogen ionization energy. Squares mark critical densities computed assuming collisions with H atoms (50\%) and H$_2$ molecules (50\%), and a temperature of 100\,K; circles mark critical densities computed assuming collisions with electrons and a temperature of 10,000\,K.}
\label{fig_ionization_energy}
\end{figure}

For FSLs arising from the ionized medium, line emission depends on the ionization structure of the gas cloud, which sets the fraction of atoms of a specific element $A$ that are in the species of interest $X$ (the term $n_X/n_A$ in Eq.~\eqref{eq_abundances}). While photoionization is generally the main mechanism responsible for setting the ionization balance in the medium, there are other processes such as shocks and X-ray/cosmic-ray heating that can play a role as well, under specific physical conditions; for more details we refer the interested reader to \citet{wolfire03, osterbrock06}. Figure~\ref{fig_ionization_energy}, \emph{top}, visualizes the ionization energies of the most common elements in the ISM, and Table~\ref{tab_ion_energy} provides a list of the first four ionization potentials for the same elements. In Fig.~\ref{fig_ionization_energy}, \emph{bottom}, we combine the information of both ionization energy and critical density for the different FSL transitions \citep[see][]{spinoglio92}.

\begin{table}[ht]
\caption{Ionization energy of relevant elements. Column (2) lists the atomic number, Z. Columns (3--6) refer to first, second, third and fourth ionization energies, respectively. All the energies are quoted in eV.}\label{tab_ion_energy}
\begin{tabular}{ccccccc}
\hline
Element  &  Z & 1st & 2nd & 3rd & 4th \\
(1)      & (2)& (3)          & (4)          & (5)          & (6)          \\
\hline
Hydrogen &  1 & 13.598      &               &              &              \\ 
Helium   &  2 & 24.587      & 54.418        &              &              \\ 
Carbon   &  6 & 11.260      & 24.383        & 47.888       & 64.494       \\ 
Nitrogen &  7 & 14.534      & 29.601        & 47.449       & 77.474       \\ 
Oxygen   &  8 & 13.618      & 35.117        & 54.936       & 77.414       \\ 
Neon     & 10 & 21.565      & 40.963        & 63.423       & 97.12        \\ 
Silicon  & 14 &  8.152      & 16.346        & 45.142       & 166.767      \\ 
Sulphur  & 16 & 10.360      & 23.338        & 34.790       & 47.222       \\ 
Argon    & 18 & 15.760      & 27.630        & 40.735       & 59.686       \\ 
\hline
\end{tabular}
\end{table}

The ionization parameter $U$ is a unitless parameter defined as the number of hydrogen-ionizing photons emitted per unit time by a photoionizing source, $Q({\rm H})=\int_{\nu_\text{ion}}^{\infty}L_\nu d\nu$, divided by $4\pi r^2$ (where $r$ is the distance between the photoionizing source and the cloud), the number density of hydrogen atoms, $n_{\rm H}$, and the speed of light, $c$:
\begin{equation}\label{eq_U}
U=\frac{Q({\rm H})}{4 \pi\,r^2\,n_{\rm H} c}.
\end{equation}

\begin{figure}[ht]
\begin{center}
\includegraphics[width=0.49\columnwidth]{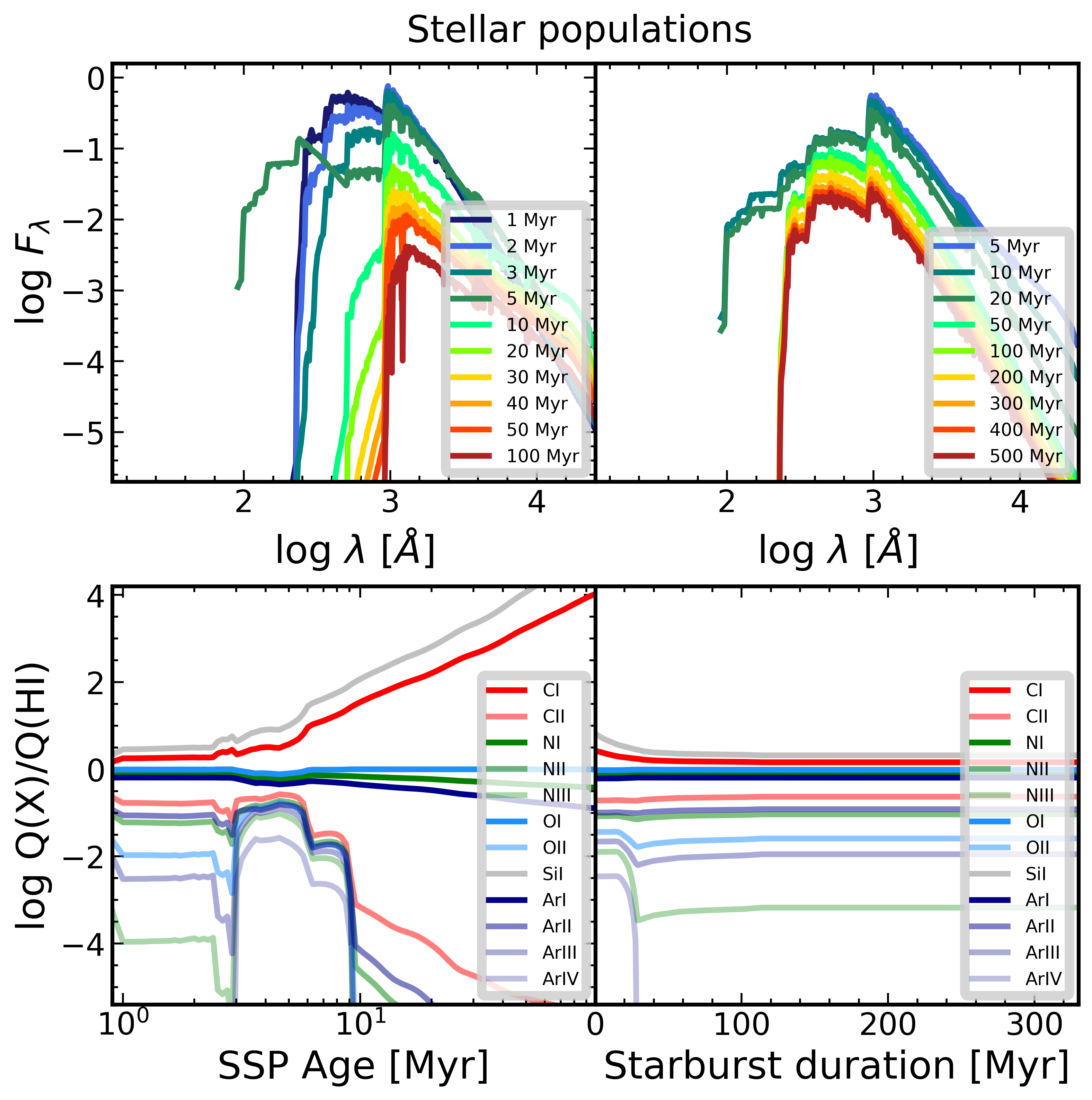}
\includegraphics[width=0.49\columnwidth]{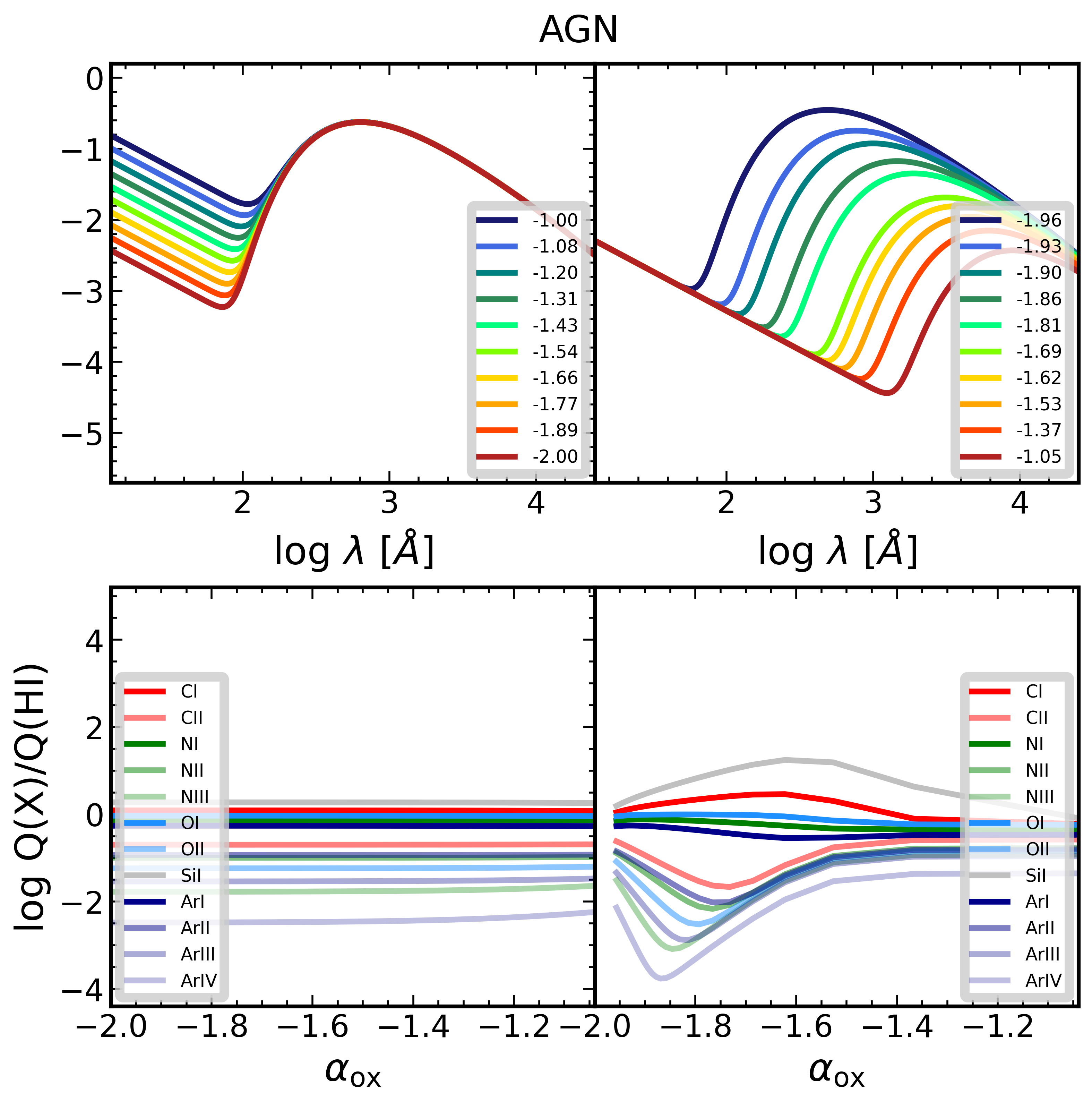}
\end{center}
\caption{\emph{Top panels:} Spectral templates of typical photoionizing sources: Stellar populations resulting from an instantaneous burst or with a fixed star formation rate (color coded by the time since the burst or the age of the stellar population; all normalized to the same stellar mass); AGN with different X-ray power-law normalizations and different temperature of the big blue bump, $T_\text{BB}$ (color coded by the corresponding $\alpha_\text{ox}$ parameter). \emph{Bottom panels:} Ratio between the number of photons that can ionize the species $Q(X)$ and the total number of hydrogen-ionizing photons $Q$(\Hi).}
\label{fig_sed_photoion}
\end{figure}

The two main sources of ionizing photons within galaxies are young stars and AGN. The top panels of Fig.~\ref{fig_sed_photoion} show the templates we adopt for each one in the following calculations. For stars, we consider single stellar population templates from \citet{bc03}\footnote{We note that different stellar population libraries have different predictions on the yield of UV photons. In particular, the treatment of binary stars carries significant repercussions on the budget of ionizing photons as a function of stellar population age. We refer the interested reader to, e.g., \citet{conroy10} for details.}, for the cases of an instantaneous burst (i.e., for a star formation history described by a delta function) and of a prolonged star formation with constant star formation rate. Only stars younger than $\sim$\,8\,Myr produce significant photoionizing radiation. If instead the star formation rate is constant, after 50--100\,Myr the global stellar emission settles on a spectral shape that, below $\lambda\,<\,4000\,{\rm \AA{}}\,\simeq\,3$\,eV, does not significantly evolve with time for several hundred Myr. For the AGN, following, e.g. \citet{ferland98}, we consider the combination of two exponential functions mimicking the big blue bump, and a power law accounting for the X-ray emission:
\begin{equation}\label{eq_agn_template}
F_\nu \propto \nu^{-0.5} \, \exp\left(-\frac{\nu}{\nu_\text{low}}\right)\,\exp\left(-\frac{\nu_\text{high}}{\nu}\right)+\left(\frac{a}{\nu}\right)^{-\alpha},
\end{equation}
with the reference cut-off frequencies being $\nu_\text{low}=k_{\rm b}T_{\rm BB}/h$, and $\nu_\text{high}=k_{\rm b}T_\text{IR}/h$. Here $T_\text{BB}$ is the big blue bump effective temperature \citep[see the original definition in][]{grandi82}, while $k_{\rm b}T_\text{IR}=0.136$ eV. The coefficient $a$ and the slope $\alpha\approx 1$ set the intercept and slope of the X-ray power-law. It is common to parametrize the relative strength of the optical vs X-ray emission via the $\alpha_\text{ox}$ parameter, defined as the ratio between the flux densities measured at 2 keV and 2500\,\AA{} \citep{zamorani81}. The bottom panels of Fig.~\ref{fig_sed_photoion} show the AGN templates for different values of power-law normalization and different temperatures $T_\text{BB}$, color-coded by the corresponding value of $\alpha_\text{ox}$.

In order to calculate the ratios between the various ionized species of each element, we compute, for every star formation and AGN template considered for photoionization, the number of emitted photons that are energetic enough to ionize a given element species, $Q(X)$, by integrating between the ionization energy required to photoionize the species $X$, and the energy needed for the following ionization level. In the bottom panel of Fig.~\ref{fig_sed_photoion} we show the yield of ionizing photons normalized to $Q$(\Hi). For a single stellar population, the yield of various ionizing photons changes dramatically within the first 10 Myr. However, if the star formation rate is constant, $Q(X)$ does not evolve significantly after the first $\sim 20$\,Myr. In the case of AGN photoionization, $Q(X)$ does not appear to be sensitive to the intercept of the X-ray power-law; this happens because the ionization energies of the key elements mostly fall in the far-UV regime. In contrast, and for the same reason, $Q(X)$ is sensitive to $T_\text{BB}$.

It is worth noting that at increasing distance from the source of the radiation field, photons are progressively absorbed, leading to a complex photoionization stratification. Hence, the emissivity of associated emission lines requires radiative transfer calculations. Although an analytical modeling is possible \citep[see, e.g.,][]{yang20oiii}, numerical methods using, e.g., \textsf{Cloudy} \citep{ferland98,ferland17} are widespread.

\subsection{Elemental abundances}\label{sec_abundances}

The last key ingredient that sets the intensity of emission lines is the relative abundance of each element in the ISM with respect to the most abundant collisional partner (the term $n_A/n_c$ in Eq.~\eqref{eq_abundances}). This is typically electrons in the ionized phase, hydrogen in the neutral phase, and H$_2$ molecules deeper into the dense cloud phase. Hydrogen does not have fine-structure transitions: In the mm/radio regime, hydrogen emits recombination lines at high quantum numbers \citep{dupree70}, however they are typically very faint, and difficult to detect in the distant Universe. Hence, to measure metallicities using only observations in the rest-frame FIR wavelengths we need to rely on models of the relative abundances of ions. This limitation can be by-passed via observations of the rest-frame optical spectra, where nebular line emission from nitrogen, oxygen, and sulfur ions can be combined with hydrogen Balmer lines to infer gas-phase metallicities \citep[see][for a review]{maiolino19}.

\begin{figure}[ht]
\begin{center}
\includegraphics[width=0.49\textwidth]{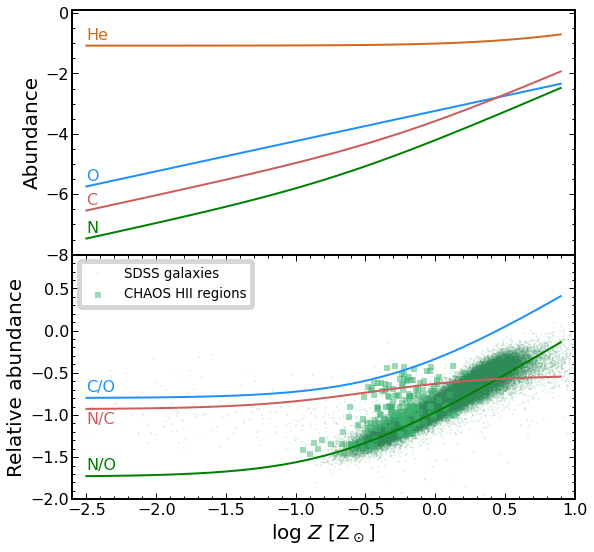}
\end{center}
\caption{Metal abundances with respect to hydrogen and relative abundances as a function of gas-phase metallicity, based on the prescriptions in \citet{nicholls17}. The impact of the secondary production of metals leads to an increase of the C/O and N/O ratio for metallicities $Z>0.1$\,Z$_\odot$. These ratios can serve to infer metallicity from IR FSLs. The N/C ratio also increases with $Z$, but only by a factor $\sim2$ over $\sim1.5$ dex of metallicity. In the bottom panel, green dots and squares show the N/O abundance as a function of metallicity for SDSS galaxies (based on calibrations by \citealt{charlot01} for $Z$ and \citealt{pilyugin16} for N/O) and for \Hii\, regions in local galaxies from the CHAOS survey \citep{berg20}. The large dispersion in the N/O ratio estimated for \Hii\, regions is the outcome of both systematic uncertainties in the calibration of metallicity estimates and intrinsic scatter. These caveats may thus limit the use of IR FSLs as absolute metallicity tracers.}
\label{fig_Z}
\end{figure}

While the abundances of most of elements in the gas phase follow, to first order, a linear scaling with the abundance of oxygen, some elements show significant deviations. In particular, nitrogen and carbon show a primary abundance sequence that scales linearly with the oxygen abundance, and a secondary super-linear abundance sequence (see Fig.~\ref{fig_Z}). Primary abundances result from the enrichment of the ISM via core-collapse supernovae in the star-forming gas clouds; secondary abundances are associated with delayed nucleo-synthesis following the evolution of intermediate-mass stars. Winds, galactic fountains, internal gas kinematics, as well as accretion from the circum-galactic medium, are processes that can displace, redistribute, or generally alter the metallicity of galaxies and the relative abundances of elements within them.

By measuring the carbon-to-oxygen or nitrogen-to-oxygen abundance ratio, one could in principle infer the gas metallicity during the secondary sequence via enrichment models; however, model uncertainties are often large ($\gsim 0.2$\,dex). We refer the interested reader to \citet{vincenzo16}, \citet{nicholls17}, \citet{maiolino19}, \citet{romano22} for in-depth discussions on the modeling of gas-phase abundances. An important caveat that interferes with a straightforward measure of relative abundances is that elements can also be locked in molecules, dust grains, or ices, and hence may not emit in their FSL transitions. Since noble gases (e.g., helium, neon) do not form molecules or are depleted onto dust grains, they could be potentially employed to construct more robust metallicity diagnostics.

\subsection{Optical depth of IR FSLs}

In the previous sections, we argued that IR FSLs typically have negligible optical depth. We can test this assumption by computing the optical depth associated with a line, $\tau_\nu$. For a gaussian broadening of the line, $\tau_\nu = \tau_0\,e^{-(v/b)^2}$; where $b=\sqrt2\,\sigma_v$ is the scale line width, $v$ is the velocity offset, $v=c (\nu-\nu_0)/\nu_0$, and $\nu_0$ is the rest-frame line frequency. The optical depth at the center of the line, $\tau_0$, can be estimated as:
\begin{equation}\label{eq_tau_line}
\tau_0 = \frac{N_l\,A_{ul}\,c^3}{8\,\pi^{3/2}\,b\,\nu_{lu}^3}\,\frac{g_u}{g_l}\,\left(1-\frac{N_u/g_u}{N_l/g_l}\right).
\end{equation}
(see Eq.~9.8 in \citealt{draine11}). The values of the statistical weights $g_l$, $g_u$, the line frequency $\nu_{lu}$, and the Einstein coefficients $A_{ul}$ are listed in Table~\ref{tab_ion_energy}. The column densities $N_u$ and $N_l$ are inferred from the local density, integrated along the line of sight: $N_{l,u}=\int n_{l,u}\,ds$, while the relative population of the upper and lower levels of the transitions can be computed with the method described in Sect.~\ref{sec_linepredict}.

\begin{figure}[ht]
\begin{center}
\includegraphics[width=0.49\textwidth]{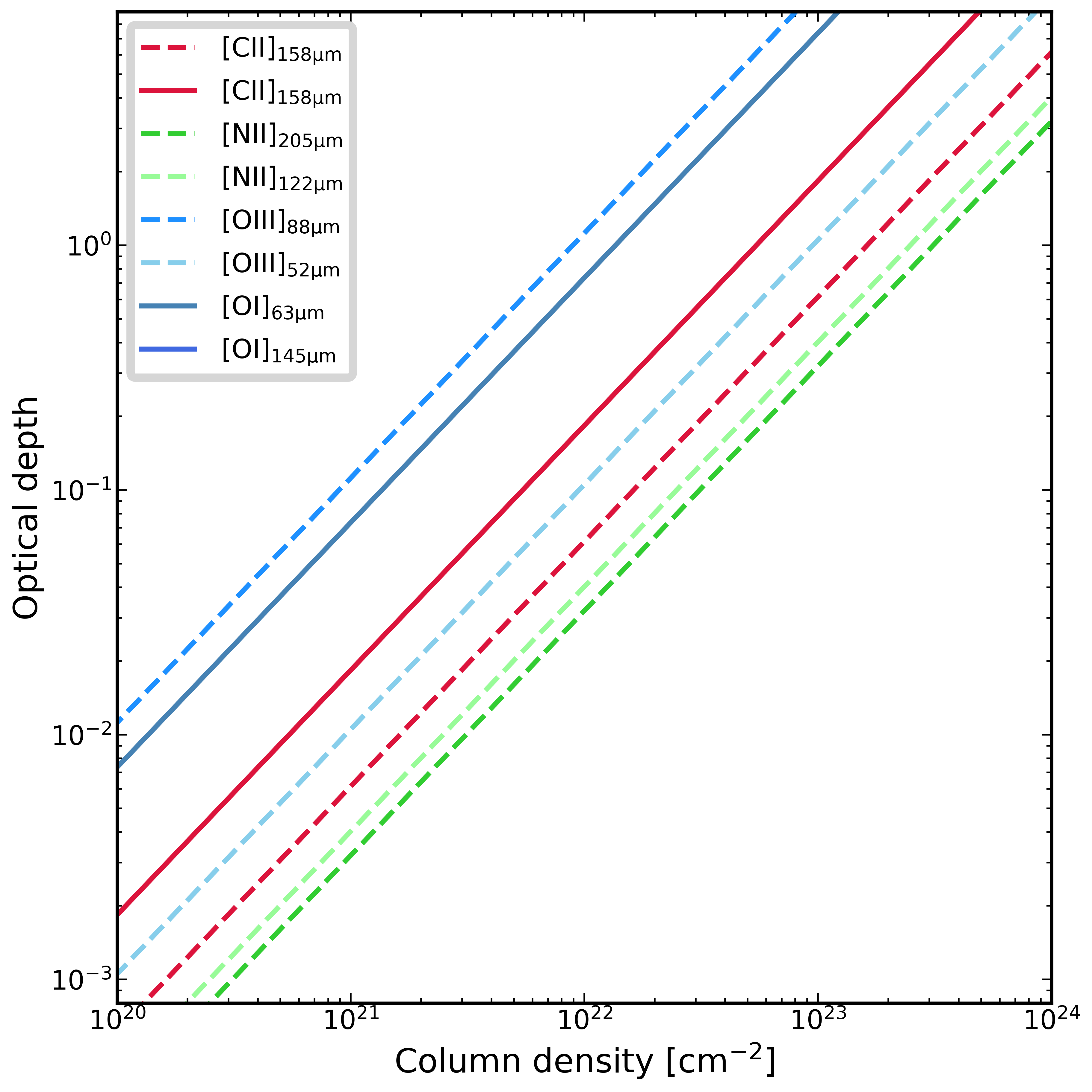}
\end{center}
\caption{Line optical depth as a function of hydrogen column densities, assuming solar abundances, $\sigma_v$\,=\,10\,\kms\, and that half of the carbon, nitrogen, and oxygen atoms are in neutral form. Dashed (solid) lines refer to transitions associated with the ionized (neutral) medium. The thickest lines are the two oxygen transitions starting from the ground level (\Oi{}\,63\,$\mu$m and \Oiii{}\,88\,$\mu$m), which become optically-thick at $N_{\rm H}>10^{22}$\,cm$^{-2}$. \Cii{} becomes only moderately thick at those column densities. Other transitions (e.g., neutral carbon, ionized neon, etc) have optical depths that are several orders of magnitude lower than the ones shown here.}
\label{fig_optical_depth}
\end{figure}

To assess whether the main IR FSLs are optically thin, we consider the case of a gas cloud with constant volume density $n=100$\,cm$^{-3}$ and kinetic temperature $T=10,000$\,K in the case of ionized gas; and $n=10,000$\,cm$^{-3}$ and $T=100$\,K for the neutral phase. In the former case, we consider electrons as collision partners, while in the latter we consider a 50\% mixture of hydrogen atoms and molecules. We assume a solar metallicity, with abundances set as in \citet{nicholls17}; and we assume that half of the carbon, nitrogen and oxygen atoms are in neutral form, with the remainder being in either singly or doubly-ionized form. We consider a line width of $\sigma_v$=10\,\kms{}. Figure~\ref{fig_optical_depth} shows the inferred line optical depths as a function of hydrogen column density for the main IR FSLs. Optical depths are always negligible, except for \Oiii{}\,88\,$\mu$m and \Oi{}\,63\,$\mu$m at $N_{\rm H}\gsim10^{22}$\,cm$^{-2}$, and \Cii{} (in the neutral medium) at $N_{\rm H}\gsim10^{23}$\,cm$^{-2}$. While these high column densities are unlikely found in the ionized gas, they are not uncommon in the neutral medium.

\subsection{Line fluxes and luminosities}\label{sec_flux2lum}

The spectra of extragalactic sources at mm wavelengths are often plotted in terms of the observed flux density, $F_\nu$, in units of mJy, as a function of the velocity shift with respect to the source rest frame, $\Delta v$. For this reason, it is common practice to quote integrated line fluxes, $F_\text{line}=\int F_\nu d\nu = \nu_\text{obs}\,c^{-1} \int F_\nu \, dv$, in units of Jy\,\kms{}. This practical solution has two unfortunately important drawbacks: 1) the quoted flux is no longer redshift invariant, and 2) for a given source, lines with the same luminosity will have different fluxes (in these units) due to the dependence on the rest-frame frequency of the line. Integrated fluxes in units of W\,m$^{-2}$ or erg\,s$^{-1}$\,cm$^{-2}$ do not have such problems, with the conversions being:
\begin{equation}\label{eq_Fline}
\frac{F_\text{line}}{\rm Jy\,km\,s^{-1}} = 10^{17}\,\frac{F_\text{line} }{\rm W\,m^{-2}} \, \left(\frac{\nu_\text{obs}}{\rm GHz}\right)^{-1} \, \frac{c}{\rm km\,s^{-1}} = 10^{20}\,\frac{F_\text{line} }{\rm erg\,s^{-1}\,cm^{-2}} \, \left(\frac{\nu_\text{obs}}{\rm GHz}\right)^{-1} \, \frac{c}{\rm km\,s^{-1}}.
\end{equation}

Most missions operating in the mid-IR bands (e.g., \textit{Spitzer}, JWST) deliver observations sampled in counts stored in pixels, $f$, in units of MJy\,str$^{-1}$. If the data include a spectral axis in wavelengths, the line flux becomes:
\begin{equation}\label{eq_Fline_MJystr}
\frac{F_\text{line}}{\rm Jy\,km\,s^{-1}}=7.046\,\sum\limits_{j}^{N_\text{chn}} \left(\sum\limits_{i}^{N_\text{pixel}}\frac{f^{\rm obs}_{\nu\,i,j} }{\rm MJy\,str^{-1}}\,\right) \, \frac{\Delta \lambda_\text{obs}}{\lambda_{j,\rm obs}}\,\left(\frac{\theta}{\rm arcsec\,pixel^{-1}}\right)^2,
\end{equation}
where $N_\text{pixel}$ is the number of pixels over which the source is measured in a given channel; $N_\text{chn}$ is the number of channels; $\theta$ is the pixel scale; $\lambda_{j, \rm obs}$ is the central wavelength at channel $j$; and $\Delta \lambda_\text{obs}$ is the width of a channel, in the same units as $\lambda_{j, \rm obs}$.

The line luminosity is then $L_\text{line}=4\pi\,D_{\rm L}^2 \,F_\text{line}$, where $D_{\rm L}$ is the luminosity distance, and can be written as:
\begin{equation}\label{eq_linelum}
\frac{L_\text{line}}{\rm L_\odot}=\frac{1.04\times10^{-3}}{1+z} \, \frac{F_\text{line}}{\rm Jy\,km\,s^{-1}} \, \frac{\nu_\text{rest}}{\rm GHz}  \left(\frac{D_{\rm L}}{\rm Mpc}\right)^2 = \frac{3.12\times10^{11}}{1+z} \, \frac{F_\text{line}}{\rm Jy\,km\,s^{-1}} \left(\frac{\lambda_\text{rest}}{\rm \mu m}\right)^{-1} \left(\frac{D_{\rm L}}{\rm Mpc}\right)^2,
\end{equation}
where $\nu_\text{rest}$ and $\lambda_\text{rest}$ are the rest frequency and wavelength of the line, respectively.

In radio and mm astronomy, it is common practice to refer to a temperature-based definition of the luminosity, $L_\text{line}'$, which corresponds to the areal integrated source brightness temperature in units of K\,\kms{}\,pc$^2$ \citep[see][]{solomon92}:
\begin{equation}\label{eq_l1}
\frac{L_\text{line}'}{\rm K\,km\,s^{-1}\,pc^2} = T_{\rm B} \, \Delta v \, \Omega_\text{source} \, D_{\rm A}^2 =  \frac{3.25 \times 10^7}{(1+z)^3} \frac{F_\text{line}}{\rm Jy\,km\,s^{-1}}\,\left(\frac{\nu_\text{obs}}{\rm GHz}\right)^{-2} \left(\frac{D_{\rm L}}{\rm Mpc}\right)^2,
\end{equation}
where $T_{\rm B}$ is the brightness temperature of the source, $\Delta v$ is the line velocity width, $\Omega_\text{source}$ is the solid angle over which the source is observed, and $D_{\rm A}$ is the angular diameter distance, with $D_{\rm A}=D_{\rm L} (1+z)^{-2}$.

\section{First principle diagnostics}\label{sec_diagnostics}

The relatively simple physics of IR FSLs enable to infer key physical quantities via the observed IR FSL luminosities. Here we first review diagnostics based on first-principle arguments. 

\subsection{Mass of the emitting species}\label{sec_masses}

Following the analysis presented in Sect.~\ref{sec_linepredict}, by inverting Eq.~\eqref{eq_linelum_theo} it is possible to infer the mass of a fine-structure line emitting species. One could compute $f(n,T)$ numerically following the approach described in Sect.~\ref{sec_linepredict}. Alternatively, one can derive the level population simply via defining an excitation temperature, $T_\text{ex}$, and thus analytically derive the corresponding mass of the emitting species:
\begin{equation}\label{eq_mass}
M_{\rm X} = m_{\rm X}\,\frac{L_{ul}\,Q(T_\text{ex})}{h \nu_{ul}\,A_{ul}\,g_u}e^{T_u/T_\text{ex}} = \mathcal{A}\,L_{ul}\,Q(T_\text{ex})\,e^{T_u/T_\text{ex}}.
\end{equation}

Table~\ref{tab_ncrit} lists the Einstein coefficients $A_{ul}$, the energy levels $E_u k_{\rm b}^{-1}=T_u$, and the conversion factors $\mathcal{A}$. The partition function, $Q(T_\text{ex})$, is the sum of the statistical weights of each level $i$ multiplied by $e^{-T_i/T_\text{ex}}$; for instance, in the three-level approximation of 6-electron ions and atoms (\Ci{}, \Nii{}, \Oiii{}), $Q(T_\text{ex})=1+3 e^{-T_{J=1}/T_\text{ex}} + 5 e^{-T_{J=2}/T_\text{ex}}$, while for the two-level approximation of 5-electron ions (\Cii{}, \Niii{}, \Oiv{}, \Nev{}), $Q(T_\text{ex})=2+4 e^{-T_{J=3/2}/T_\text{ex}}$. Equation~\eqref{eq_mass} can be written in terms of temperature luminosity ($L'$, based on Eq.~\eqref{eq_l1}). For instance, in the case of neutral carbon, in the optically-thin case, Eq.~\eqref{eq_mass} becomes: 
%
\begin{equation}
\frac{M_{\rm C^0}}{\rm M_\odot} = \mathcal{C}\,m_\text{C}\,\frac{8 \pi k_{\rm b} \nu_\text{rest}^2}{h c^3 A_{10}}\,Q(T_\text{ex})\frac{1}{3}\,e^{T_1/T_\text{ex}}\,\frac{L_{\rm [CI]_{1-0}}'}{\mathrm{ K\,km\,s^{-1}\,pc^2}},
\end{equation}
where $T_1=23.6$\,K and $T_2=62.5$\,K are the energy levels above ground state; $\mathcal{C}=9.52\times10^{41}$ is the conversion factor between pc$^2$ and cm$^2$ and between \kms{} and cm\,s$^{-1}$; $m_\text{C}$ is the mass of a single carbon atom in grams, $A_{ul}$ are the Einstein coefficients in units of s$^{-1}$, and $\nu_\text{rest}$ is the line rest-frame frequency in Hz \citep[see][for reference]{weiss03,weiss05,papadopoulos04}.

\subsection{Electron density}\label{sec_ne}

We can benefit from the differences in critical densities between various energy levels of a given species to construct density diagnostics. By using the line ratios of the same species (e.g., a pair of \Nii{} or \Oiii{} transitions), one circumvents dependencies on other gas conditions, such as the relative elemental abundances or the knowledge of their fractional ionization stages. Because IR FSLs arise from low-energy levels, which are collisionally populated even in the low-density and cold ISM, their use is particularly effective as density tracers \citep{rubin89b}. Hence, atoms and ions with a three-level fine-structure splitting of the ground level are particularly suited: \Ci{}, \Nii{}, \Oiii{}, \Oi{}, \Nev{}, \Neiii{}, \Sili{}, \Siii{}, or \Arv{}.

\begin{figure}[ht]
\begin{center}
\includegraphics[width=\textwidth]{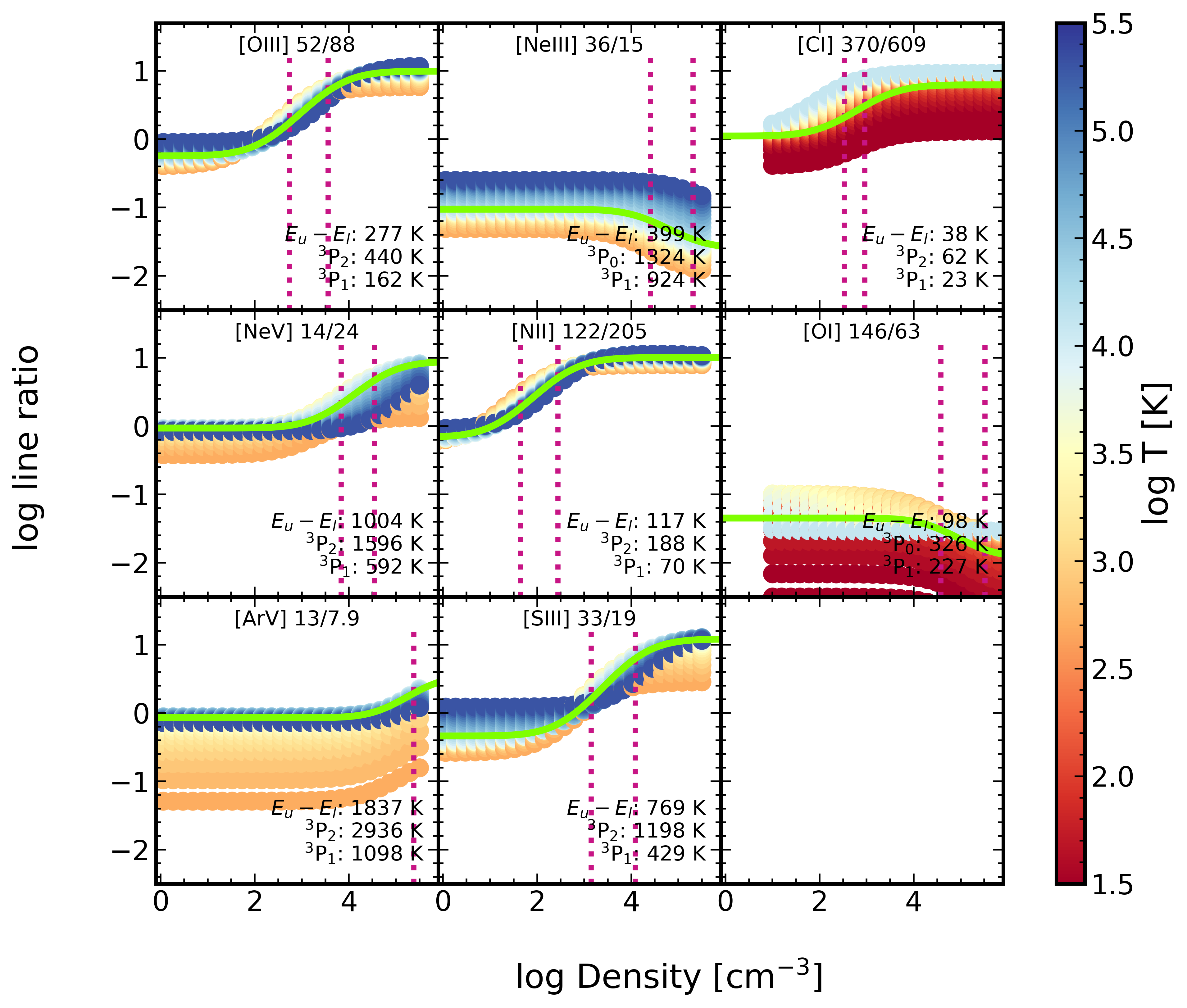}
\end{center}
\caption{Density diagnostics involving IR fine-structure lines, based on the analytical prescription described in the text, and color-coded based on the temperature of the collision partners. Panels on the left-hand and middle columns refer to ionized gas tracers, panels on the right-hand to neutral gas tracers. Vertical dotted lines mark the critical densities of the two transitions, while labels in the bottom-right corners list the temperatures corresponding to the excitation energies of the $^3$P$_1$ and $^3$P$_2$ levels. The best fit curves computed at $T$=10,000\,K (100\,K) for the ionized (neutral) medium tracers are shown in green (see Table~\ref{tab_density_params}).}
\label{fig_density_diagnostics}
\end{figure}

Figure~\ref{fig_density_diagnostics} shows the density dependence of various IR FSL ratios. The different values of critical densities imply that the associated diagnostics are sensitive to different regimes of density: 3--3$\times10^{3}$\,cm$^{-3}$ for \Nii{} 205/122\,$\mu$m;  30--$10^{4}$\,cm$^{-3}$ for \Oiii{} 88/52$\mu$m; $10^{3}$--$10^{5}$\,cm$^{-3}$ for [Ne\,{\sc v}] 24/14$\mu$m; $>10^{4}$\,cm$^{-3}$ for [Ar\,{\sc v}]. We fit the line ratios as a function of $n$ with the functional form: log ratio = $C \times {\rm erf}[(\log n - A)/B] + D$. We consider the case of \iTkin\,=\,10,000\,K for tracers of the ionized gas, and \iTkin\,=\,100\,K for the neutral medium. Table~\ref{tab_density_params} provides the best fit values of the fitted parameters.

\begin{table}[ht]
\caption{Parametric fits of the ratio of various fine-structure lines as a function of density, parametrized as: log ratio = $C \times {\rm erf}[(\log n - A)/B] + D$, where $n$ is in units of cm$^{-3}$. These fits are shown as green lines in Fig.~\ref{fig_density_diagnostics}. Note however that the ratios depend on $T$ for $T\lsim E_2/k_{\rm b}$, where $E_2$ is the energy of the ${}^3$P$_2$ level.}
\label{tab_density_params}
\begin{tabular}{cccccc}
\hline
Element  & Ratio & $A$ & $B$ & $C$ & $D$  \\
(1)      & (2)   & (3) & (4) & (5) & (6)  \\
\hline
 \Oiii{}  &   52/88  & 1.16	 &  2.94   &  0.62  &	0.38 \\
 \Neiii{} &   36/15  & 0.97	 &  4.82   & $-$0.28  &  $-$1.33 \\ 
 \Nev{}   &   24/14  & 0.99	 &  4.09   &  0.47  &   0.46 \\ 
 \Arv{}   &  7.9/13  &$-$0.86	 &  5.26   & $-$0.29  &	0.24 \\ 
 \Nii{}   & 122/205  & 1.13	 &  1.91   &  0.58  &	0.42 \\ 
 \Siii{}  &   19/33  &$-$1.18	 &  3.39   & $-$0.72  &	0.37 \\ 
 \Ci{}    & 370/609  & 1.05	 &  2.78   &  0.37  &	0.42 \\ 
 \Oi{}    & 146/63   &  1.02	 &  4.91   & $-$0.33  &  $-$1.63 \\ 
\hline
\end{tabular}
\end{table}

Assuming that $f_u(n_e,T_\text{kin})$ is negligible for any level beyond the fine-structure splitting of the ground level, the line ratio converges to:
\begin{equation}\label{eq_lowdens_ratio}
\frac{j_{21}}{j_{10}}\approx\frac{\Omega(2,1)\,\exp(-E_{21}/k_{\rm b} T_\text{kin})}{\Omega(1,0)+\Omega(2,1)\,\exp(-E_{21}/k_{\rm b} T_\text{kin})}\,\frac{E_{21}}{E_{10}}
\end{equation}
in the low-density regime; and to:
\begin{equation}\label{eq_highdens_ratio}
\frac{j_{21}}{j_{10}}\approx\frac{g_2\,A_{21}\,E_{21}}{g_1\,A_{10}\,E_{10}}e^{-E_{21}/k_{\rm b} T}
\end{equation}
in the high-density regime. Here $\Omega(u,l)$ is the collisional strengths of the transition. The higher energy levels might be underpopulated for low \iTkin\, values. While this is rarely the case for the ionized medium (as all of the IR FSLs arise from levels with energies $E_2/k_{\rm b}\,<\,3000$\,K), this might not be the case for the transitions associated with the neutral medium. E.g., the energies of the first two excited fine-structure levels of the neutral oxygen atom are $E_1/k_{\rm b}\,=\,98$\,K and $E_2/k_{\rm b}\,=\,326$\,K, which may be higher than the temperature of the collision partners.

\subsection{Electron temperature}\label{sec_Te}

\begin{figure}[ht]
\begin{center}
\includegraphics[width=0.74\textwidth]{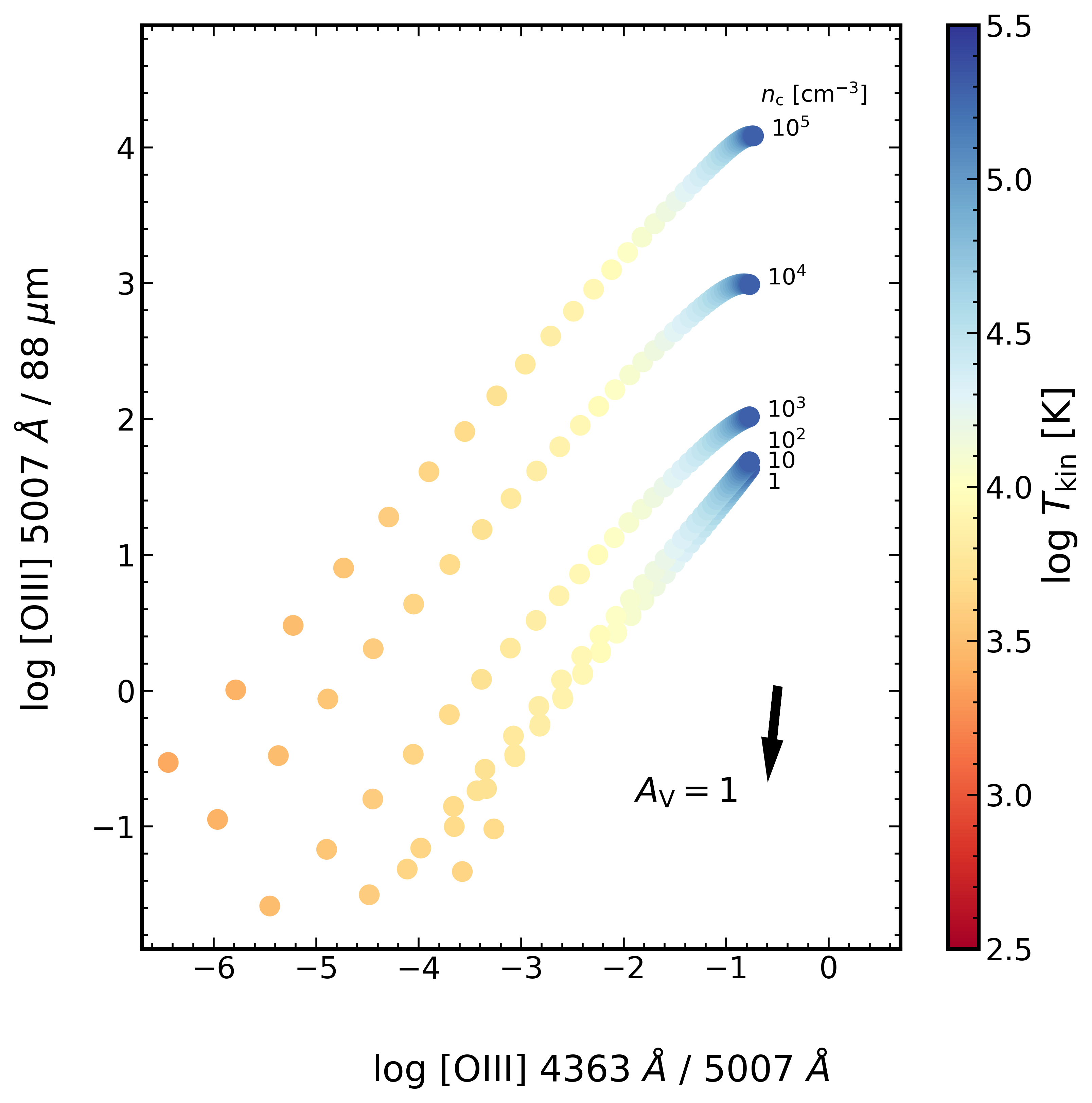}
\end{center}
\caption{Optical and FIR line ratios of various \Oiii{} transitions. The auroral (4363 \AA{}) vs nebular (5007 \AA{}) line ratio, and the nebular to fine-structure (88 $\mu$m) line ratios, are commonly used to infer the kinetic temperature of electrons, $T_\text{kin}$. At high densities ($n_{\rm c}\gtrsim 500$ cm$^{-3}$, corresponding to the critical density of the transition), the 88\,$\mu$m line is collisionally--suppressed; however, these densities are not common in the ionized medium. The arrow shows how the diagnostics would be affected by extinction due to a gas column density corresponding to an extinction of $A_{\rm V}=1$.
}
\label{fig_T_opt_fir}
\end{figure}

In the working framework of thermal equilibrium, levels are populated by collisions. We can infer their kinetic temperature of the collision partners by comparing the luminosities of transitions originating from levels at different energies. Once again, by focusing on atoms and ions of the same element and ionization stage, we remove any dependence on relative elemental abundance or the intensity and hardness of the photoionizing field. For instance, for 6- and 8-electron atoms and ions, the ratios between the ${}^1$S$_0$ $\rightarrow$ ${}^1$D$_2$ transition and the ${}^1$D$_2$ $\rightarrow$ ${}^3$P or between the latter and one of the fine-structure transitions of the ${}^3$P levels provide strong constraints on the temperature of collision partners. Examples of temperature-sensitive line ratios are \Oiii{}\,4363\,\AA{}/5007\,\AA{},  and \Oiii{}\,5007\,\AA{}/88\,$\mu$m (see Fig.~\ref{fig_T_opt_fir}).

Optical lines however are sensitive to extinction. The combination of auroral, nebular and FSL luminosities can simultaneously constrain electron temperature and extinction, if the extinction law is known \citep[see Fig.~\ref{fig_T_opt_fir} and other diagnostics in][]{spinoglio92}. 

\begin{figure}[ht]
\begin{center}
\includegraphics[width=\textwidth]{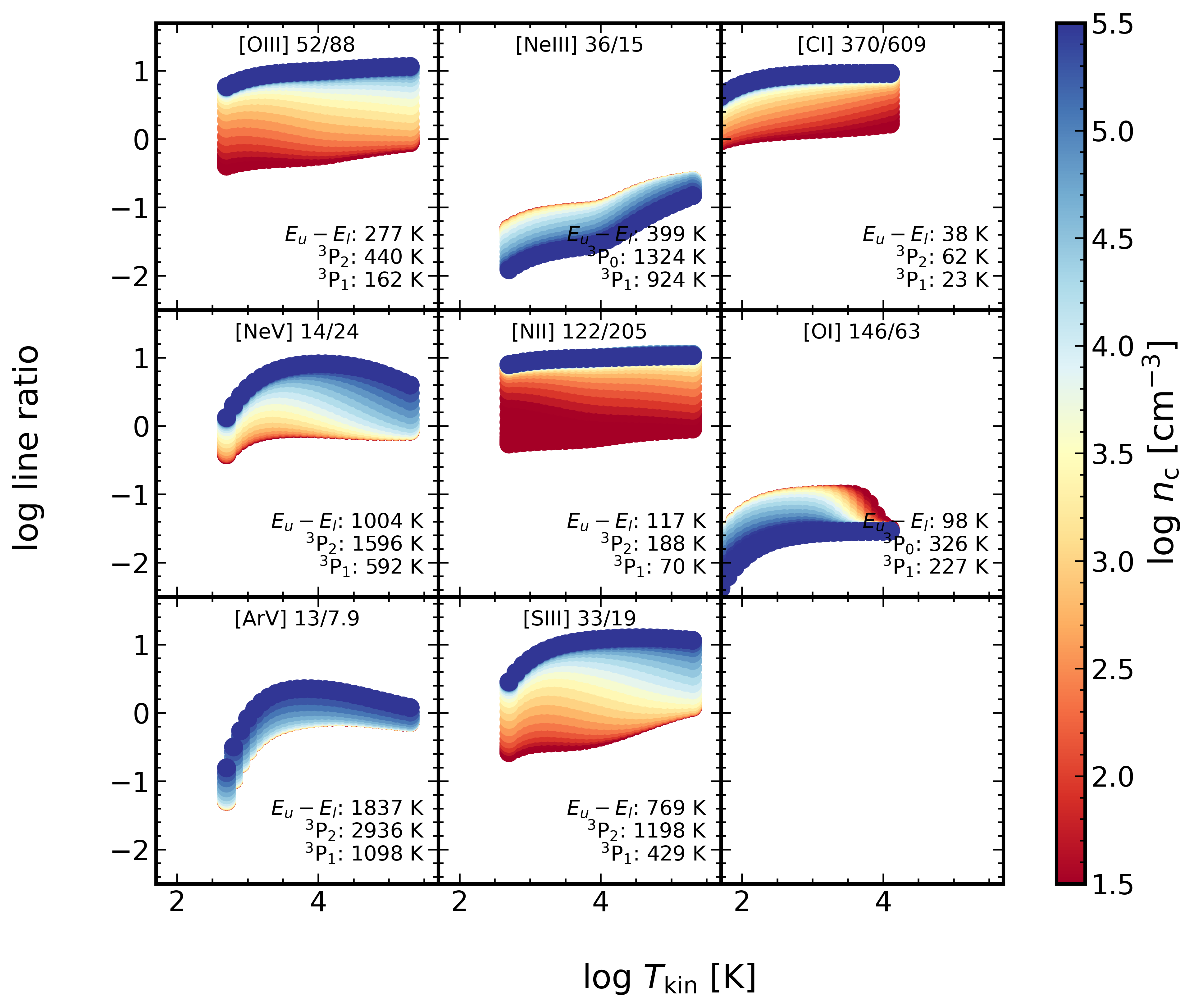}
\end{center}
\caption{Same as Fig.~\ref{fig_density_diagnostics}, but with explicit dependence on the kinetic temperature of the collision partners. The line ratio diagnostics are accurate density tracers when $T_\text{kin}\gg E_u/k_{\rm b}$. In these regimes, the residual dependence on \iTkin\, is due to the oscillator strengths, which also are weakly dependent on \iTkin\, (see Appendix~F in \citealt{draine11}).}
\label{fig_n_vs_T}
\end{figure}

\begin{figure}[ht]
\begin{center}
\includegraphics[width=0.87\textwidth]{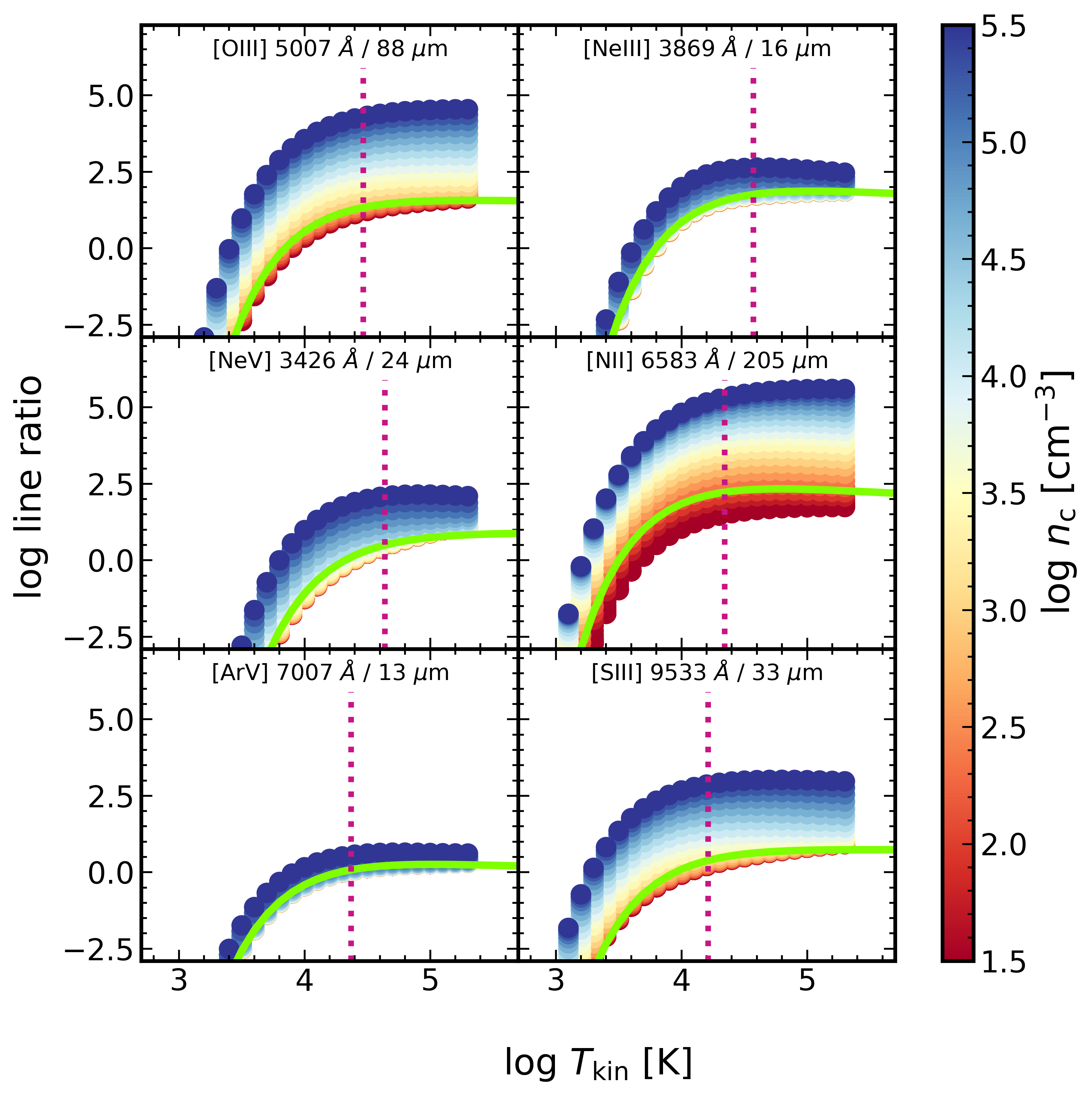}
\end{center}
\caption{Temperature diagnostics involving IR fine-structure lines, based on the analytical prescription described in the text, and color-coded based on the density. Vertical dotted lines mark the temperature corresponding to the energy of the $^1$D$_2$, the higher level involved in the ratios shown here. The lower values are outside of the plotted range, to the left. The green curves show the best fit to the function ratio = A\,$\exp(T/B)\,(1+\sqrt{C/T})^{-1}$, with the best fit parameters listed in Table~\ref{tab_temperature_params}.
}
\label{fig_temperature_diagnostics}
\end{figure}

While the ratios presented in Fig.~\ref{fig_temperature_diagnostics} are particularly useful for the ionized medium (where typical $T_\text{kin}\gtrsim 10^4$\,K), in the low temperature regime of the neutral medium the ${}^1$D$_2$ level is scarcely populated, and hence temperature measurements of the collision partners mostly rely on measuring the excitation temperature via the ${}^3$P$_2$ $\rightarrow$ ${}^3$P$_1$ vs.~${}^3$P$_1$ $\rightarrow$ ${}^3$P$_0$ line ratio \citep[see Fig.~\ref{fig_n_vs_T}; see also, e.g.,][]{walter11}, although this ratio is also sensitive to $n_{\rm c}$, as discussed in Sect.~\ref{sec_ne}.

Figure~\ref{fig_temperature_diagnostics} shows nebular-to-FSL luminosity ratios as a function of temperature, color coded by density. We fit the case of $n=100$\,cm$^{-3}$ using the functional form: ratio = $A\,\exp(T/B)\,(1+\sqrt{C/T})^{-1}$ (green lines in the plot) and provide the best-fit parameters in Table~\ref{tab_temperature_params}.

\begin{table}[ht]
\caption{Parametric fits of the ratio of various nebular-to-fine-structure lines as a function of temperature, parametrized as: ratio = $A\,\exp(T/B)\,(1+\sqrt{C/T})^{-1}$, computed for a density $n=100$\,cm$^{-3}$. These fits are shown as green lines in Fig.~\ref{fig_temperature_diagnostics}.}
\label{tab_temperature_params}
\begin{tabular}{ccccc}
\hline
Element  & Ratio & $A$ & $B$ & $C$  \\
(1)      & (2)   & (3) & (4) & (5)  \\
\hline
 \Oiii{}  &  5007\,\AA{}/88\,$\mu$m &27.57 & 31938 &  43098 \\
 \Neiii{} &  3869\,\AA{}/16\,$\mu$m &30.20 & 35645 &  516950 \\
 \Nev{}   &  3426\,\AA{}/24\,$\mu$m & 7.77 & 49733 &   2895 \\
 \Nii{}   &  6583\,\AA{}/205\,$\mu$m &77.84 & 21777 & 457774 \\
 \Arv{}   &  7007\,\AA{}/13\,$\mu$m & 0.30 & 23021 & 865063 \\
 \Siii{}  &  9533\,\AA{}/33\,$\mu$m & 4.72 & 19959 &   5452 \\
\hline
\end{tabular}
\end{table}

\subsection{Ionization parameter}\label{sec_U}

In the case of an ionized medium, the intensity of the photoionizing radiation field is often expressed in terms of the ionization parameter $U$, which is defined as the number of ionizing photons per unit gas density (Eq.~\eqref{eq_U}). It is important to stress that $U$ is a spectrally-integrated quantity and does not carry information regarding the frequency-dependent radiation intensity of the photoionizing source; as a consequence, $U$ does \emph{not} trace the hardness of the radiation field per se. Instead, that information is encapsulated in the spectral shape and slope of the ionizing source itself (Fig.~\ref{fig_sed_photoion}). However, an increase in $U$ can have an effect that is \emph{similar} to that of having a ionization source with a harder spectrum, even when the spectral shape of the photoionizing source may be the same. This is because, for higher values of $U$, in addition to having more photons available at all energies, there are more hard photons (especially close to the source of radiation) capable of ionizing a larger fraction of the limited amount of metals to higher levels. Indeed, the ionization parameter varies with the distance from the source. \citet{yang20oiii} demonstrate how one can model this effect analytically. The stratification of ionization has been long measured in Galactic \Hii\, regions \citep[e.g.,][]{simpson86}. However, when studying distant sources, one needs to assume that the observed ratios reflect galaxy-integrated, luminosity-weighted average conditions. One should bear in mind this caveat when interpreting observed line ratios.

\begin{figure}[ht]
\begin{center}
\includegraphics[width=0.49\textwidth]{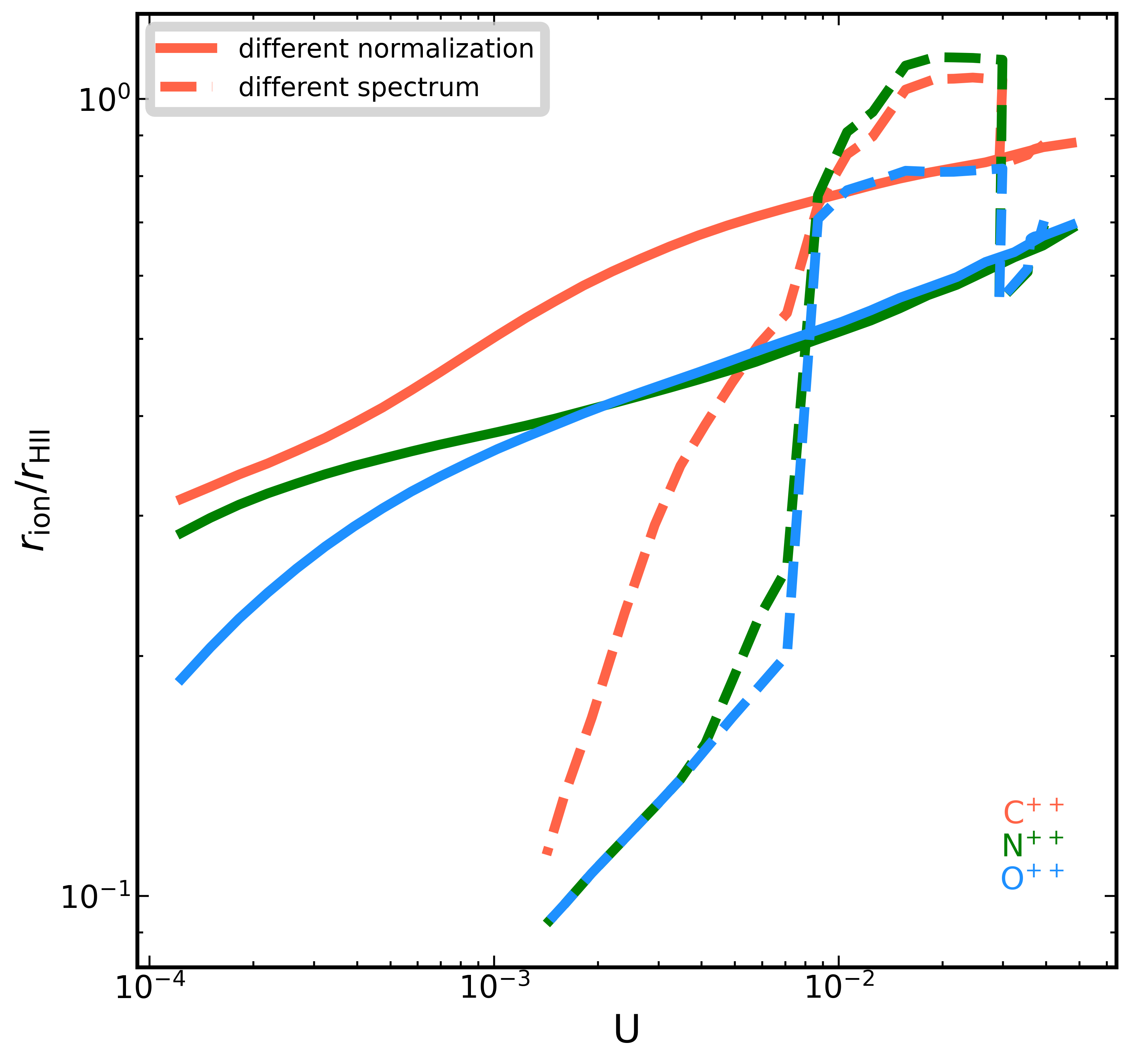}
\includegraphics[width=0.49\textwidth]{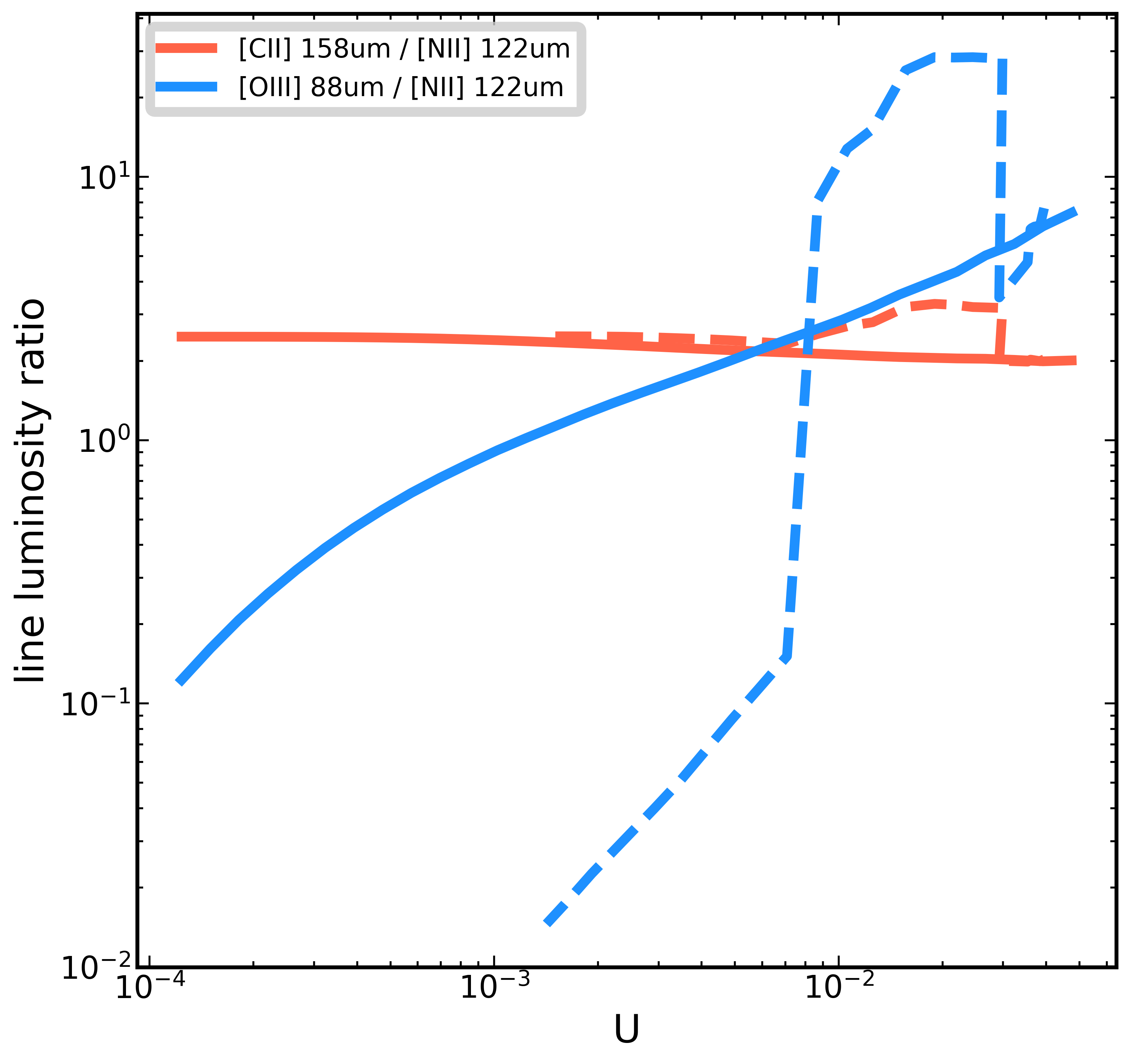}
\\
\end{center}
 \caption{\emph{Left:} The radius of ionized bubbles, normalized by the size of \Hii\, region, as a function of ionization parameter $U$. Solid lines show the case of a fixed spectrum of a single stellar population of 2\,Myr for the range of $U$; dashed lines show the case of a single stellar population aging in time from 0.1 to 150\,Myr, normalized to fixed stellar mass, and $U$ is computed at $R_\text{out}=\left(3\, Q({\rm H})/(4 \, \pi\,n^2 \alpha_{\rm B}({\rm H},T)\right)^{1/3}$. The gas cloud has a density $n$\,=\,100\,cm$^{-3}$ and solar metallicity. Both trends result in a change in $U$: As $U$ grows, the size of the ionized bubbles of higher-ionization species increases with respect to that of hydrogen. \emph{Right:} The corresponding luminosity of \Cii{}\,158\,$\mu$m and \Oiii{}\,88\,$\mu$m, normalized to the luminosity of \Nii{}\,122\,$\mu$m, as a function of $U$. We assume constant hydrogen density $n$\,=\,100\,cm$^{-3}$ and temperature $T$\,=\,10,000\,K. 
 The luminosity ratios of lines with different ionization energies provides constraints on the ionization parameter $U$. In order to infer the origin of $U$ (hardness of the ionization spectrum vs.\ normalization), different line ratios are typically required.
}\label{fig_Urad}
\end{figure}

As an example, Fig.~\ref{fig_Urad} shows the relative size of the bubbles of ionized hydrogen, nitrogen (singly and doubly ionized) and oxygen (singly and doubly ionized) as a function of $U$. We consider the case of a single stellar population with age of 2 Myr (blue line in Fig.~\ref{fig_sed_photoion}, \emph{top left}), with varying normalization (intensity), as well as the case of an evolving single stellar population at fixed normalization (using the entire set of spectral libraries shown in Fig.~\ref{fig_sed_photoion}, \emph{top left}). For the gas cloud, we assume a fixed temperature $T$\,=\,10,000\,K, a density $n$\,=\,100\,cm$^{-3}$, and solar metallicity. A change in either the normalization (intensity) or the hardness of the photoionizing spectrum results in a change of the value of $U$; but irrespectively of the origin of such change, an increase in $U$ leads to a larger fraction of the \Hii\, volume being filled with higher-ionization stages of carbon, nitrogen, and oxygen. This is why in general $U$ is, to first order, degenerate (correlated) with the hardness of the radiation field, despite the fact that the former is by definition not a direct tracer of the latter. In other words, increasing the budget of photoionizing radiation while keeping the hardness constant has a \emph{similar} but not \emph{equal} effect as increasing the hardness while keeping the ionizing photon rate constant.

Measuring the ionization parameter alone usually requires the knowledge of just two emission lines with similar critical densities but different ionization potentials. If relative elemental abundances can be estimated (or at least reasonably assumed), one could adopt lines from different elements, e.g., \Oiii{} 1$\rightarrow$0 and \Nii{} 2$\rightarrow$1 transitions, which have relatively similar critical densities (534 and 272\,cm$^{-3}$, respectively), but different ionization energies (35.12\,eV vs.~14.53\,eV). Breaking the degeneracy between the intensity and hardness of the radiation field, however, is significantly more challenging, as it requires observing transitions of different ions of the same element for various elements (e.g., \Niii{}/\Nii{}, \Oiv{}/\Oiii{}, \Neiii{/\Neii{}}, etc.) after controlling for density, due to the different critical densities of the involved lines. Therefore, because at least three or more lines are required to achieve this (including at least one with a high ionization potential), such combinations are typically observationally expensive; although potentially very useful to distinguish between ages of young stellar populations.

\subsection{Abundances and metallicity}\label{sec_metal}

Measuring metallicity requires comparing the emission of metal and hydrogen lines. Hydrogen atoms and molecules however do not emit in the FIR bands, where a number of the FSLs discussed here reside. Thus, metallicity estimates via IR FSL emission often have to rely on tracking the evolution of the relative abundance of different elements as a function of metallicity. When considering gas-phase metallicity, it is common practice to pin it to that of oxygen: $Z$\,=\,log\,[O/H]\,+\,12. This is commonly normalized to the solar value, which in this case we assume to be log\,Z$_\odot$\,=\,8.69 \citep{asplund09}. In Fig.~\ref{fig_Z} we show the abundance patterns of carbon, nitrogen, and helium as a function of gas-phase metallicity, following \citet{nicholls17}:
\begin{equation}\label{eq_C_O}
\log {\rm [C/O]}=\log \left(10^{-0.8}+10^{\log {\rm [O/H]}+ 2.72}\right)
\end{equation}
\begin{equation}\label{eq_N_O}
\log {\rm [N/O]}=\log \left(10^{-1.732}+10^{\log {\rm [O/H]}+ 2.19}\right)
\end{equation}
\begin{equation}\label{eq_He_H}
\log {\rm [He/H]}=-1.0783 + \log \left(1+0.1703\,\frac{\rm Z}{\rm [Z_\odot]}\right),
\end{equation}

Equations~\eqref{eq_C_O}--\eqref{eq_He_H} imply that, for $Z\lsim 0.1$\,Z$_\odot$, abundances of all elements scale linearly with $Z$, and relative elemental abundances are therefore constant, implying that FLSs are not sensitive to metallicity variations in this regime. Beyond 0.1\,Z$_{\odot}$, the onset of secondary production of metals leads to an increase of C/O and N/O (and, to lesser extent, N/C), providing us with diagnostics of metallicity based on IR FSLs. Hence, IR FSLs can act as a proxy of metallicity that is virtually uneffected by reddening. 

The main drawback of such metallicity estimates is the relatively small dynamical range of relative elemental abundance ($\simeq$\,0.5\,dex) over a larger range in metallicities (0.1\,$\lsim Z$\,/\Zsun\,$\lsim$\,1), once compared to the intrinsic scatter and the methodological unknowns. Uncertainties on the measured line intensities due to dependencies on the gas density, or on the intensity and hardness of the ionization field can hamper robust derivations of metallicity. Access to various transitions of the same species can alleviate some of these degeneracies. For instance, having access to the \Oiii{}\,52 and 88\,$\mu$m transitions, in addition to the \Niii{}\,57\,$\mu$m line, can account and control for the electron density and reduce scatter in the N/O ratio down to $\lsim$\,0.2\,dex \citep{chartab22}. In the case of carbon, the fraction of \Cii{} arising from the ionized or neutral medium adds to the uncertainty in the interpretation of the observed line ratios. This fraction is usually constrained via the \Cii{}\,158\,$\mu$m / \Nii{}\,205\,$\mu$m ratio, since the two lines have similar critical densities, but the latter only arises from the ionized medium. However, the correction does rely on an assumption on the relative abundance of [C$^+$/N$^+$] in the ionized gas.

The calibration of different recipes adopted in the derivation of both $Z$ and the relative elemental abundances also introduce systematic offsets. In Fig.~\ref{fig_Z}, the [N/O] ratio is shown as a function of $Z$ for emission line galaxies in the SDSS catalog, where the [N/O] is computed based on \citet{pilyugin16} and $Z$ is inferred following \citet{charlot01}; the figure also displays \Hii\, regions in local galaxies measured in the CHAOS survey \citep{berg20}. The large scatter and systematics in the calibration make metallicity estimates based on relative abundances challenging, at least on absolute grounds. To reduce the impact of these uncertainties, the use of optical-band diagnostics, directly involving hydrogen lines, is advisable. We point the interested reader to dedicated reviews on this topic \citep[e.g.,][]{asplund09, tolstoy09, nomoto13, maiolino19, romano22}.

\section{Brief history of IR FSL observations}\label{sec_history}

In the following, we briefly review how the field of IR FSL investigations progressed from an observational point of view.

\subsection{The nearby Universe}
\label{sec_lowz}

\subsubsection{The Milky Way}
\label{secsec_mw}

The opacity of the Earth atmosphere hinders the study of the FIR sky. The first theoretical models investigating the IR FSL emission in the Milky Way came in the late 1960s and early 1970s \citep[e.g.,][]{petrosian69,petrosian70,dalgarno72,simpson75}. The first line detections became possible thanks to the pioneering work with a 30\,cm telescope mounted on a Lear Jet and with a 91\,cm telescope mounted on the Kuiper Airborn Observatory (KAO): \citet{ward75} first reported an 8-$\sigma$ detection of the \Oiii{}\,88\,$\mu$m line in the Omega Nebula (Messier 17), soon after confirmed at higher spectral resolution by \citet{baluteau76}. In the following years, a number of other line transitions were reported: \Siii{}\,18\,$\mu$m \citep{baluteau76}, \Oiii{}\,52\,$\mu$m \citep{melnick78},  \Oi{}\,63\,$\mu$m \citep{melnick79,baluteau81}, \Niii{}\,57\,$\mu$m \citep{moorwood80}, \Ci{}\,609\,$\mu$m \citep{phillips80}, \Cii{}\,158\,$\mu$m \citep{russell81}, \Oi{}\,146\,$\mu$m \citep{stacey83}, \Ci{}\,370\,$\mu$m \citep{jaffe85}, \Nii{}\,122\,$\mu$m \citep{colgan93}. Curiously, the \Cii{}\,158\,$\mu$m line that twenty years later opened the study of the high-redshift FSL emission, represented a major challenge to these early-days observations of Galactic nebulae, partially because of the intrinsically very extended distribution, which was subtracted off by relatively small chopping in the observations; partially due to lack of sensitivity of available detectors; and in some extent to uncertainties in the expected rest-frame line frequency. Nonetheless, these first observations, aided by campaigns in the sub-mm and radio regimes \citep[e.g.,][]{crawford85}, helped shaping models for ``Photon-dissociation regions'' \citep[PDR; see][for reviews]{hollenbach97,wolfire22}.

The Far Infrared Absolute Spectrophotometer (FIRAS) on board the Cosmic Background Explorer (COBE) provided the first all-sky spectral line survey at FIR wavelengths, albeit with an angular resolution of $\sim 7^\circ$ (which limited its mapping capabilities to large-scale Galactic features). This enabled the first Galactic-scale assessment of the distribution of dust, \Cii{}, and \Nii{} \citep{wright91}. These maps revealed that the \Cii{} emission has a spatial distribution similar to that of the dust, and accounts for $\sim 0.3$\,\% of the total dust luminosity, whereas the \Nii{}\,122\,$\mu$m and \Nii{}\,205\,$\mu$m lines emit 0.04 and 0.03\,\%, respectively.

The \textit{Infrared Space Observatory} by the European Space Agency \citep{kessler96} led to fundamental steps forward in the study of IR FSLs in star-forming regions within the Milky Way. The Short-Wavelength Spectrometer allowed to sample the wavelength range 2.38 to 45.2\,$\mu$m at a spectral resolution 1000--2000 \citep{Leech03}, and delivered IR FSL detections for ions of argon, neon, sulfur, and silicon \citep[e.g.,][]{rosenthal00}. The Long-Wavelength Spectrometer enabled spectroscopy at 43--198\,$\mu$m \citep{clegg96}, thus yielding systematic observations of \Nii{}, \Niii{}, \Oiii{}, \Cii{}, \Oi{} in (proto)planetary nebulae \citep[e.g.,][]{liu01}, in
star-forming regions \citep[e.g.,][]{leeks04}, in
young stellar objects \citep[e.g.,][]{liseau06}, and other Galactic environments.

It was not until the launch of the \textit{Herschel Space Observatory} \citep{pilbrat10}, however, when Galactic FSL science experienced an extraordinary boom. Hundreds of individual sources and extended regions in the Milky Way were observed during the telescope's $\sim$\,4 years of continued operations, before the coolant ran out. One of the most important surveys performed was the Galactic Observations of Terahertz C+ (GOT C+) survey, which targeted the \Cii{} line to study the diffuse ISM across more than a hundred lines-of-sight (LOS) through the Galaxy using the HIFI instrument \citep{degraauw10}. Early results from this program analyzed optically thin clouds characterized by having \Cii{} and \Hi\, emission, but no detectable CO, and found that \Cii{} emission is generally stronger than expected for diffuse \Hi\, clouds.
This suggested that the excess \Cii{} emission identified in these clouds is best explained by the presence of a significant diffuse warm H$_2$, `dark gas' component \citep{langer10}. This CO-dark H$_2$ gas accounts, on average, for $\sim$\,30\% of the molecular mass of the Milky Way \citep{pineda13}. Further studies of the \Nii{}\,122\,$\mu$m and \Nii{}\,205\,$\mu$m lines along around a hundred LOS through the inner Galaxy yielded relatively large electron densities, n$_e \simeq 10-50$\,cm$^{-3}$ \citep{goldsmith15, pineda19}, larger than those expected from the diffuse warm interstellar medium ($\sim$\,0.1\,cm$^{-3}$) but lower than those typical of compact \Hii\, regions ($>$\,$5\times\,10^{3}$\,cm$^{-3}$).

The Orion Nebula complex was also investigated in detail with \textit{Herschel}. Studies based on velocity-resolved \Cii{} observations revealed that the stellar wind originating from the massive, O7V star $\theta^1$ Orionis C has created a `bubble' of roughly two parsec radius, which is slowly expanding at $\sim$\,13\,km\,s$^{-1}$, indicating that the mechanical energy from the stellar wind is being converted very efficiently into kinetic energy of the enveloping neutral shell, and can cause more disruption in the surrounding ISM than do photo-ionization and evaporation, or future supernova explosions \citep{pabst19}. Other nebulae associated with later-type massive, B-type stars also showed bubbles expanding at slower velocities $\lesssim$\,5\,km\,s$^{-1}$, but in this case likely caused by the thermal expansion of the over-pressurized, ionized gas \citep{pabst20}. Spatially resolved studies of the Orion Bar showed that most ($\sim$\,80\%) of the \Cii{} emission originates from PDRs \citep{bernard-salas12}. Finally, \cite{goicoechea13} showed that the spectrum toward Sgr\,A* is dominated by strong \Oiii{}, \Oi{}, \Cii{}, \Nii{}, and \Ci{} fine-structure lines arising in gas irradiated by UV photons from the central stellar cluster.

\subsubsection{Star-forming galaxies and AGN}
\label{secsec_lirgs}

In the 1980s and 1990s, the KAO first, and the ESA's Infrared Space Observatory (ISO) later, really opened up the field of extragalactic IR FSL observations \citep[see, e.g.,][]{watson84, crawford85, lugten86, stacey91, madden93, lord96, malhotra97, malhotra01, fischer99, luhman03}. Line detections included \Oiii{}\,52 and 88\,$\mu$m, \Oi{}\,63 and 146\,$\mu$m, \Nii{}\,122\,$\mu$m, \Niii{}\,57\,$\mu$m, besides obviously \Cii{}\,158\,$\mu$m. The rapidly growing samples (encompassing a few hundred local galaxies by the mid 2000s; see \citealt{brauher08}) enabled the development of diagnostics of the ISM physics based on IR FSLs \citep[see, e.g.,][]{sturm02, fischer14}.  

Local star-forming galaxies were also systematically targeted by \textit{Herschel} and the \textit{Stratospheric Observatory for Infrared Astronomy} (\textit{SOFIA}) later on. The Magellanic clouds \citep{lebouteiller12, meixner13, chevance16, pineda17}, M\,33 \citep{mookerjea11, kramer20}, M\,51 \citep{pineda20}, or those included in the KINGFISH sample \citep{kennicutt11} where observed in a spatially resolved manner, reaching physical scales of at least one to a few kpc, or even sub-kpc resolution in the most nearby sources. Most galaxies were targeted in the main FSLs, including \Oi{}\,63\,$\mu$m, \Oiii{}\,88\,$\mu$m, \Nii{}\,122 and 205\,$\mu$m, and \Cii{}\,158\,$\mu$m, which allowed to perform for the first time a detailed analysis of the resolved ISM in nearby, star-forming and Seyfert-host galaxies and study their gas properties as a function of their resolved energy sources \citep{contursi13, smith17, herrera-camus18a, herrera-camus18b}.

Starburst galaxies and Luminous Infrared Galaxies (LIRGs; $L_{\rm IR, 8-1000\mu m}\,=\,$\,10$^{11-12}$\,\Lsun) were also the case study of several large \textit{Herschel} programs. Observations of the most prominent FSLs in these galaxies provided an unprecedented view on the impact of strong AGN activity on the properties of their ISM and the process of star-formation, and the environmental effects associated to dynamical instabilities such as galaxy interactions and mergers \citep{luhman03, gracia-carpio11, diaz-santos13, spinoglio22b}. These observations allowed to connect the small-scale physics associated to PDRs with the large-scale phenomena that characterize the integrated properties of galaxies, such as their star-formation efficiency, their dust temperature and infrared luminosity surface density, \SigmaIR, or their merger stage \citep{diaz-santos17}. When mid- and FIR FSLs are combined together, they also have the power to disentangle between galaxy populations (dwarfs, starbursts, Seyferts, \Hii-region dominated galaxies) and their nuclear energy source (star formation vs. AGN) based on their sensitivity to the intensity of the ionization field and the gas density \citep{fernandez-ontiveros16, spinoglio22b}.

Studies of Ultra-luminous IR Galaxies (ULIRGs; $L_{\rm IR, 8-1000\mu m} > $\,10$^{12}$\,\Lsun) showed that gas cooling via FSLs is dominated by \Cii{}\,158\,$\mu$m and \Oi{}\,63\,$\mu$m emission. Both lines have similar strength and their combined contribution can account for up to $\simeq$\,60--80\% of the total IR line emission budget \citep{rosenberg15}. It was also found that, while these sources have on average larger contributions from their central, quasar-like engine, the emission from FIR lines mostly arise nonetheless from gas heated by starlight, with no significant influence from their AGN activity, except in the most extreme cases \citep{farrah13}. On the other hand, it was also discovered that gas heated by low velocity shocks due to injection of mechanical energy from jets generated by Seyfert nuclei can boost \Cii{} emission \citep{guillard15, appleton18}. Moreover, \Cii{} can also be enhanced in large-scale turbulent regions induced through galaxy collisions, where the energy is dissipated to small scales and low velocities, via shocks and turbulent eddies \citep{appleton13}.

\subsubsection{Dwarf galaxies and metallicity tracers}
\label{secsec_lowmet}

In recent years, FIR FSLs have also been used to measure accurate gas-phase metallicities, avoiding some of the challenges and biases attached to the use of optical emission lines, such as temperature dependencies and dust attenuation, which can be specially severe in compact and/or dusty star-forming galaxies. In particular, the (\Oiii{}\,52\,$\mu$m + \Oiii{}\,88\,$\mu$m) / \Niii{}\,57\,$\mu$m ratio, due to the secondary production of nitrogen by intermediate-mass stars relative to oxygen (see Sect.\ref{sec_abundances}), is a good tracer of metallicity at absolute abundances 12\,+\,log(O/H)\,$\gtrsim$\,7.7, as this ratio only depends weakly on density and ionization parameter \citep{nagao11}. A number of studies have put this method to work and estimated the metallicity of dusty galaxies in the local universe, finding that these sources follow the mass-metallicity relation and that the lower metallicities derived from optical lines is a result of heavily obscured metal-rich gas, which has a negligible effect when using FSL diagnostics \citep{pererira-santaella17, fernandez-ontiveros21, spinoglio22a}. However, discrepancies in the calibration of the relative N/O abundance as a function of $Z$ can introduce artificial, $\gtrsim$\,0.3\,dex offsets between absolute metallicity values derived from optical and FIR lines (see Sect.~\ref{sec_metal}). Thus, attention should be paid before comparing metallicities obtained from different line ratios \citep{chartab22}.

Regarding low metallicity galaxies specifically, in a systematic survey of dwarf galaxies in the nearby Universe \citep{madden13} spanning oxygen abundances in the 7.0\,$\lesssim$\,12\,+\,log(O/H)\,$\lesssim$\,8.5 range, \cite{cormier15} showed that compared to metal-rich galaxies, moderate far-UV fields and a lower PDR covering factor were necessary to explain some of the FIR emission line ratios observed in low metallicity systems. However, it was also found that the amount of \Cii{}\,158\,$\mu$m arising from ionized gas was still small, typically less than 30\% \citep{cormier19}, and comparable to normal star-forming galaxies and dusty LIRGs \citep{croxall17, diaz-santos17}. Exploring the main effects of interstellar dust on the chemical abundances measured in galaxies as a function of redshift is an ongoing endeavor that also requires advances in our theoretical knowledge and modeling \citep{calura25}.

\subsection{Observations at high redshift}\label{sec_highz}

\begin{figure}[ht]
\begin{center}
\includegraphics[width=0.47\columnwidth]{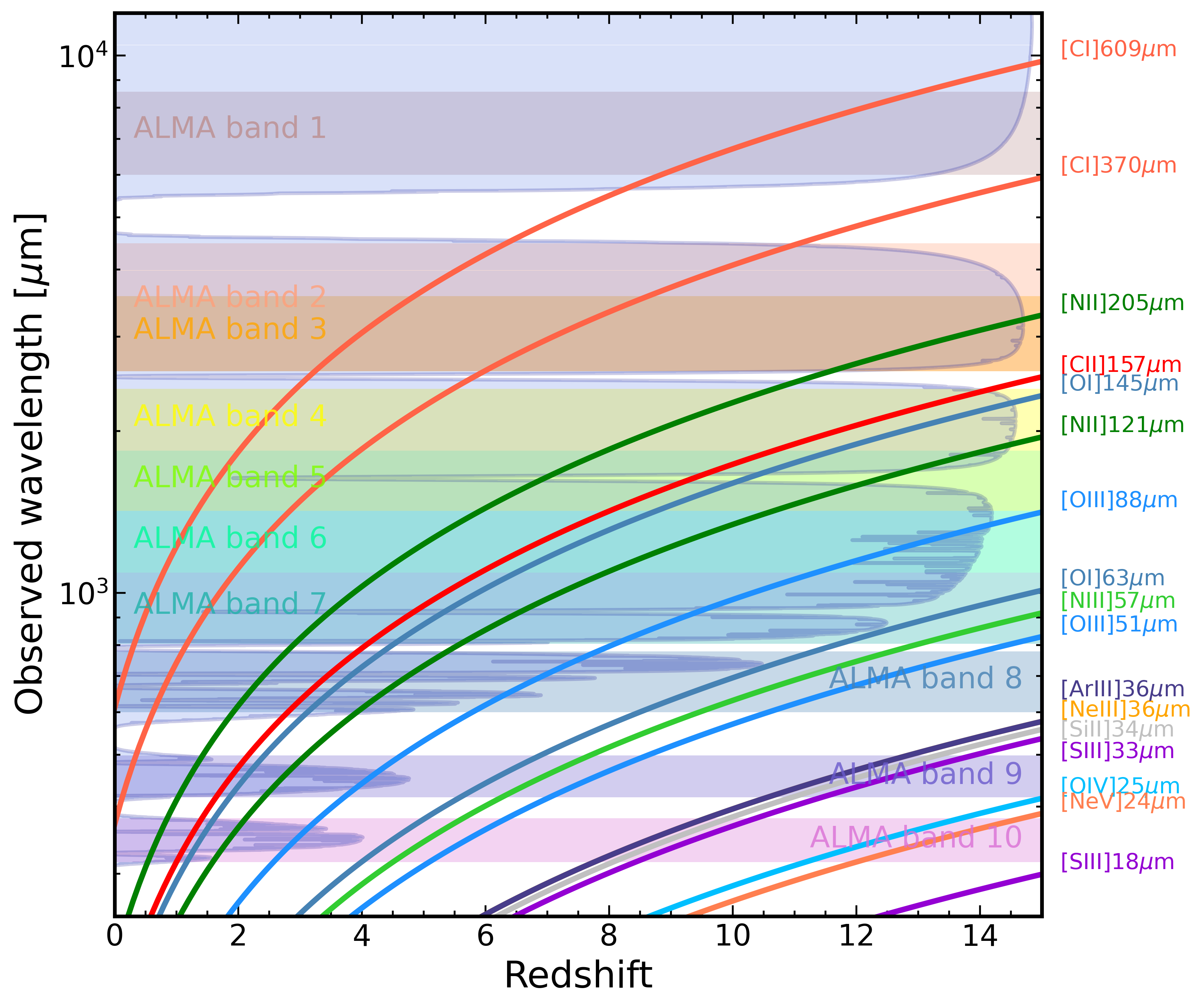}
\includegraphics[width=0.51\columnwidth]{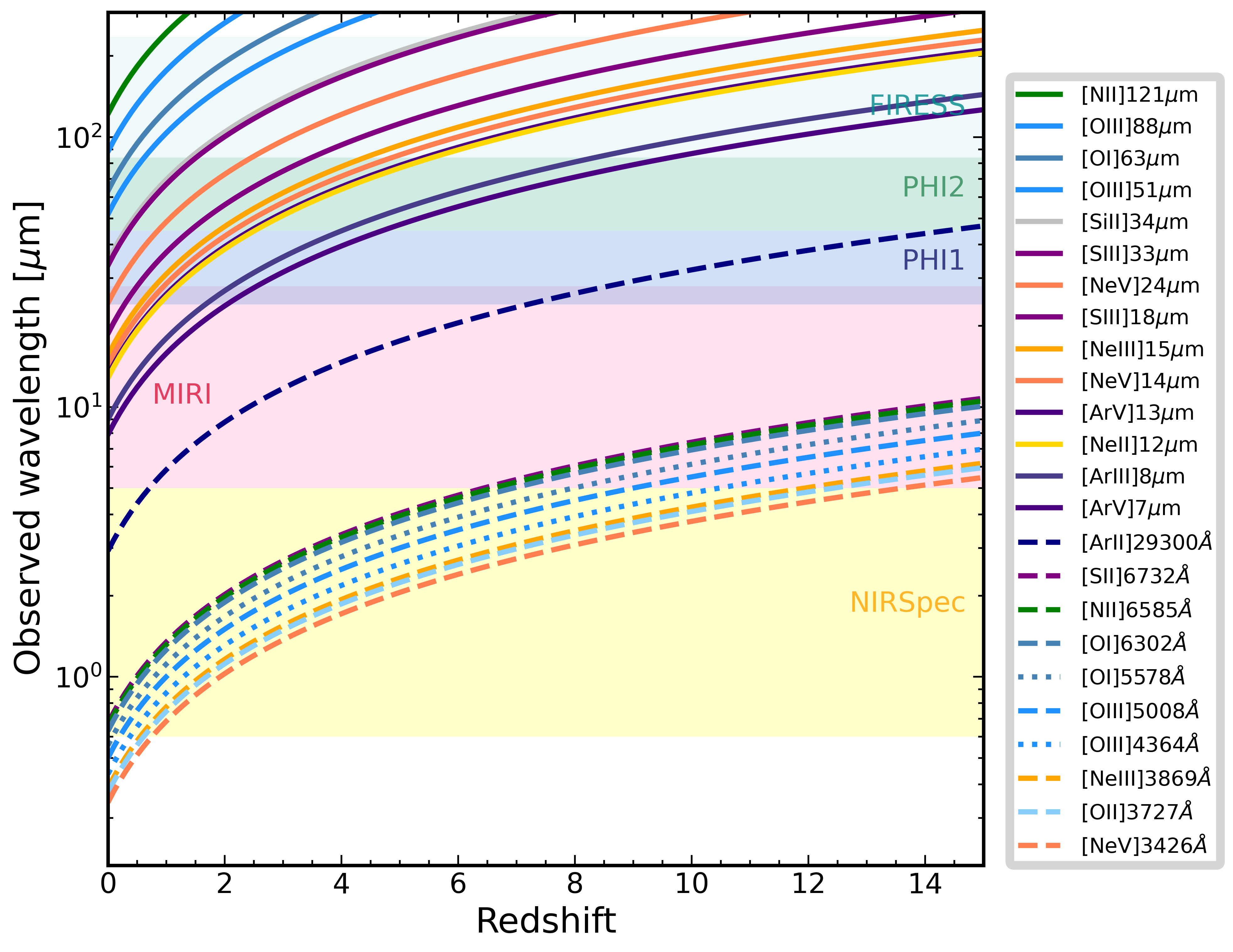}
\end{center}
\caption{Observed line wavelengths of various emission lines as a function of redshift, in the mm (\emph{left}) and mid-IR (\emph{right}) wavelength regimes. Far-IR FSLs are plotted with solid lines. For comparison, the transparency of the mm sky is shown in the horizontal axis in the left-hand panel. The ALMA, JWST, and PRIMA bands are also shown in colors. Most of the IR FSLs are observable at mm wavelength in the early cosmic ages; at these redshifts, rest-frame optical lines are observable with MIRI.}
\label{fig_z_nu}
\end{figure}

Because of their ubiquity in the ISM, their luminosity, and because they are virtually unaffected by dust reddening, IR FSLs at high redshift are pivotal in order to explore the evolution of the ISM in galaxies as a function of cosmic time. A peculiar advantage of studying the IR emission of high redshift sources is that, because of the cosmological redshift, the lines are shifted in the transparent windows of the atmosphere at sub-mm and mm wavelengths (see Fig.~\ref{fig_z_nu}).

\subsubsection{Brief historical overview}

The IRAM 30m telescope and the IRAM Plateau de Bure Interferometer played a crucial role in opening the window of IR FSL observations at high redshift ($z>1$). \citet{weiss03} reported the detection of the \Ci{}\,370\,$\mu$m in the Cloverleaf, a highly magnified, IR luminous quasar at $z\sim 2.5$, and wrote: ``Together with the detection of the lower fine structure line \citep{barvainis97}, the Cloverleaf quasar is now only the second extragalactic system, besides M~82, where both carbon lines have convincingly been detected.'' --- a testament to the pioneering role of such observations. More \Ci{} detections came in the next few years \citep{pety04, weiss05, wagg06, ao08, casey09, riechers09}. These studies provided independent measurements of the gas mass to compare with the ones based on CO, and led to first insights on the gas excitation properties. 
\citet{maiolino05} reported the first detection of \Cii{}\,158\,$\mu$m at $z>0.1$, this time for a quasar at $z\sim 6.4$, J1148+5251; this was followed by \citet{iono06} and \citet{walter09nature}. These works established the key role of \Cii{} in measuring gas morphology and kinematics in high-$z$ galaxies. \citet{walter09} set the first tight limits on the \Nii{}\,205\,$\mu$m in a high-$z$ galaxy, opening the way to multi-tracer investigations of the ISM conditions. 

The year 2010 was a watershed moment for the study of IR FSLs at high redshift, as multiple facilities started to contribute to this quest. \citet{haileydunsheath10} and \citet{stacey10} presented the first \Cii{} detections secured via ZEUS on the Caltech Submillimeter Observatory (CSO). The high frequency capabilities of ZEUS, paired with the low water vapour column densities on Mauna Kea, enabled the search for even higher frequency lines, such as \Oiii{}\,88\,$\mu$m \citep{ferkinhoff10}.
\citet{wagg10} reported the first detection of \Cii{} using APEX, thus opening up the Southern sky to IR FSL studies at high redshifts.
\citet{ivison10} used \textit{Herschel}/FTS to secure a multi-line study of SMM J2135--0102 at $z=2.33$, including coverage of the \Cii{}\,158\,$\mu$m, \Nii{}\,122\,$\mu$m, \Oiii{}\,88\,$\mu$m, \Oi{}\,146 and 63\,$\mu$m transitions, albeit only the former line was detected. 
The advent of the Atacama Large Millimeter/sub-millimeter Array (ALMA) in 2012 marked the definitive change of gear in the study of IR FSLs at high redshift. The unprecedented sensitivity offered by ALMA, in particular at high frequencies ($>$300 GHz), triggered an exponential growth in the number of IR FSLs detected in the distant Universe: by the end of 2012, 93 IR FSL observations of $z>1$ galaxies had been reported in the literature. By the end of 2024, they reached 1500 for more than 600 individual sources!

\begin{figure}[ht]
\begin{center}
\includegraphics[width=0.75\columnwidth]{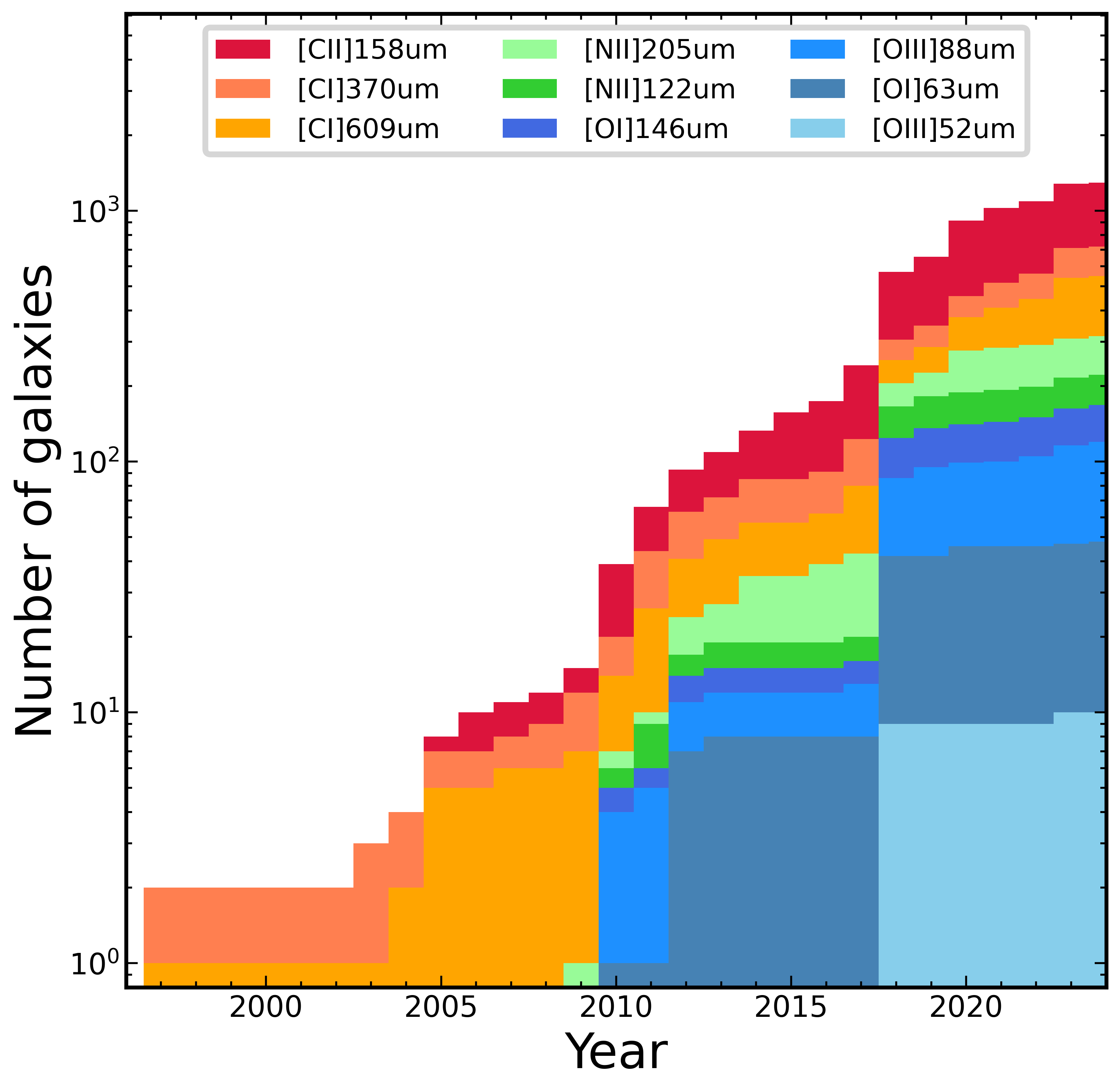}
\end{center}
\caption{Cumulative distribution of $z>1$ galaxies observed in fine-structure lines as a function of publication year, color coded by emission line. More than 500 individual galaxies at $z>1$ have been observed in at least one IR FSL.}
\label{fig_year_distr}
\end{figure}

\begin{figure}[ht]
\begin{center}
\includegraphics[width=0.75\columnwidth]{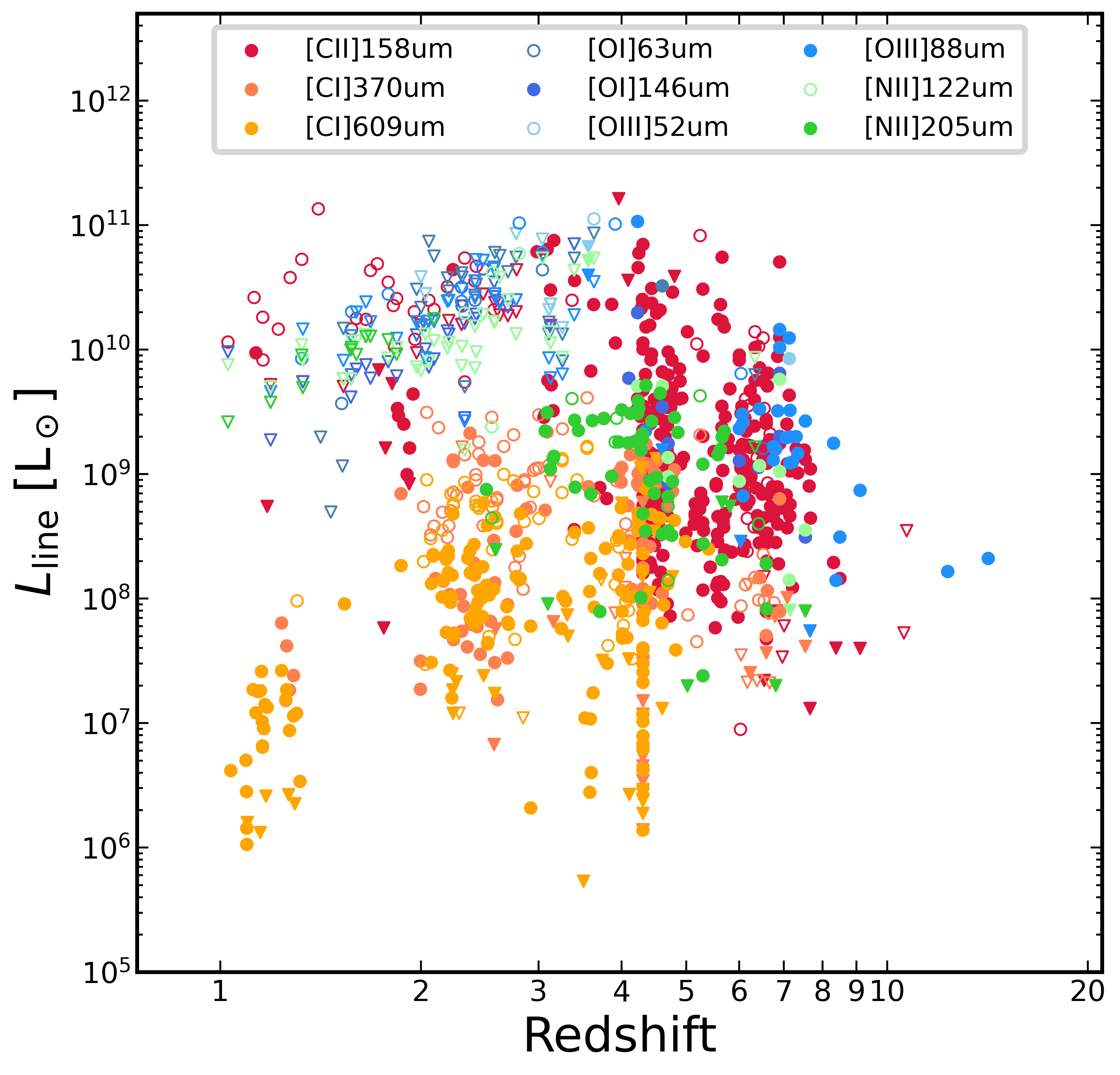}
\end{center}
\caption{IR fine-structure line luminosity as a function of redshift reported in the literature. The color code reflects the emission line. Only one entry per line per galaxy is shown. Filled symbols refer to ALMA observations. Circles mark detections, triangles mark 3-$\sigma$ upper limits. Observations so far have covered more than 2 dex in luminosity for many of these FSLs at high redshift.}
\label{fig_lum_z}
\end{figure}

\begin{figure}[ht]
\begin{center}
\includegraphics[width=0.75\columnwidth]{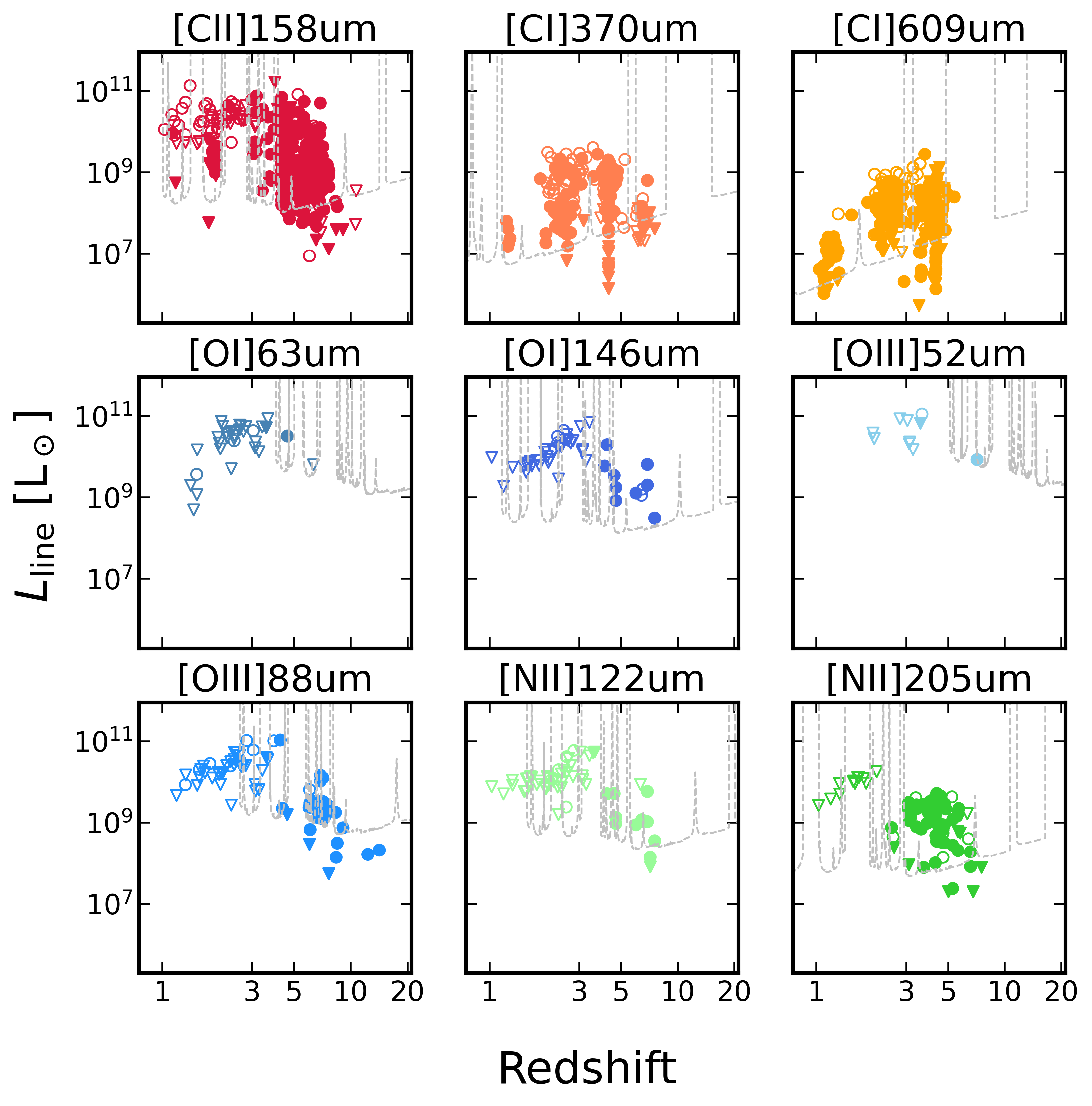}
\end{center}
\caption{Same plot as in Fig.~\ref{fig_lum_z}, but broken down in different emission lines. Overplotted in grey is the 5-$\sigma$ sensitivity that can be reached with 1\,hr of integration with ALMA, assuming a line width of 300 \kms{}. These IR FSLs are now accessible with relatively modest integration time even at very high redshift.}
\label{fig_lum_z_perline}
\end{figure}

Figure~\ref{fig_year_distr} shows the number of galaxies at $z\,>\,1$ with FSL observations compiled for this review as a function of the year of the observations\footnote{The galaxy catalog was compiled from an exhaustive literature search that is complete to be best of the authors' knowledge, expanding up to the end of 2024. For a similar effort conducted independently in parallel to our work, see \citet{peng25}.}. Figures~\ref{fig_lum_z} and \ref{fig_lum_z_perline} present the distribution of galaxies as a function of redshift, all together and per emission line, respectively. For reference, Fig.~\ref{fig_lum_z_perline} also shows typical ALMA sensitivity limits (see figure caption for details).

\begin{figure}[htbp]
\begin{center}
\includegraphics[width=0.49\columnwidth]{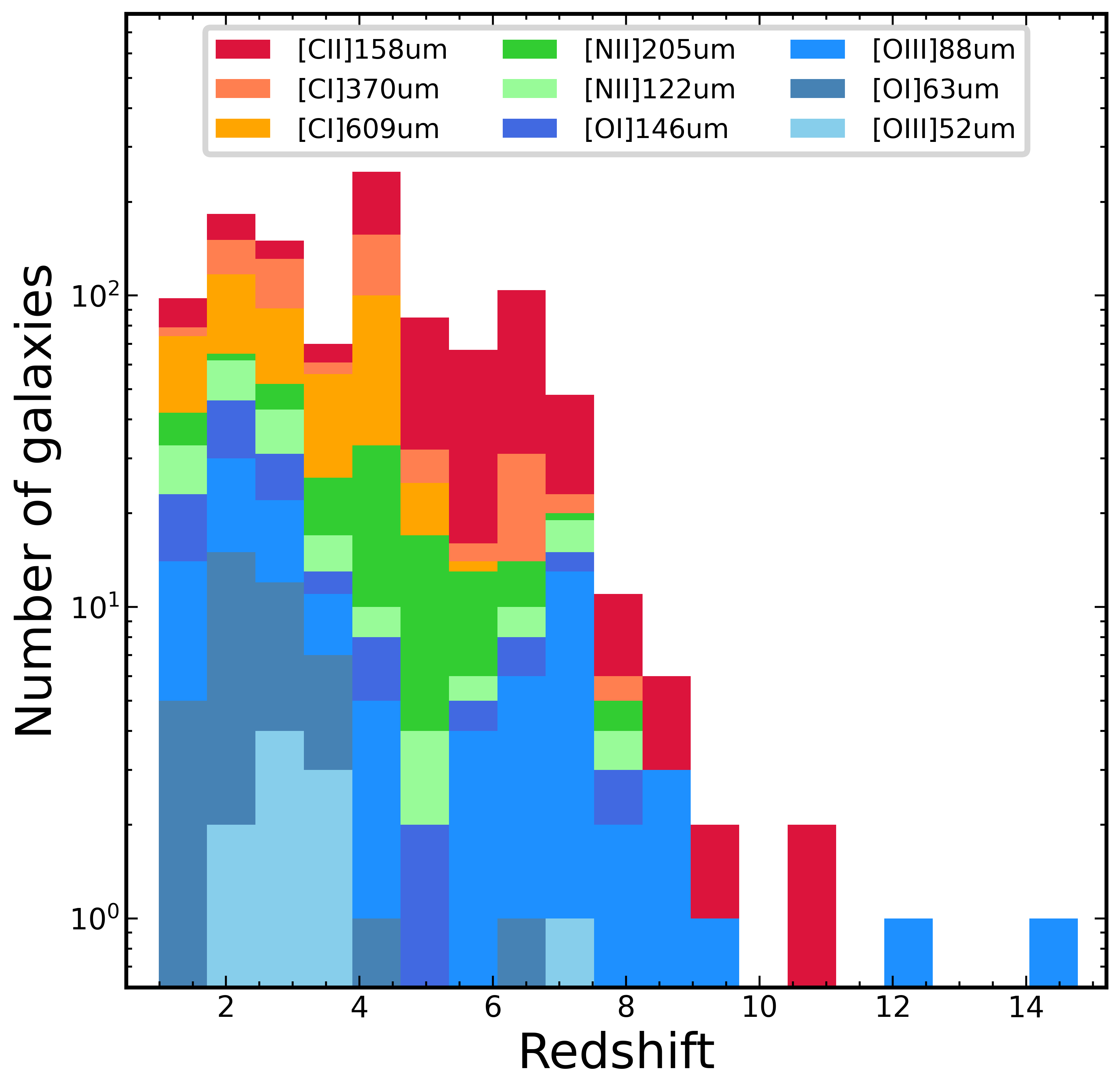}
\includegraphics[width=0.49\columnwidth]{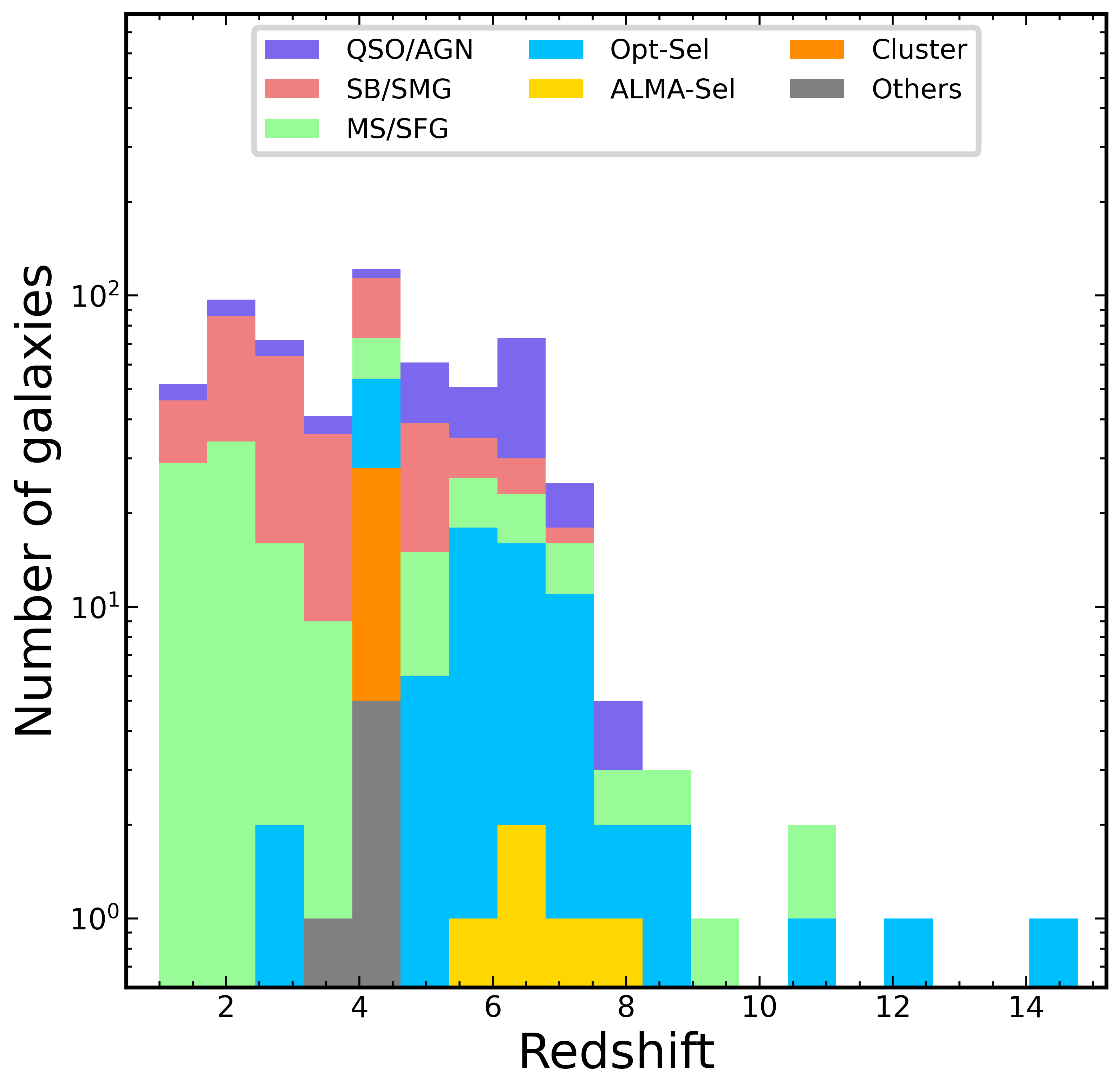}
\includegraphics[width=0.49\textwidth]{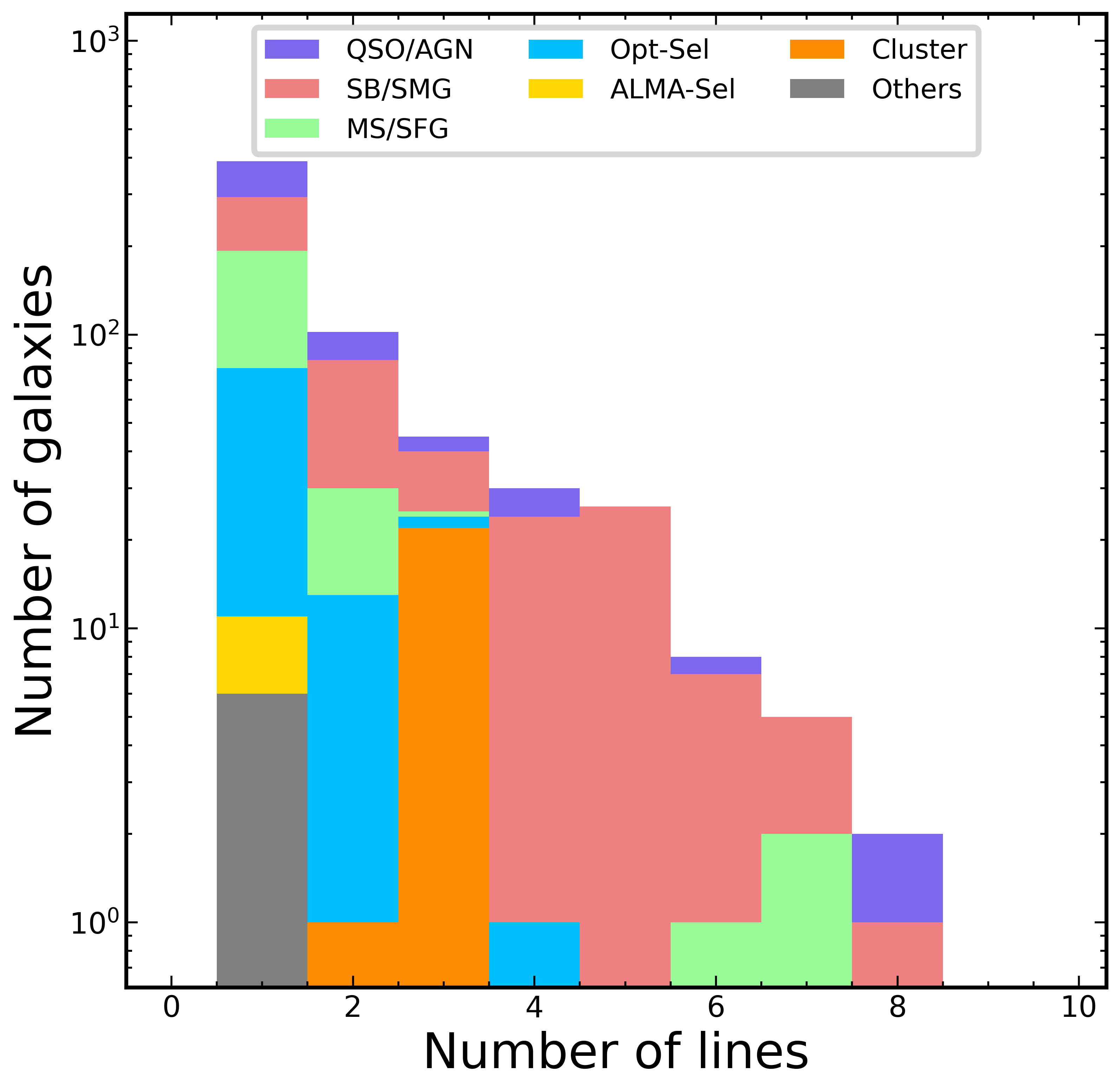}
\end{center}
\caption{\emph{Top panels:} Redshift distribution of $z>1$ galaxies observed in their IR FSLs, color-coded by emission line (\emph{left}) and galaxy type (\emph{right}). Here `Opt-Sel' refers to optically-selected galaxies, such as Ly$\alpha$ emitters and Lyman break galaxies; `ALMA-Sel' refers to serendipitous line emitters detected with ALMA in the field of other sources; `Cluster' refers to cluster members; and `Others' is a label for a number of less represented types (quiescent galaxies, DLA host galaxies, etc). Atmospheric windows and target selection functions impact the shape of these distributions. For example, hundreds of quasars are known at $z>6$, where multiple emission lines are accessible; on the other hand, blind selection of SMGs based on the dust continuum biased the samples to cosmic noon, where these sources are more common. \emph{Bottom panel:} Number of $z>1$ galaxies observed in their IR FSLs, as a function of the number of IR FSLs, color coded by galaxy type. The number of galaxies with observations in multiple lines is rapidly growing.
}
\label{fig_z_distr}
\end{figure}

Figure~\ref{fig_z_distr} shows the histograms of the number of galaxies in the catalog as a function of redshift, color-coded by emission line (top-left panel) and source type/selection (top-right panel). For the galaxy types, we consider the main class used in the source selection from the literature: quasars and AGN from various classification techniques; sub-mm galaxies and starbursts, mostly selected via their luminous dust continuum emission in the infrared wavelengths; main sequence and more `typical' star-forming galaxies; optically--selected galaxies (primarily Lyman Break Galaxies and Ly$\alpha$ emitters); line emitters identified with interferometric observations; cluster members of diverse type. Boundaries among these classes are sometimes loose (e.g., some SMGs may host AGN); a more rigorous zoology however transcends the scope of this review. The bottom panel of the figure displays the number of galaxies as a function of the number of lines observed for each source.

The leap in sensitivity offered by ALMA, together with the improved capabilities of the NOrthern Extended Millimeter Array (NOEMA) and the Jansky Very Large Array (JVLA) enabled to switch from extremely IR-luminous sources to more ``normal'' classes of galaxies. After multiple attempts with PdBI \citep[e.g.,][]{walter12grb,gonzalezlopez14,ota14}, \citet{willott15} and \citet{capak15} succeeded in measuring the \Cii{} emission in the first samples of optically-selected galaxies at $z>5$ using ALMA, while \citet{pavesi16} secured the first \Nii{}\,205\,$\mu$m detections in `typical' star-forming galaxies at high redshift with ALMA. \citet{inoue16} and \citet{carniani17} presented the first \Oiii{}\,88\,$\mu$m detections in `normal' galaxies at cosmic dawn.

Thanks to its intrinsic luminosity, and to observability reasons, \Cii{} quickly arose as the workhorse for studying high-redshift galaxies, to the point that about half of the IR FSL observations at $z>1$ in the literature concern \Cii{}. In the following, we discuss the main scientific goals of studying IR FSLs in high-$z$ galaxies over the last few years.

\subsubsection{Measuring redshifts}

\begin{figure}[htbp]
\begin{center}
\includegraphics[width=0.89\columnwidth]{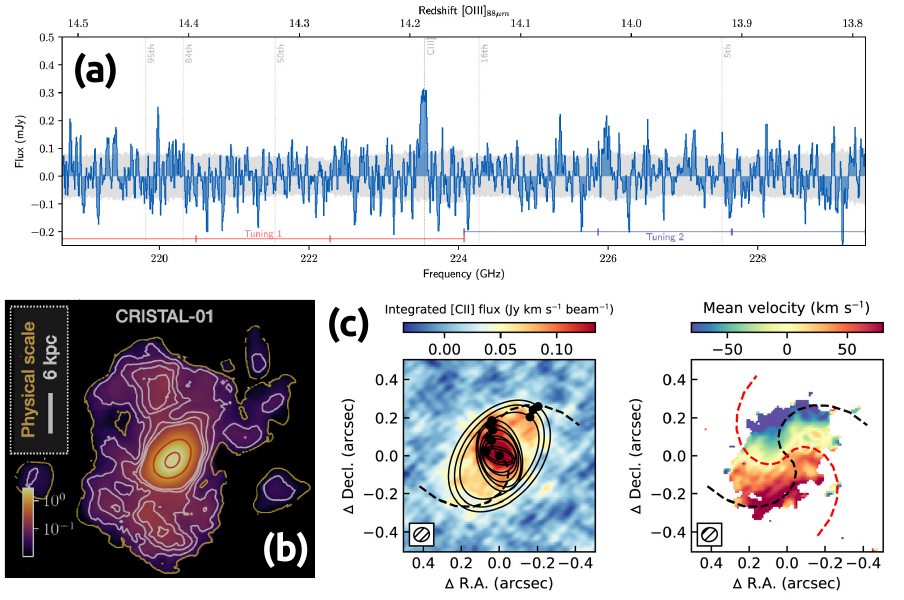}
\end{center}
\caption{Examples of recent results obtained via FSL observations of high--$z$ galaxies. Panel (a): the detection of \Oiii{} 88\,$\mu$m in GS-z14 enabled a precise redshift measurement ($z=14.193$) and enabled constraints on the ISM physics in the most distant galaxy known at that time. Figure adapted from \citet{schouws24}. Panel (b): \Cii{} map of the CRISTAL-01 galaxy, at $z\approx4.5$, showing a complex morphology extending over tens of kpc. Adapted from \citet{herrera-camus25}. Panel (c): Maps of \Cii{} emission and velocity field in the host galaxy of the quasar PJ036+03 at $z\approx6.5$. The morphology and kinematics of the gas emission reveal the presence of a warped disk in this quasar host galaxy. Figure adapted from \citet{neeleman23}.}
\label{fig_literature}
\end{figure}

The typical galaxy at $z>5$ is identified via its emission in the rest-frame UV/optical bands, which are shifted toward the edge of the optical band and into the near-IR at these redshifts. The sharp drop in observed flux associated Ly$\alpha$ absorption due to the Gunn--Peterson effect \citep{gp65} leads to photometric redshift measurements as inaccurate as $\delta z \gsim 0.1$. Spectroscopic follow-up at optical/near-IR wavelengths is often challenging from the ground, even for 8-m-class telescopes \citep[e.g.,][]{vanzella14}, due to the heavy contamination from atmospheric lines (mostly from OH) and the intrinsic faintness of the targets. Observations of the \Cii{} or \Oiii{} lines with NOEMA and ALMA have represented a viable alternative to secure precise redshifts \citep[e.g.,][]{inoue16,knudsen16,laporte17}. These campaigns allowed to push the redshift frontier of spectroscopically confirmed galaxies at first to $z=9.1$ \citep{hashimoto18} and more recently to $z=14.2$ \citep[][see Fig.~\ref{fig_literature}a]{schouws24,carniani24}. Similarly, FSLs have been instrumental in order to secure the redshift of high--$z$ galaxies that are heavily obscured \citep[e.g.,][]{walter12,swinbank12,weiss13,sun25}.

Because of the intrinsic faintness of the sources (possibly worsened by the small metal content of these high redshift galaxies), and because of the modest redshift-space coverage of a single frequency tuning (with ALMA, $\Delta\nu/\nu\approx3\%$ at 1.2\,mm, corresponding to a \Cii{} coverage of $\Delta z \sim$\,0.1 at $z$\,=\,6.6; with NOEMA, this is about twice as wide), these efforts came at the cost of significant investment of observing time. In the last few years, the extremely sensitive spectroscopic capabilities offered by JWST and its wide simultaneous wavelength coverage ($\Delta \nu/\nu \gsim 1$ with NIRSpec) have made more observationally efficient to secure redshifts of typical high-$z$ galaxies via their rest-frame optical emission, unless they are heavily dust-obscured.

Nonetheless, observations of \Cii{} are still pivotal in order to measure very precise redshifts ($\sigma_z\lsim 50$\,\kms{}) in high-$z$ galaxies. Such accuracy is mandatory for specific science cases; e.g., in the study of kinematics of galaxies within structures, or in order to measure the near-zone radii of bubbles photoionized by quasars at cosmic dawn \citep[e.g.,][]{eilers20}. In the last years \Oiii{} has also become a crucial line to accurately determine the redshifts of the most distant galaxies known to date (at $z \gsim 12$), where the brightest optical lines fall far beyond the coverage of JWST.

\subsubsection{Measuring gas masses and/or star formation rates} 

The \Cii{} and \Ci{} lines have been repeatedly used to infer the total gas mass and the neutral gas mass, respectively. The relatively simple physics of IR FSLs allows us to infer gas masses from first principles, via assumptions on the excitation temperature of the gas, the carbon abundance, and the fraction of carbon atoms that are in different species; usually assumed to be in neutral or singly-ionized form (see Sect.~\ref{sec_masses} and \citealt{weiss03,weiss05,walter11,venemans17}). 

Alternatively, one can empirically calibrate the luminosity-to-mass conversion factor by comparing with independent estimates of the gas mass, derived either via the dust mass or the CO emission (see, e.g., \citealt{zanella18,dunne22,decarli22,friascastillo24}). Because none of these methods directly targets hydrogen, all these empirical methods rely on conversion factors (the carbon abundance, the gas-to-dust ratio, the CO-to-H$_2$ conversion factor $\alpha_{\rm CO}$, etc.) that are generally unknown or were calibrated on previous CO or dust measurements, and that also strongly depend on various galaxy properties (in particular, the gas-phase metallicity). For instance, Eq.~\eqref{eq_linelum_theo} can be re-written as: $M_{\rm gas}=\alpha_{\rm [CII]}\,L_{\rm [CII]}$. The analytical approach depicted in Sect.~\ref{sec_linepredict}, assuming $n=10,000$\,cm$^{-3}$, $T_\text{kin}=100$\,K, a neutral medium at solar metallicity (implying [C/H]=$2.6\times10^{-4}$, based on Eq.~\eqref{eq_C_O}), with half of the carbon in singly-ionized form, yields $\alpha_{\rm [CII]}=27.2$ \Msun{}/\Lsun{}, whereas the empirical calibration by \citet{zanella18} results in $\alpha_{\rm [CII]}=30$ \Msun{}/\Lsun{}.

A similar approach can lead to estimates of the ionized gas mass using, e.g., \Nii{} or \Oiii{}. This is a bit more challenging due to the relatively fainter emission of nitrogen lines and the high frequency of doubly-ionized oxigen lines. Furthermore, in the ionized gas the collision partners have a temperature $T\gg E_u/k_{\rm b}$, hence the fraction of the ions populating high-energy levels may be non-negligible. A robust estimate of the ionized gas fraction would thus require the combination of both IR FSLs and nebular lines.

Spatially-resolved maps of the \Cii{} emission in local galaxies show that its morphology closely resembles that of star formation rate tracers such as H$\alpha$ or the 24\,$\mu$m continuum \citep[see, e.g.,][]{rodriguezfernandez06} once images are degraded to a common angular resolution. This paved the way to the use of \Cii{} as a SFR tracer in distant galaxies \citep[e.g.,][]{delooze11,delooze14,herrera-camus16}. A qualitative explanation for \Cii{}--SFR relations is that the \Cii{} emission arises from the outer layers of molecular clouds as well as from the ionized gas within \Hii\, regions around young, massive, UV-luminous stars. In the local Universe, the \Cii{}--SFR relations are calibrated using a suite of SFR tracers, including H$\alpha$, 24\,$\mu$m continuum, UV and/or IR luminosities. Beyond the local Universe, studies have mostly focused on the $z>4$ Universe due to the easier access to the \Cii{} line in the (sub-)mm bands. At these redshifts, SFR estimators are typically limited to the rest-frame UV and IR luminosities; only very recently, JWST has provided access to the H$\alpha$ recombination line at high redshift, which can be used to trace instantaneous ($<$ a few Myr) star formation \citep[see review in][]{kennicutt12}. 

There are two main complications in the use of \Cii{} as SFR tracer. First, in metal-poor galaxies, \Cii{} emission tends to be lower than in solar-metallicity galaxies \citep{delooze14,vallini15,sutter19,casavecchia24}, as a natural consequence of the lower metal content. Second, a large fraction of IR-luminous galaxies appears deficient in \Cii{} emission (see Sect.~\ref{sec_deficit}), leading to a saturation in \Cii{} luminosity at growing IR luminosity (and hence, in the SFR estimates). The reliability of \Cii{} as a SFR tracer is thus subject to a general understanding of what type of galaxy is under consideration, and more specifically, which physical conditions dominate in the \Cii{}-emitting region, both in terms of the general heating mechanism and the density and dust content of the gas in and around the star forming regions \citep{diaz-santos17}. Some authors \citep[e.g.,][]{delooze14,herrera-camus18a} have proposed different recipes or multi-parameter scaling relations in order to account for these variations in the \Cii{}--SFR relation.

Clearly, the existence of `star formation laws' implies that the \Cii{} luminosity should display some degree of correlation with both $M_{\rm gas}$ and SFR; examples of either use are abundant in the literature \citep[e.g.,][]{heintz23,decarli23,bethermin23,hygate23,liang24,vanleeuwen24,zanella24}. Which is the most physically meaningful scaling relation reminds us of a `chicken or the egg' problem. Ultimately, the use of \Cii{} luminosity as a proxy for SFR builds on the response of the gas reservoir to the source of external radiation and to the physical conditions set by the local environment, while its use as a mass tracer reflects an intrinsic property of the gas itself.

\subsubsection{Line deficits}\label{sec_deficit}

To zero order, the luminosity of IR FSLs scales linearly with the FIR luminosity output of galaxies, \iLFIR\footnote{The FIR flux, $F_{\rm FIR}$, is defined as the integral of the spectral energy distribution of the source in the wavelength range 42.5--122.5\,$\mu$m. Then, \iLFIR\,=\,$4\pi\,D_{\rm L}^2\,F_{\rm FIR}$, where $D_{\rm L}$ is the luminosity distance.}. At high FIR luminosities, however, the line emission `saturates', leading to reduced line-to-IR luminosity ratios. These are the so-called emission line deficits, which have been since long observed in large samples of galaxies in the nearby Universe \citep{malhotra97} --- in particular in local LIRGs \citep{luhman03} and in general in systems where dust opacity is significant, regardless of which is the galaxy's dominant energy source (star-formation or AGN). In these systems, and depending on the emission line, the deficits can reach values of two or three orders of magnitude lower than those seen in normal star-forming galaxies \citep{gracia-carpio11, farrah13, diaz-santos17}. However, even if the outcome is effectively the same, the physical origin of these deficits is not unique for all the lines. Given the large range in ionization potentials (IP) and critical densities (\ncrit) of FSLs, different processes, gas conditions, or simply varying contributions from different phases of the ISM can lead to the same effect: a suppression of the line compared to the FIR dust continuum emission.

For instance, the galaxy-integrated deficit measured in the \Nii{}\,205\,$\mu$m emission line is, in most sources, likely caused by saturation due to high electron densities, since the line has a critical density of only \ncrite\,$\simeq$\,40\,\ncmmm{}. Densities close to this threshold are common in nearby, normal, main-sequence galaxies \citep{herrera-camus16} as well as in dusty starbursting galaxies \citep{diaz-santos17}, in which values of tens of electrons per cm$^{-3}$ are usually observed. Therefore, galaxies that simply contain on average a large volume filling-factor or number of \Hii\, regions (within which electron densities easily exceed 100\,\ncmmm) will show stronger line deficits, even if the physical processes occurring within these sources are the same as those happening in least star-forming galaxies.

The deficit observed in the \Oi{} 63\,$\mu$m emission likely has a different origin, as this line may suffer from self-absorption along the line of sight caused by in-situ opacity or due to foreground cold gas clouds.
The \Oi{}\,63\,$\mu$m emission/absorption arises from the $^3P_1 \rightarrow ^3P_2$ transition to/from the ground-state level, and giving the large abundance of oxygen in the ISM (the most common metal in the solar neighborhood; \citealt{asplund09}), any cold cloud crossing the line of sight towards the background source(s) of \Oi{}\,63\,$\mu$m emission will have a large impact on the transmitted radiation \citep{poglitsch96}. In addition, the \Oi{}\,63\,$\mu$m transition has a large optical depth (see Fig.~\ref{fig_optical_depth}) and may be affected by in-situ self-absorption. Statistical equilibrium predicts that the \Oi{}\,145\,$\mu$m/63\,$\mu$m line ratio should not exceed a value of $\simeq$\,0.1 in the optically thin limit (see Fig.~\ref{fig_density_diagnostics}). And yet, values exceeding that threshold have been observed in some of the most luminous, obscured galaxies in the local Universe \citep{rosenberg15}, suggesting the \Oi{}\,63\,$\mu$m line is self-absorbed compared to the optically thin \Oi{}\,145\,$\mu$m transition. Knowing whether the absorption is caused locally or by foreground sources requires spatially resolving the sources of emission/absorption, which is in most cases unattainable in extra-galactic observations, especially at high redshift.

The \Oiii{}\,52\,$\mu$m and 88\,$\mu$m lines are mostly sensitive to the intensity and hardness of the ionization field per unit gas mass, parametrized as $\langle U\rangle$\,$\propto$\,$G/n_{\rm H}$, as well as to metallicity \citep{cormier15}, and their ratio is practically insensitive to temperature at the level of the hot ionized gas phase. This is because the excitation energies of these lines ($E_{u}\,k_{\rm b}^{-1}$\,$\simeq$\,440 and 160 K, respectively) are low compared to the usual gas temperature in an \Hii\, region ionized by OB stars, which is typically larger than 10,000\,K. Thus, both lines are virtually always in the high-temperature saturation limit. Additionally, the \Oiii{}-to-IR luminosity ratio, similarly to all the line-to-dust continuum ratios involving ionized lines, is sensitive, to zero-order, to the luminosity-weighted relative contribution of the ionized vs.~dusty phases of the ISM, which in turn can depend on, e.g., the gas-phase metallicity and/or the metal-to-dust content of the source.

The origin of the \Cii{}\,158\,$\mu$m deficit is more complex and historically it has been attributed to a large variety of different physical processes. Due to its relatively low IP ($\sim$\,11.26\,eV) the emission can arise from both the ionized and neutral phases of the ISM, where the critical densities are \ncrite\,$\simeq$\,50\,\ncmmm\, and \ncritH\,$\simeq$\,400\,\ncmmm, respectively. Some of the most commonly invoked causes for the \Cii{}\,158\,$\mu$m deficit are: (1) saturation due to high density of the neutral gas, (2) temperature saturation due to high temperatures in PDRs, (3) progressive ionization of very small dust grains and thus reduced yield of photo-electric heating in environments with high far-UV field per gas density (high $U$), and (4) high dust-to-gas opacity caused by an increase of the average ionization parameter; a scenario in which dust grains preferentially absorb UV photons before they reach the gas in the PDRs --- the so-called dust-bounded regions. In this scenario, the line deficit is mainly driven by the scaling of the FIR luminosity roughly as \iTdust$^{4+\beta}$ accordingly to the Stefan-Boltzmann's law, where $\beta$ is the emissivity index of dust in the FIR, which ranges between 1--2; meanwhile, the \Cii{}\,158\,$\mu$m emission scales instead with \iTdust\, following a power law dependence on the gas kinetic temperature that is typically of the order unity or less.

\begin{figure}[htbp]
\begin{center}
\includegraphics[width=0.325\columnwidth]{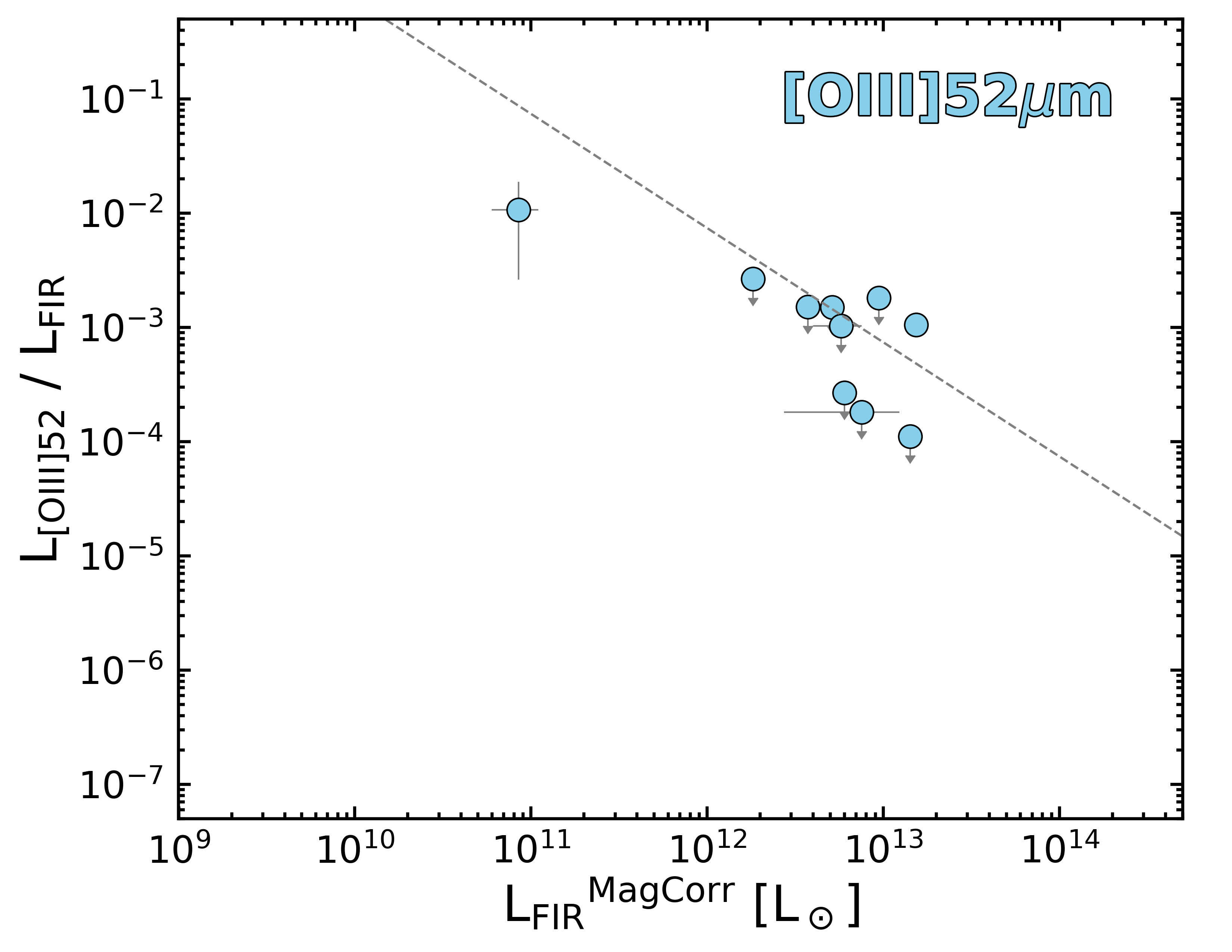}
\includegraphics[width=0.325\columnwidth]{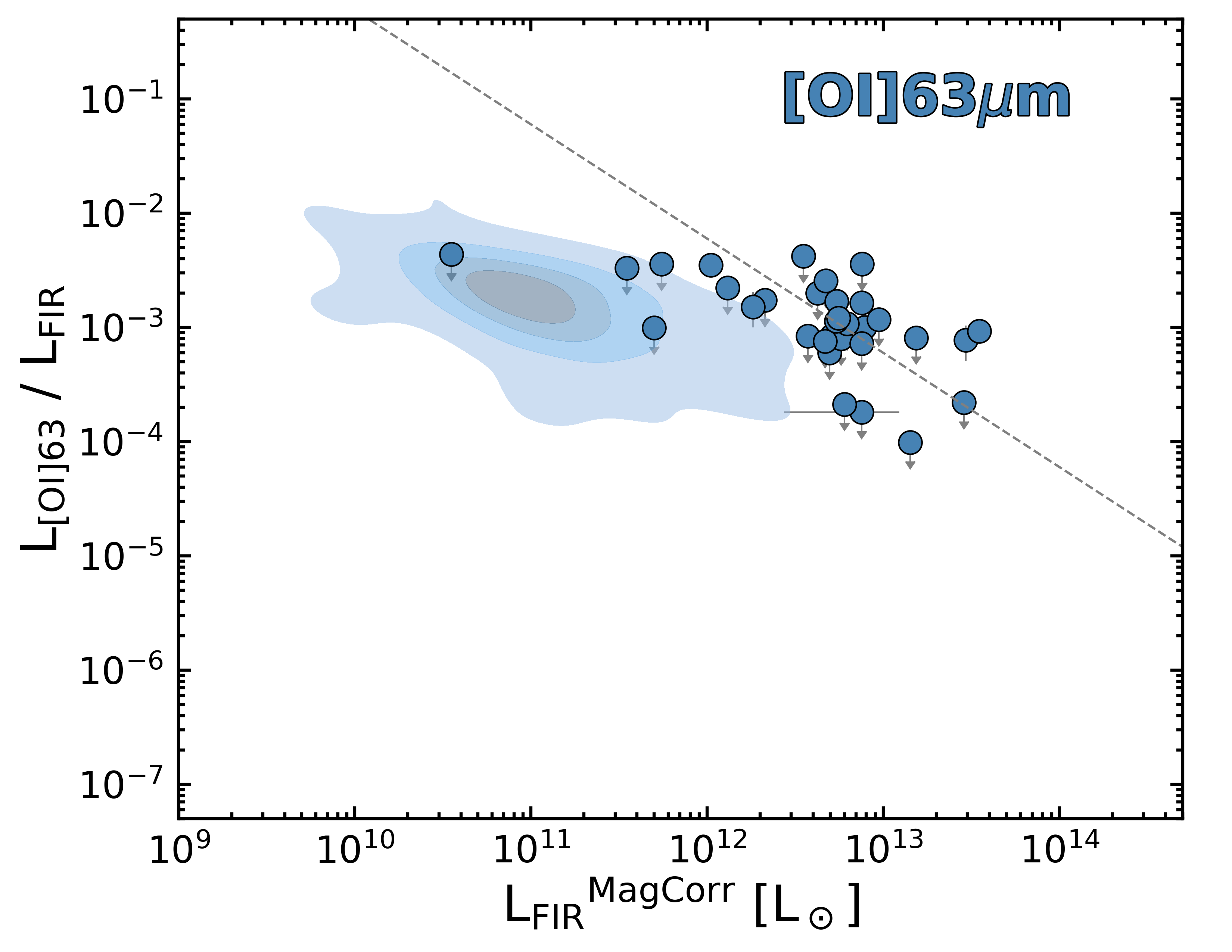}
\includegraphics[width=0.325\columnwidth]{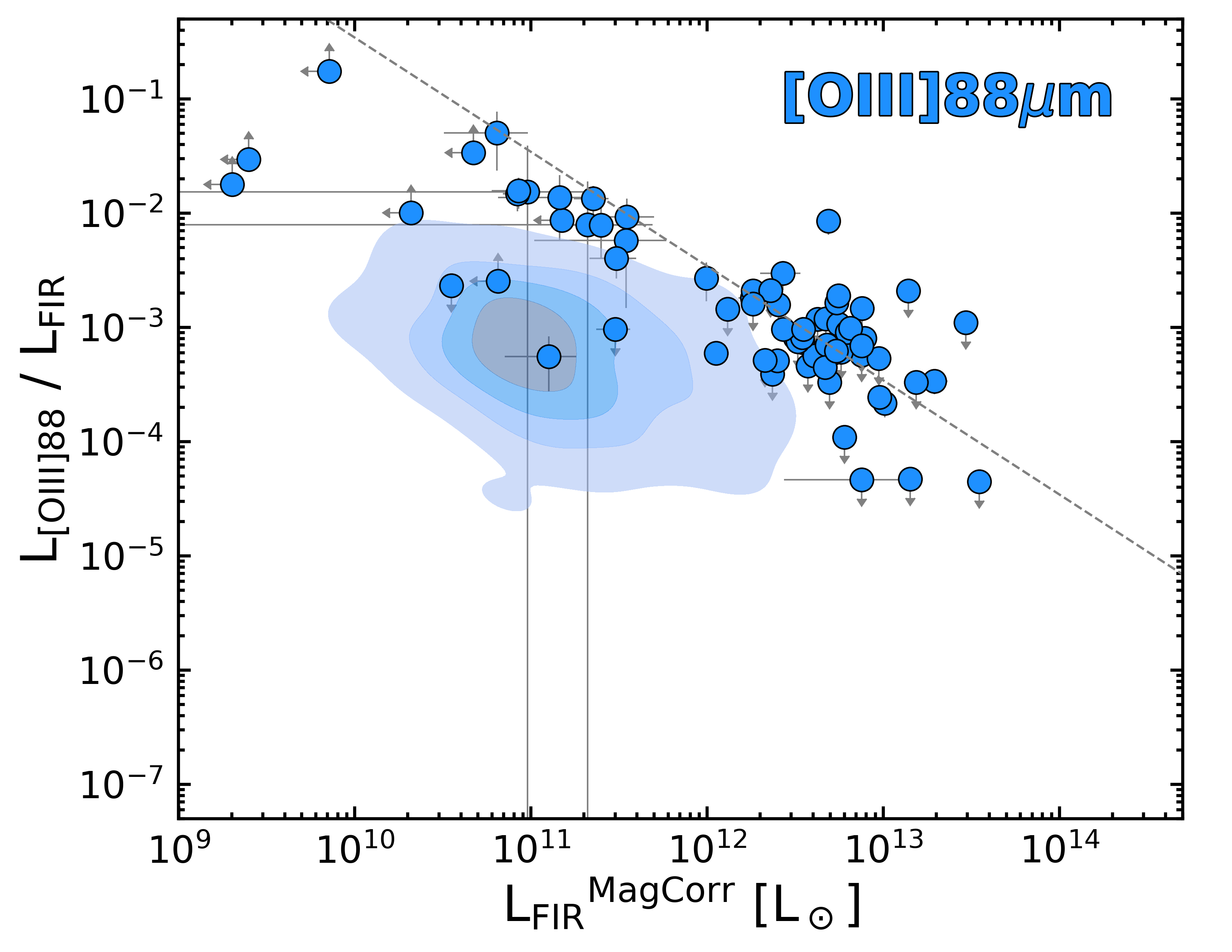}
\includegraphics[width=0.325\columnwidth]{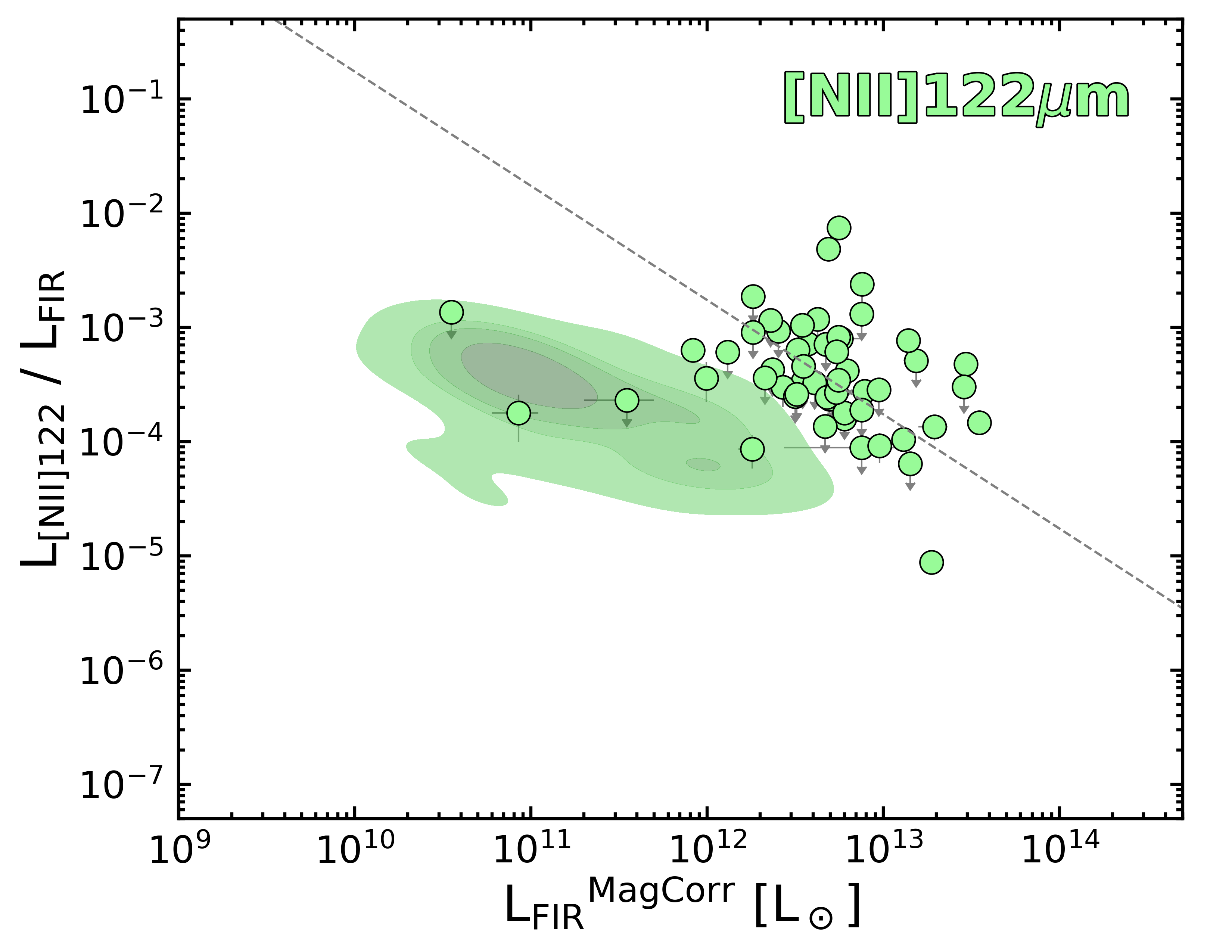}
\includegraphics[width=0.325\columnwidth]{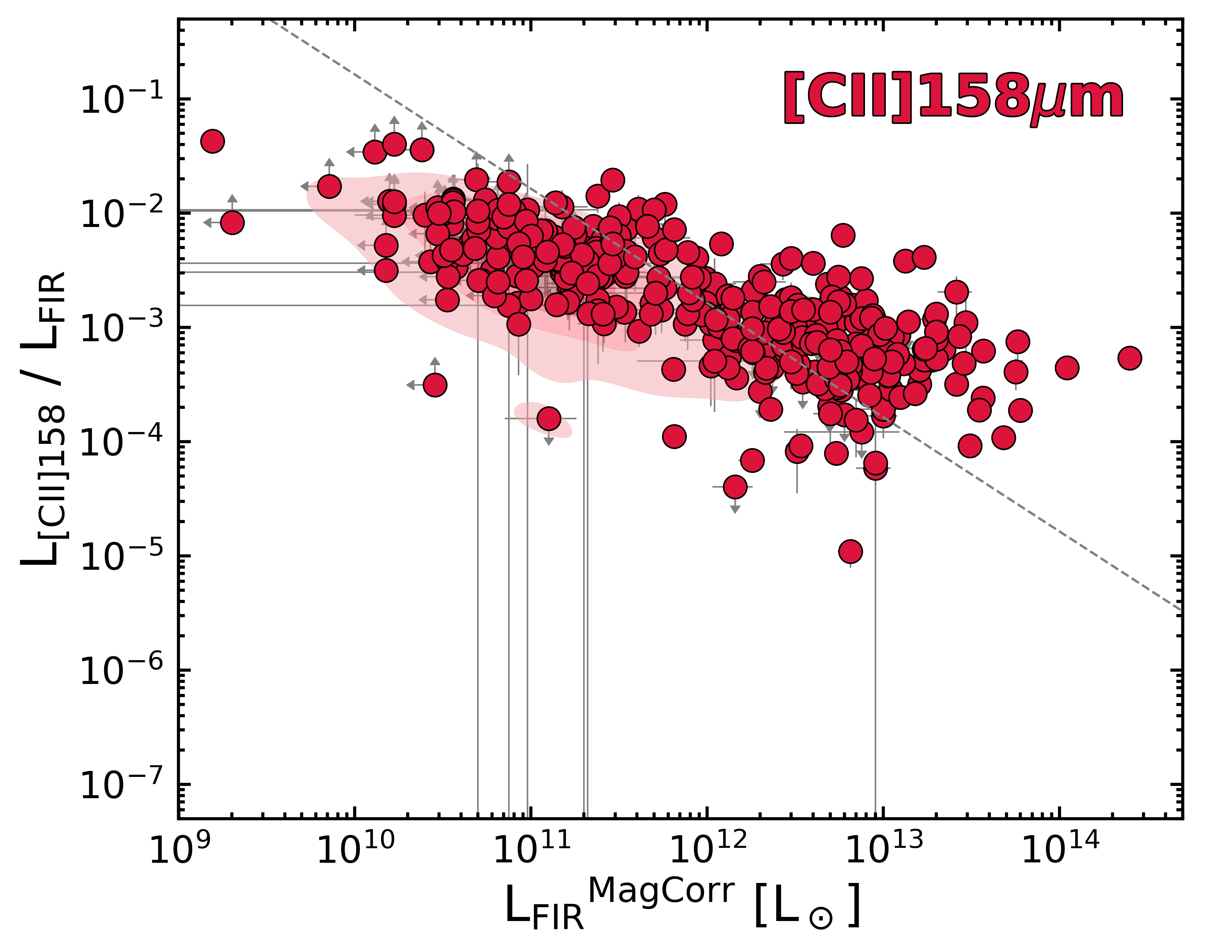}
\includegraphics[width=0.325\columnwidth]{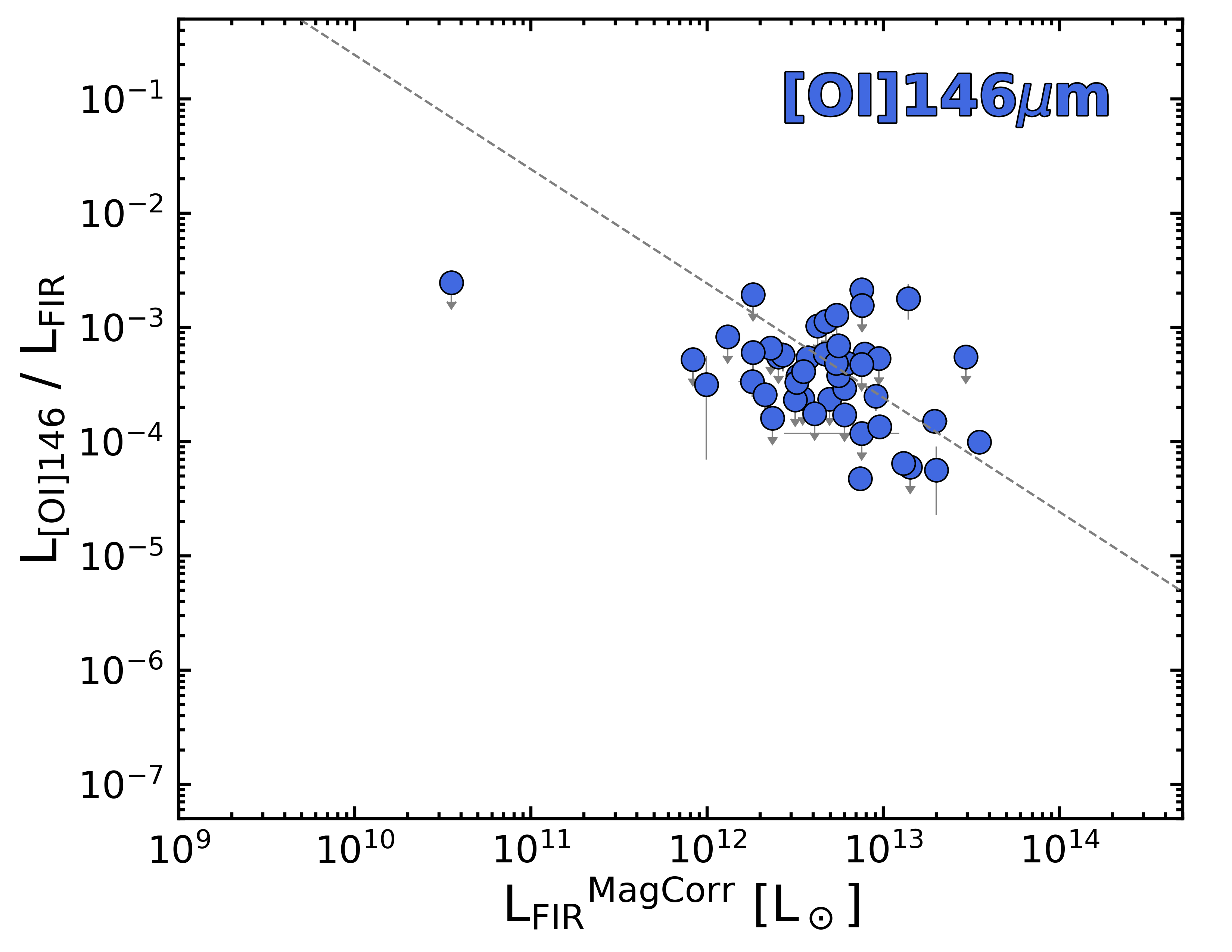}
\includegraphics[width=0.325\columnwidth]{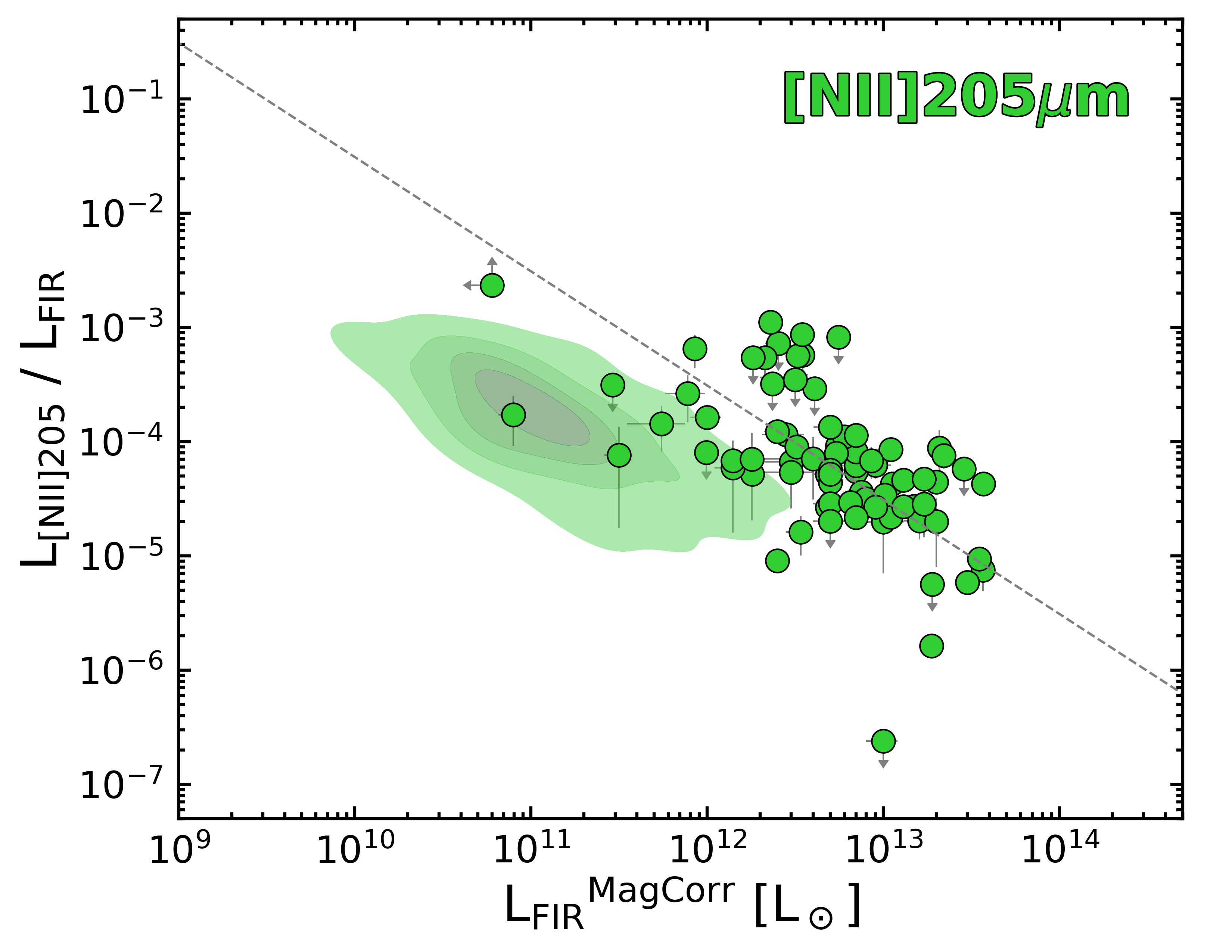}
\includegraphics[width=0.325\columnwidth]{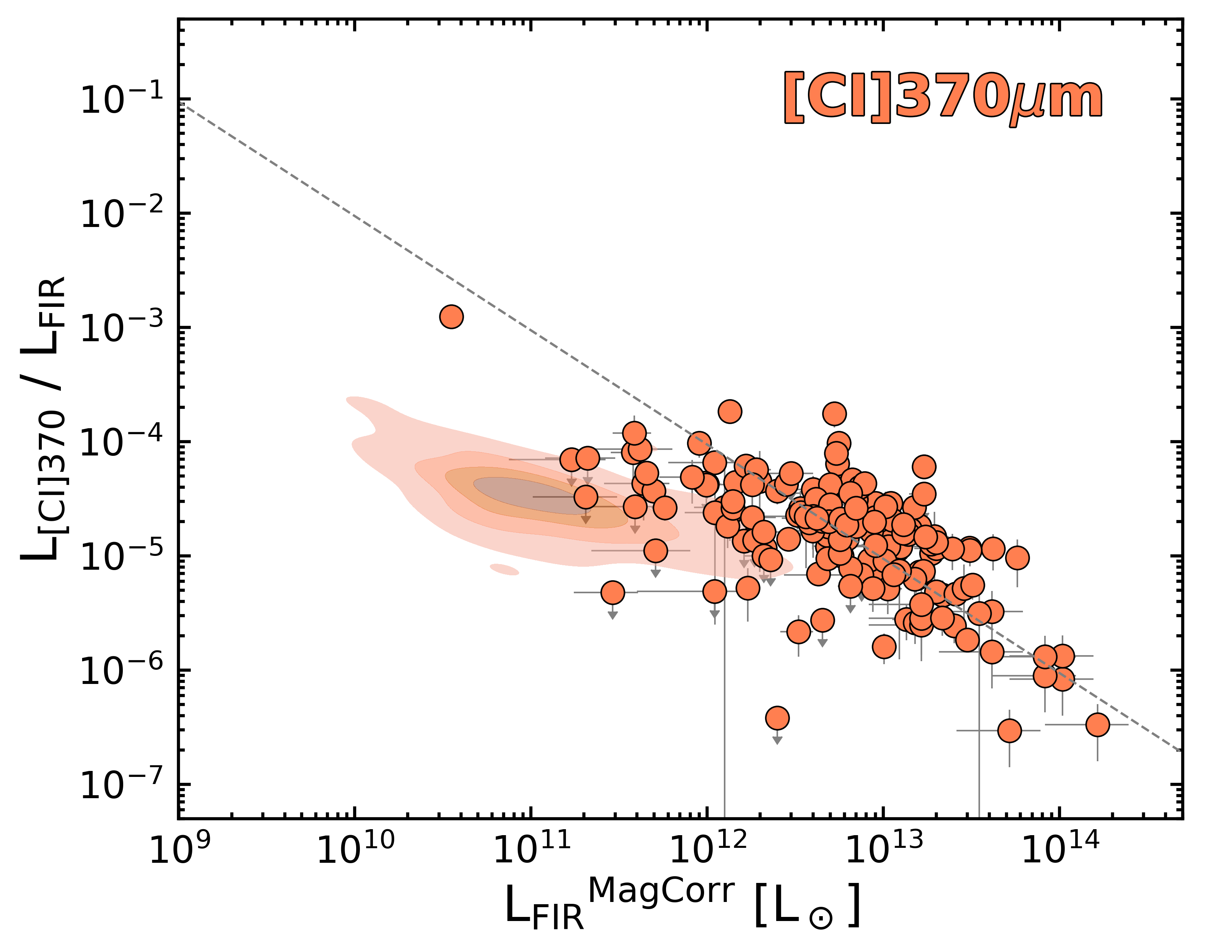}
\includegraphics[width=0.325\columnwidth]{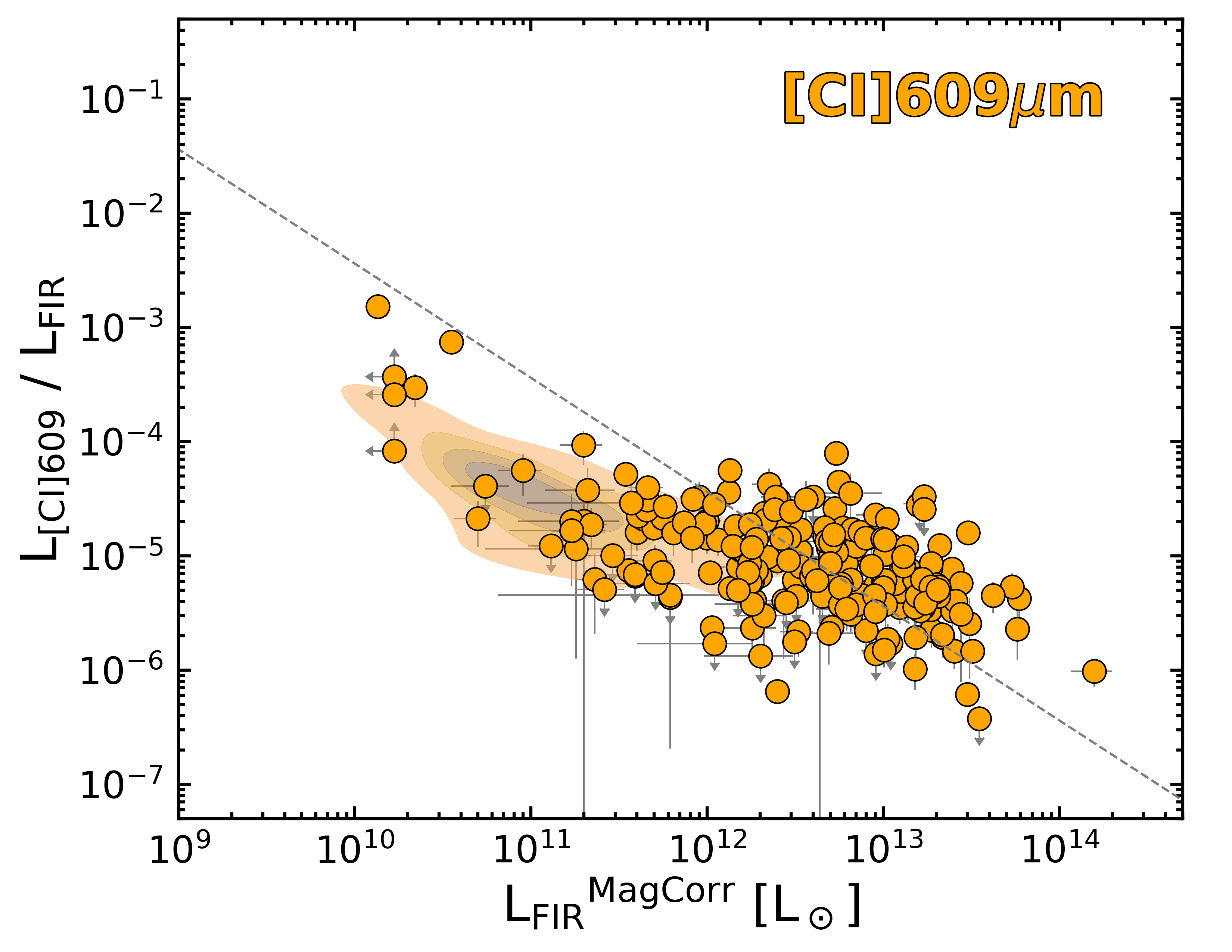}
\end{center}
\caption{Emission line to FIR continuum luminosity ratios for the FSL catalog compiled for this work as a function of FIR luminosity of the sources (corrected for magnification) are shown as points. The contours represent the location in the parameter space where nearby, starbursting, dusty luminous infrared galaxies lie. There is an overall trend for all the IR fine-structure lines to have relatively lower line-to-\iLFIR\, ratios (i.e., to show larger line deficits) as the FIR luminosity of the galaxies increases. The most pronounced deficits are seen in the \Oiii{}\,88\,$\mu$m line, which span a downturn of three orders of magnitude in the line-to-FIR ratio across nearly four decades in FIR luminosity, suggesting the line emission has reached a plateau. The \Cii{}\,158\,$\mu$m and \Nii{}\,205\,$\mu$m shows an approximately two orders magnitude deficit across four decades in \iLFIR, similar to what is observed in the two transitions of neutral carbon, \Ci{}\,370\,$\mu$m and 609\,$\mu$m. The remaining lines also exhibit deficits, albeit smaller and with weaker significance.}
\label{fig_deficit}
\end{figure}

Figure~\ref{fig_deficit} shows the emission line deficits as a function of FIR luminosity (corrected for magnification) for the FSL catalog of galaxies compiled for this review. Each emission line displays a unique behavior of the deficit. For instance, \Oiii{}\,88\,$\mu$m shows a decrease in the line-to-FIR ratio of nearly three orders of magnitude across four decades in FIR luminosity. \Cii{}\,158\,$\mu$m and \Ci{}\,609\,$\mu$m display a smaller downturns, of around two orders of magnitude, over the same FIR dynamic range as \Oiii{}\,88\,$\mu$m. \Nii{}\,205\,$\mu$m and \Ci{}\,370\,$\mu$m show similar deficits, around two orders of magnitude, but spread over only two decades in \iLFIR, closely following a 1-to-1 trend. Finally, for other lines like \Oiii{}\,52\,$\mu$m, \Nii{}\,122\,$\mu$m or \Oi{}\,146\,$\mu$m there are very few measurements at $z>1$ to date, and it is difficult to identify whether they show any deficit or follow any trend. 

The contours displayed in the figure depict, for reference, the locus of nearby ($z\,\lesssim\,0.1$), dusty, luminous infrared galaxies (LIRGs). In general, high-redshift galaxies reside in regions of the parameter space characterized by deficits roughly similar to local LIRGs, but at higher FIR luminosities $\gsim10^{12}$\,\Lsun. This implies that the FSL emission in galaxies increases at a similar pace as their \iLFIR\, as we move towards earlier times, when the average FIR luminosity of galaxies was higher.

\begin{figure}[htbp]
\begin{center}
\includegraphics[width=0.325\columnwidth]{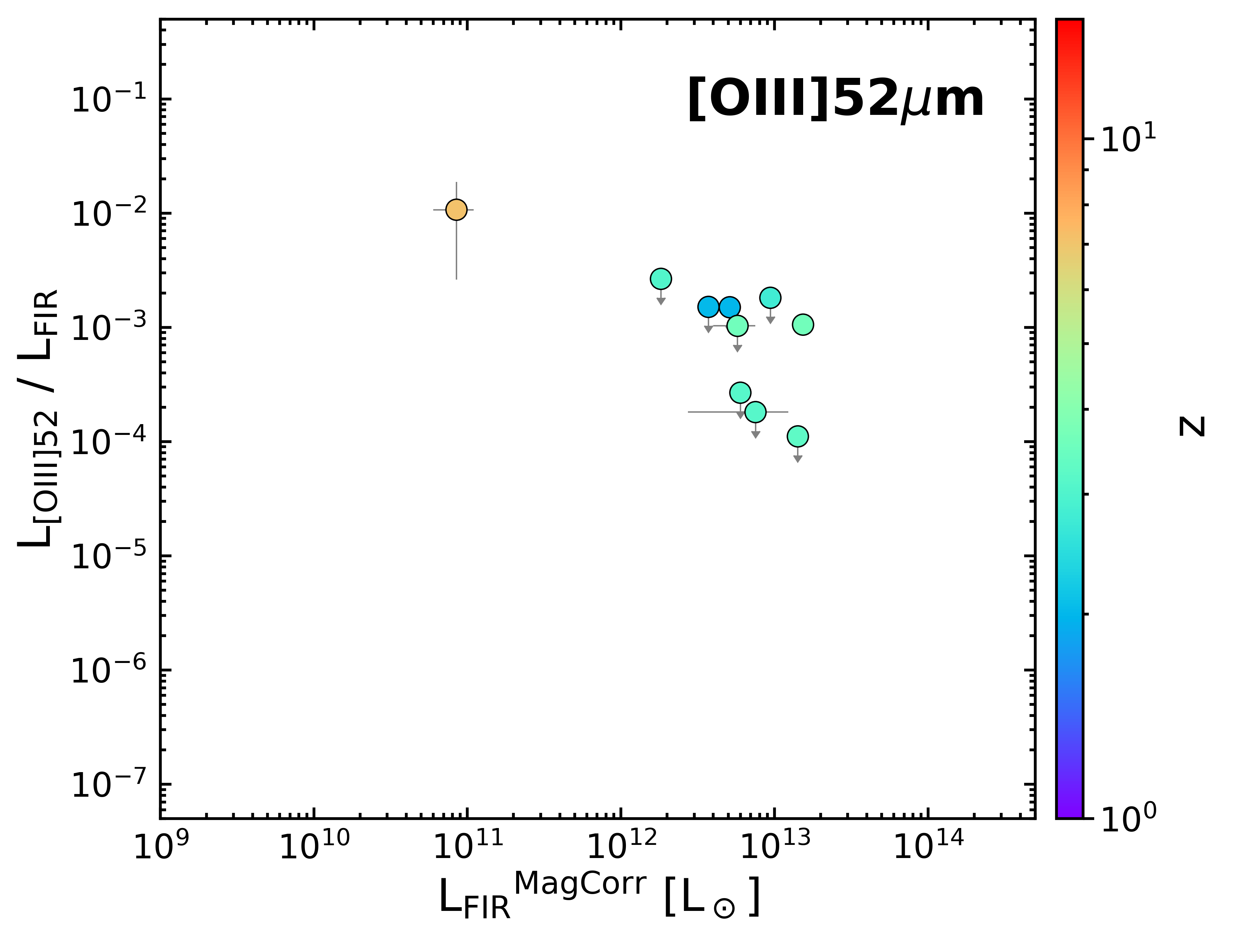}
\includegraphics[width=0.325\columnwidth]{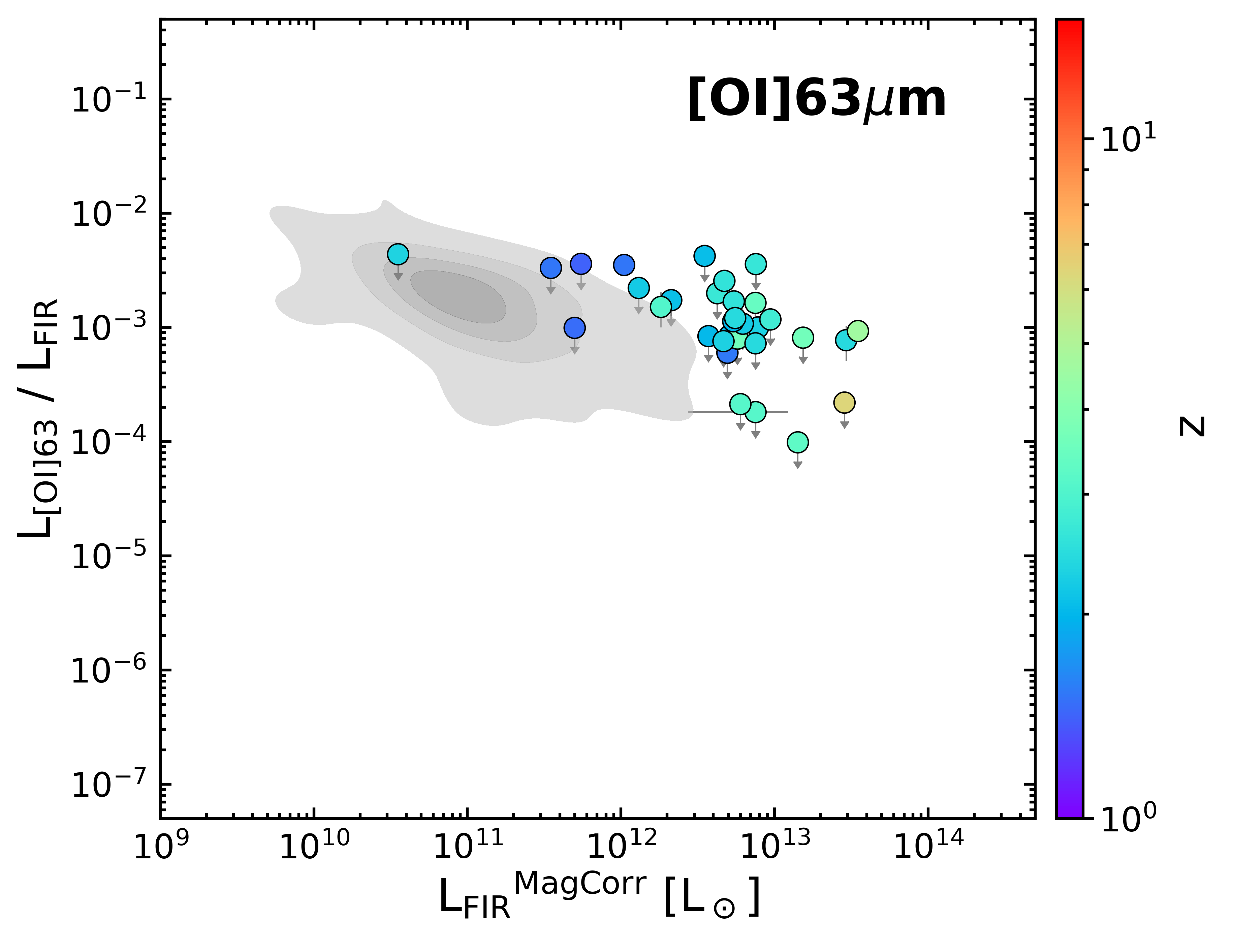}
\includegraphics[width=0.325\columnwidth]{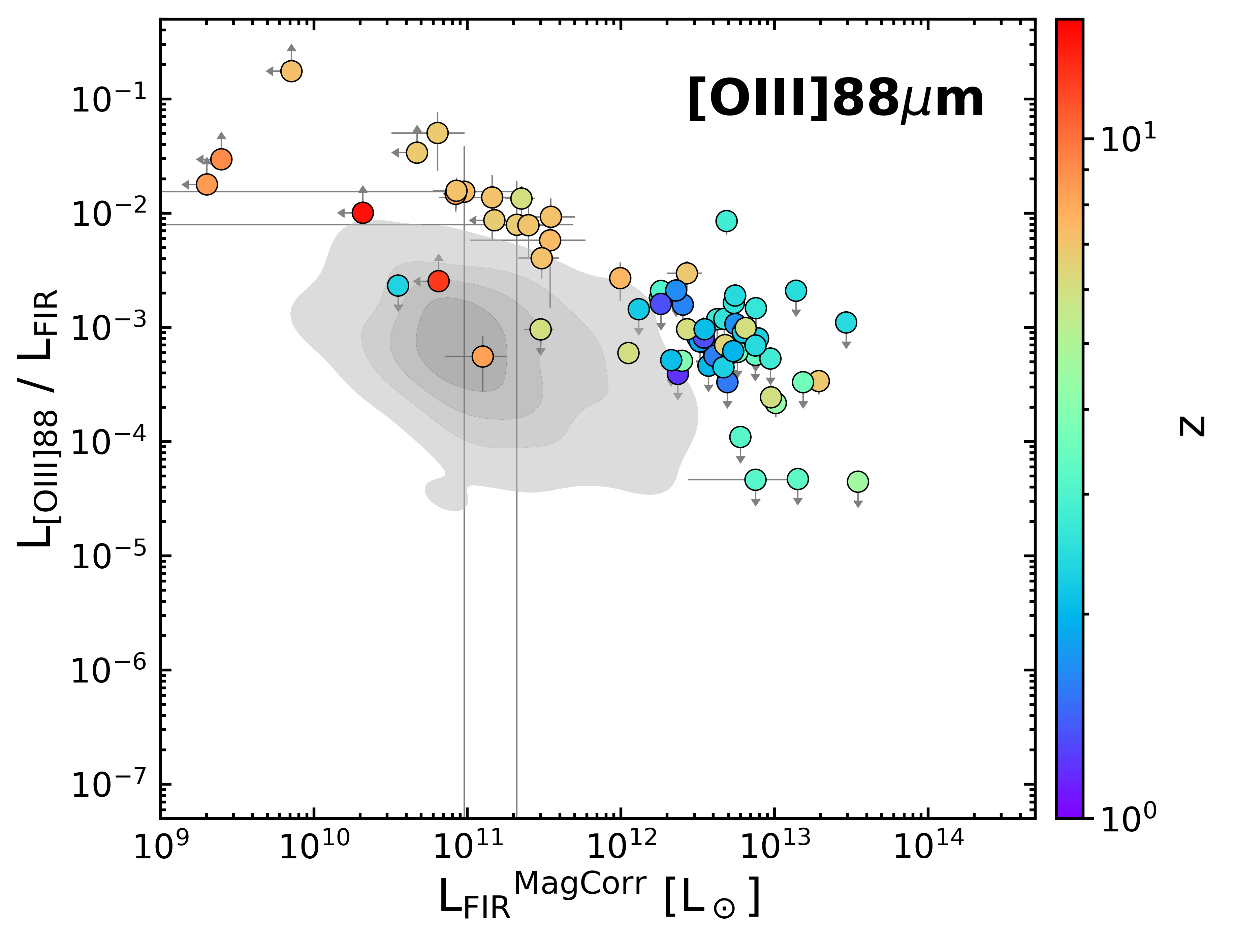}
\includegraphics[width=0.325\columnwidth]{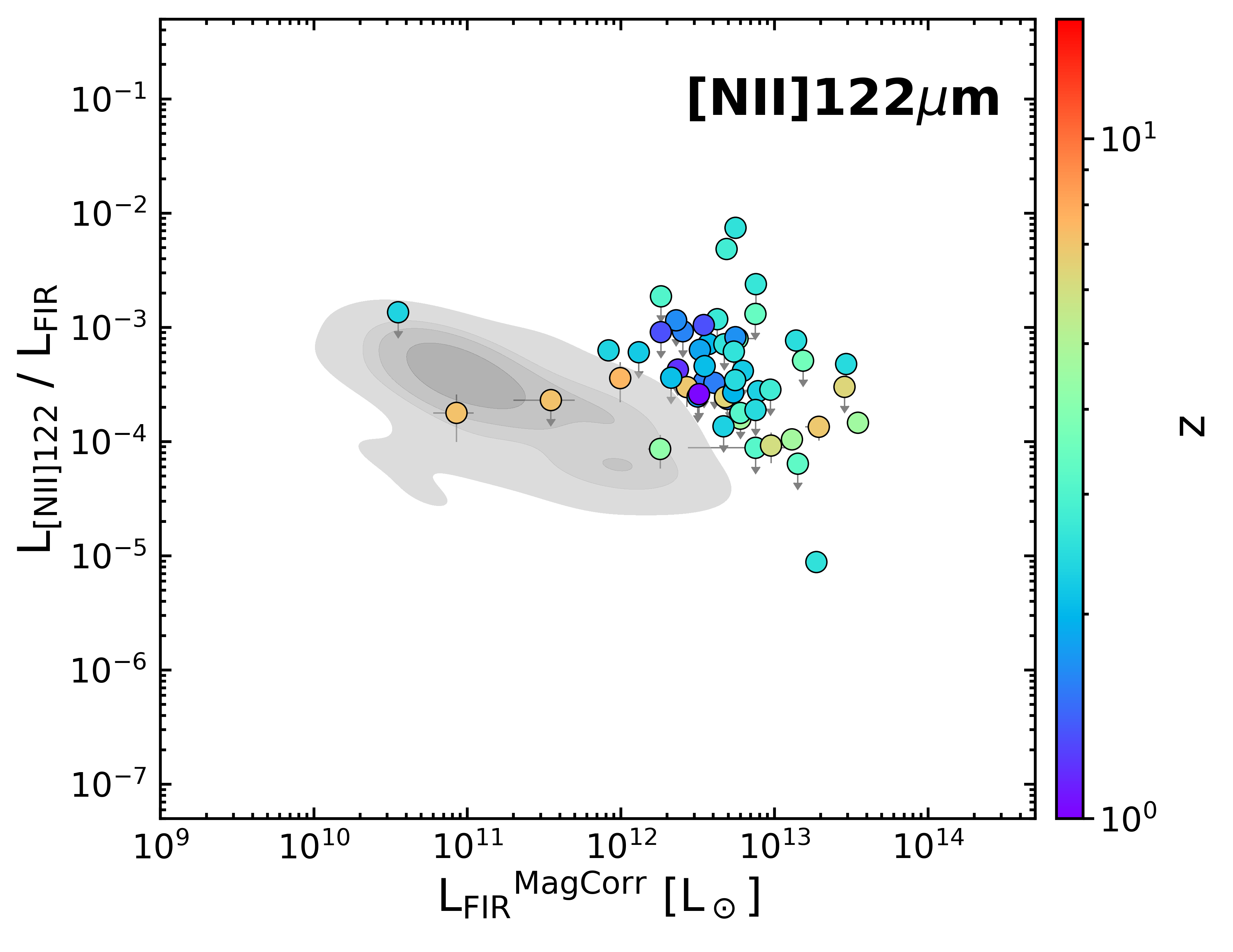}
\includegraphics[width=0.325\columnwidth]{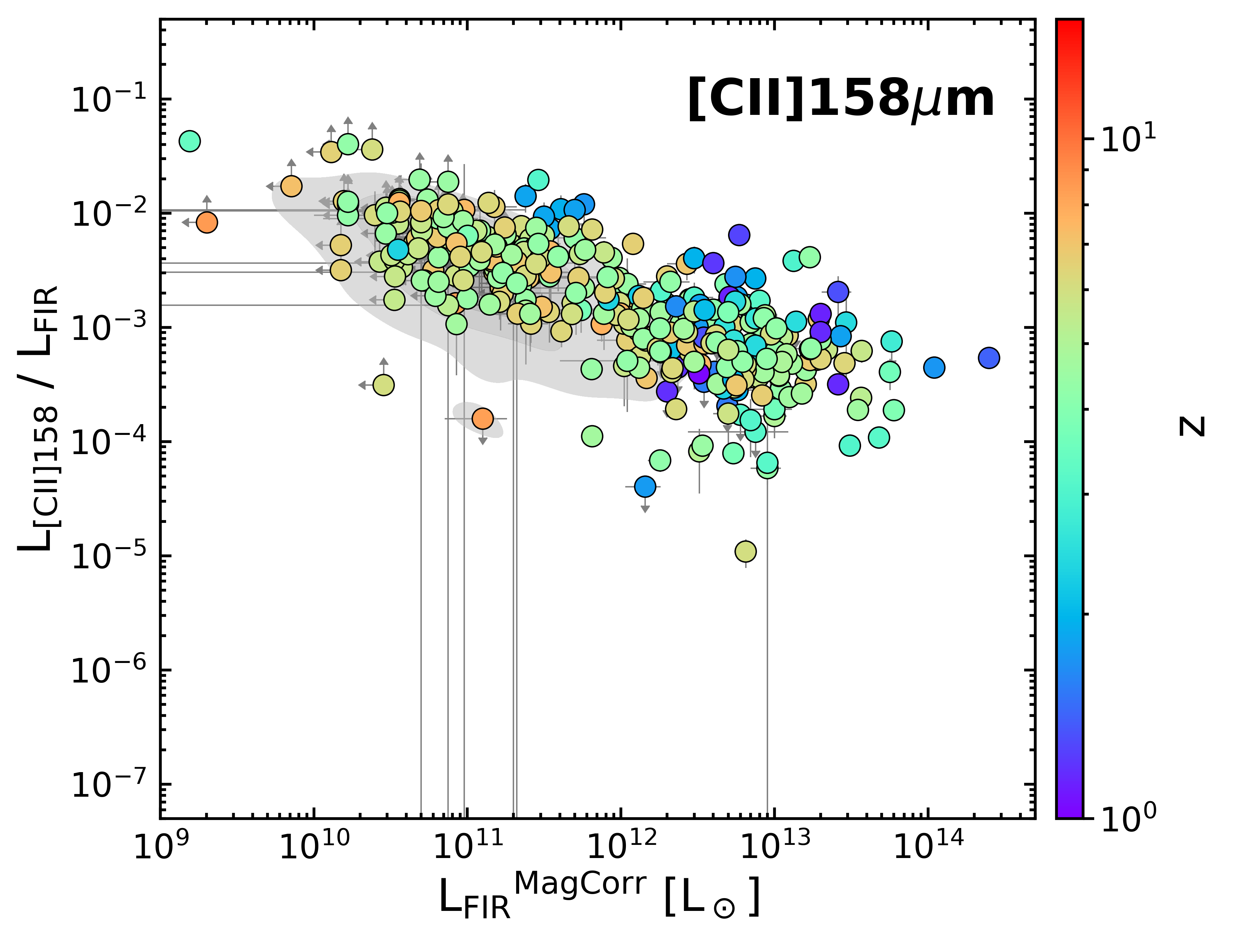}
\includegraphics[width=0.325\columnwidth]{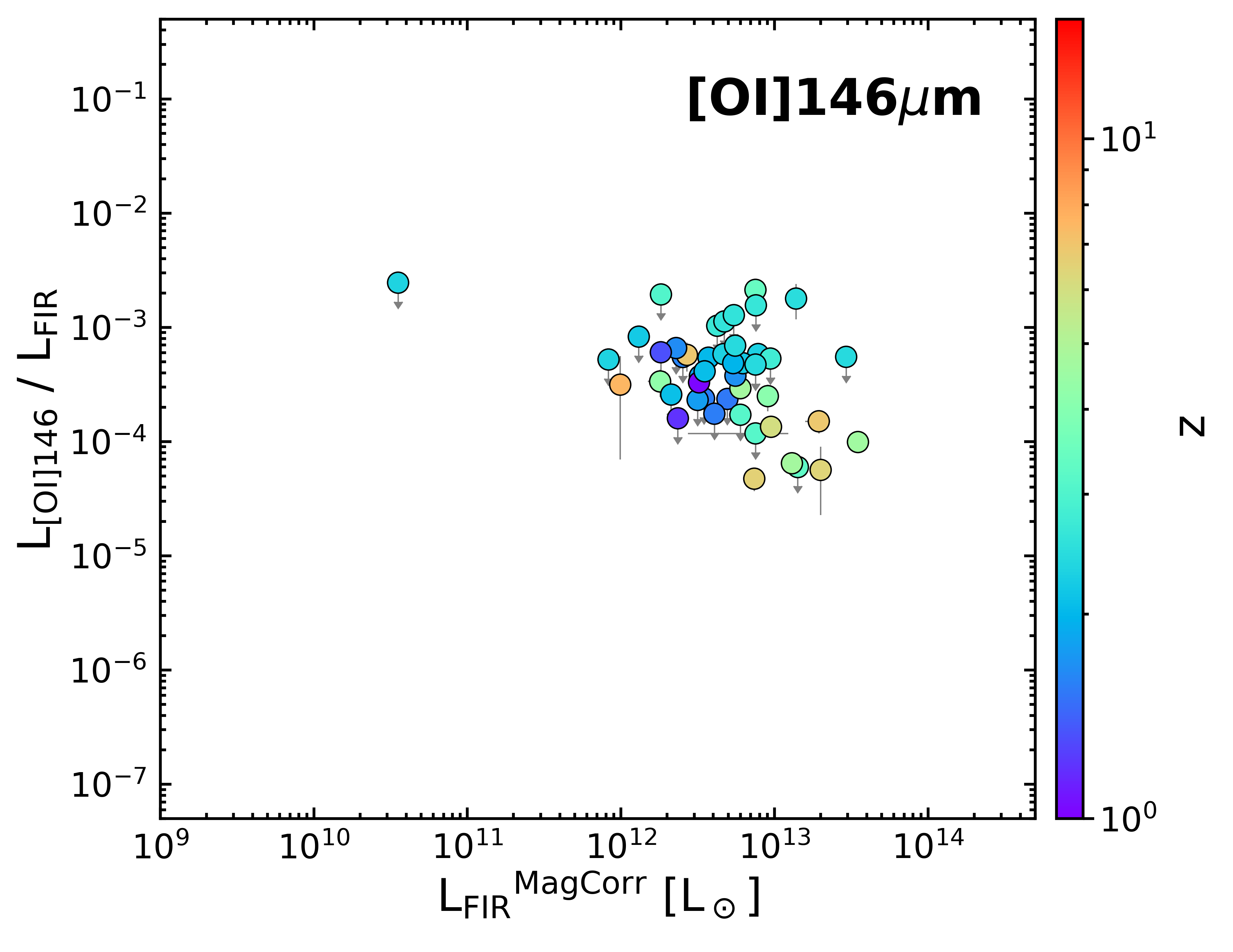}
\includegraphics[width=0.325\columnwidth]{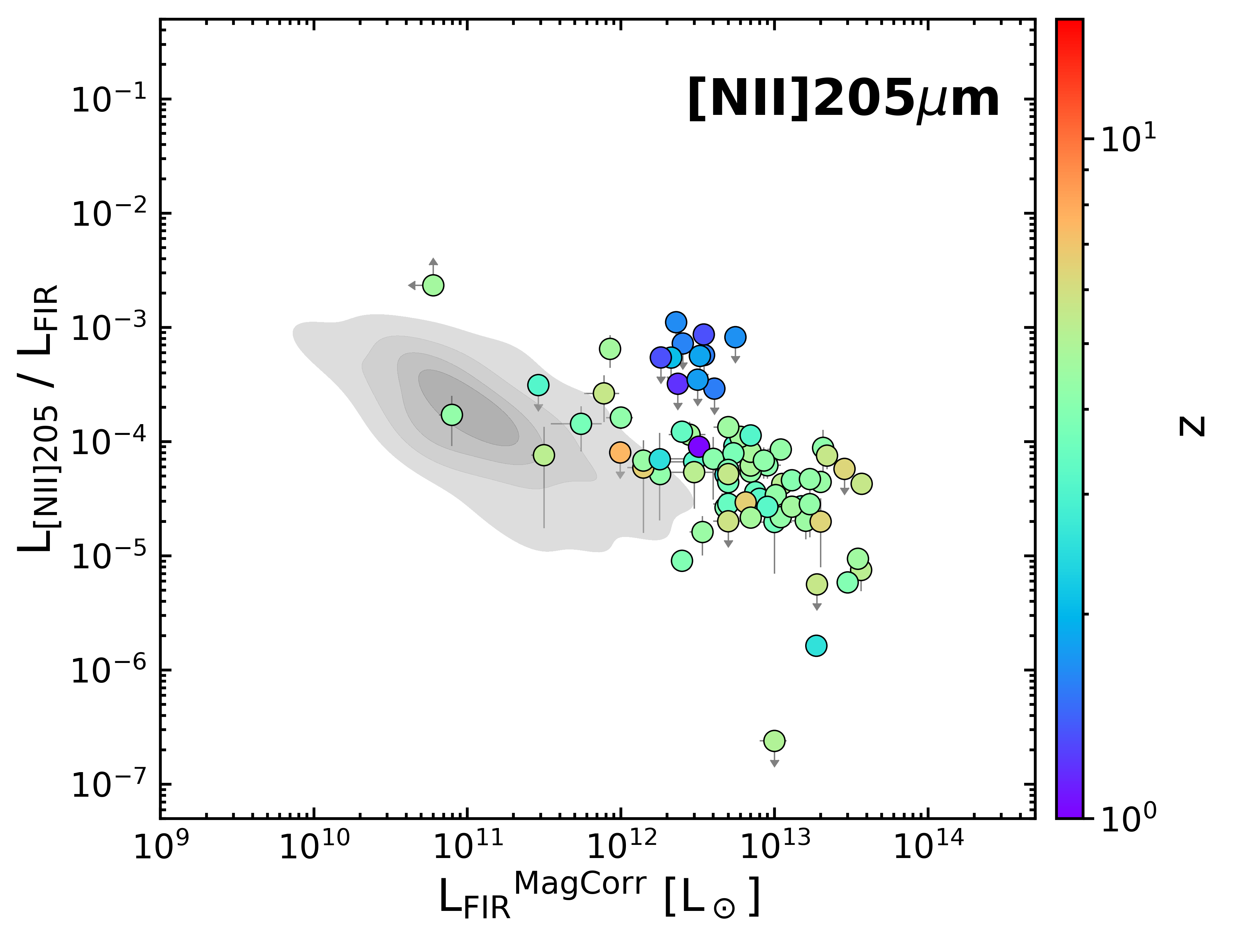}
\includegraphics[width=0.325\columnwidth]{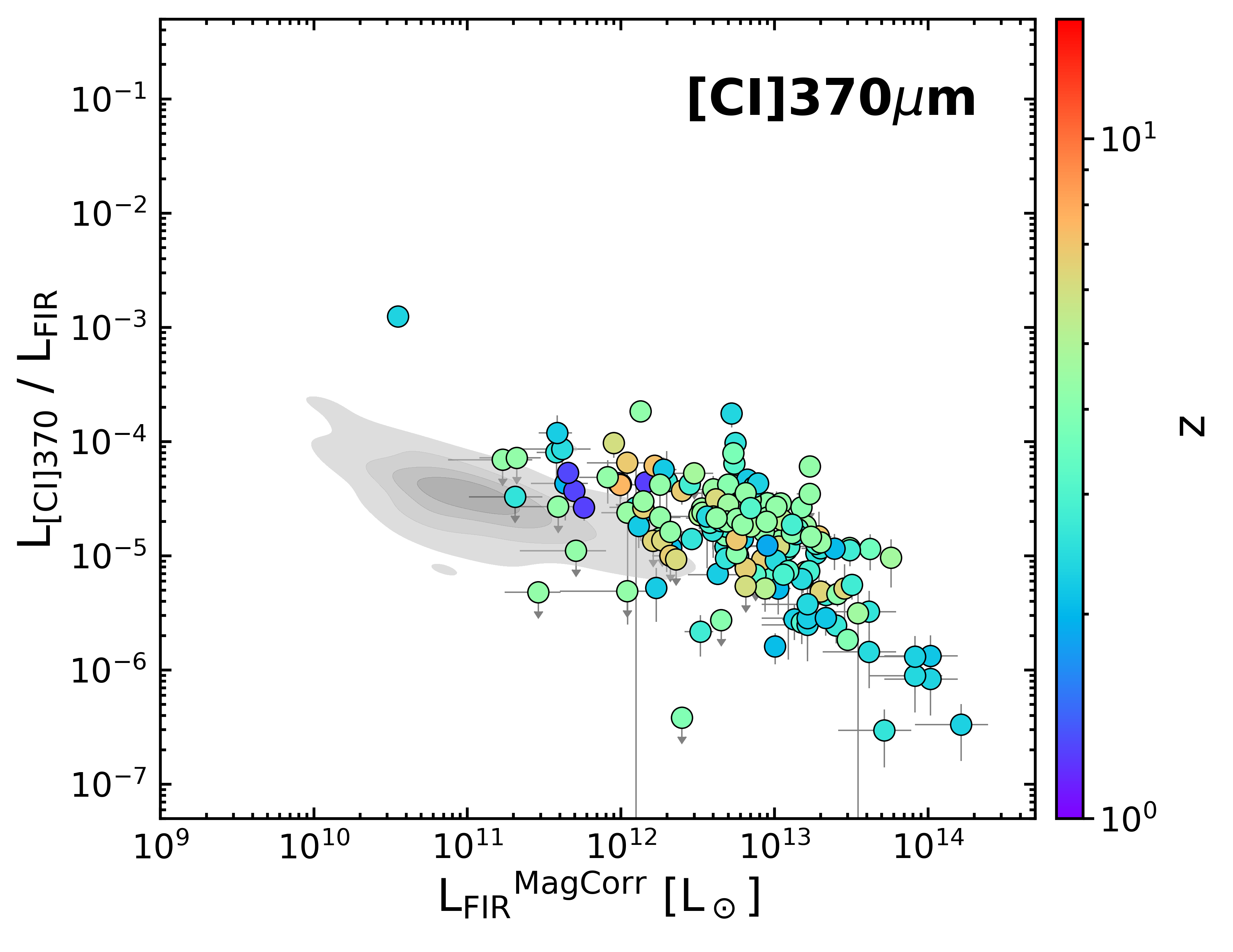}
\includegraphics[width=0.325\columnwidth]{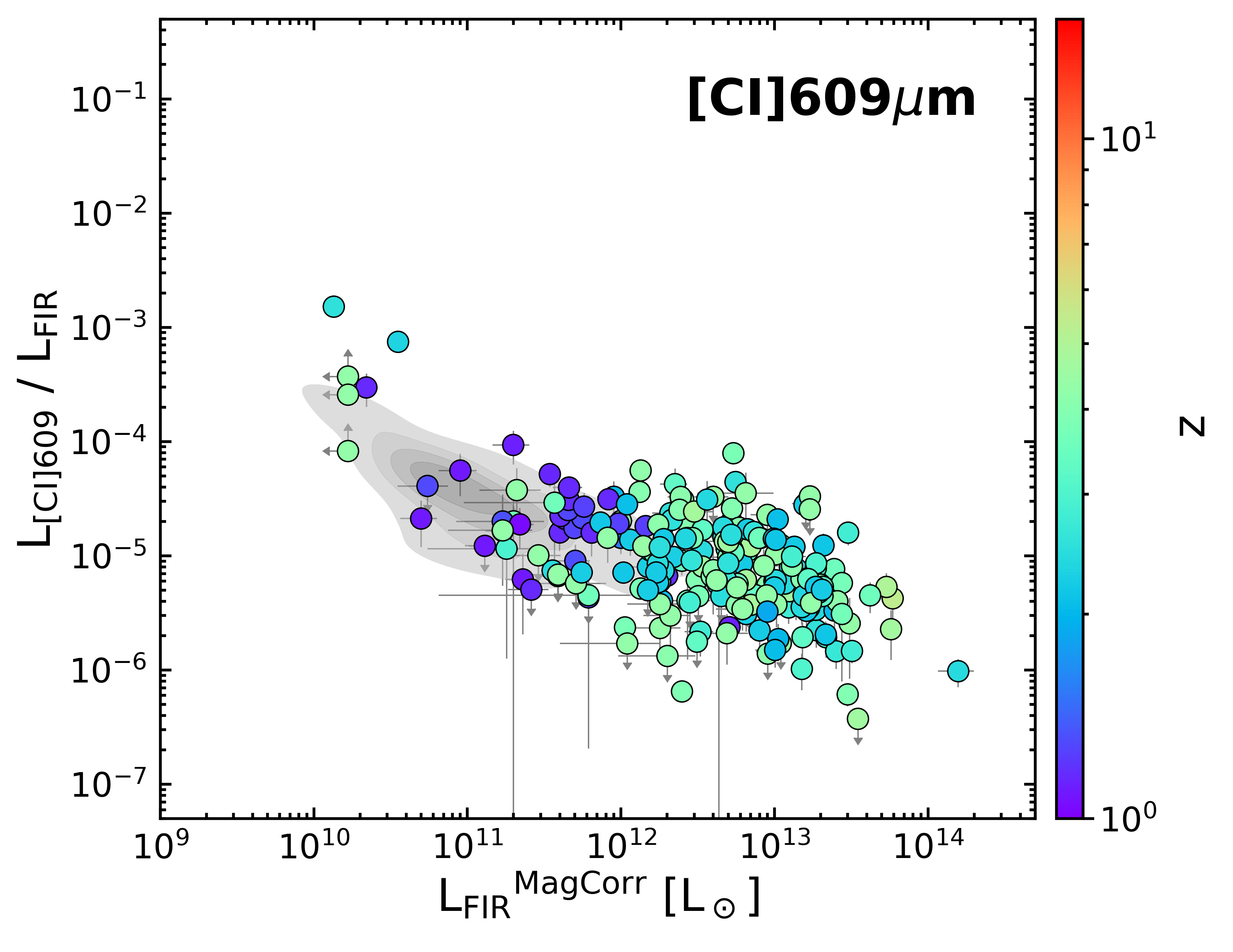}
\end{center}
\caption{Emission line to FIR lumionsity ratio for the FSL catalog as a function of FIR luminosity (corrected for magnification), color-coded by redshift. The gray contours represent the location in the parameter space where nearby, starbursting, luminous infrared galaxies lie. To first order, similar trends are observed for FSLs at high redshift as in the local universe; however, some offsets are apparent, due to either selection effects or intrinsic differences in the conditions of the dominant phases of the ISM in galaxies nearby and at high redshift.
}
\label{fig_deficit_z}
\end{figure}

This is better showcased in Fig.~\ref{fig_deficit_z}, which displays the same panels as Fig.~\ref{fig_deficit} but with sources color-coded as a function of redshift. It can be seen that high redshift galaxies with FIR luminosities between $10^{12-14}$\,\Lsun\, show similar line deficits to nearby LIRGs, which in general have \iLFIR\,$\lesssim$\,$10^{12}$\,\Lsun. Moreover, for emission lines with enough observations, high-$z$ galaxies seem to follow similar slopes/trends as local, dusty galaxies but shifted to higher \iLFIR. This is likely consequence of the cosmic evolution of the star-formation rate and molecular gas content in galaxies as a function of look-back time \citep{madau14, decarli19}, where galaxies at high-$z$ have on average boosted SFRs and molecular gas reservoirs at a given stellar mass when compared to nearby galaxies; and yet, both the nearby and high-$z$ galaxy populations seem to have similar line-deficit trends, suggesting the physical processes they are undergoing are likely the same and are having the same impact on the sources despite the order of magnitude difference in IR luminosities. The shift in \iLFIR\, is also important for practical reasons, since it means that for a given \iLFIR, the line-to-FIR ratio may be multi-valued, depending on the redshift of the population. This implies that the derivation of any scaling relation (e.g., SFR calibrations) from a sample or samples of galaxies that include sources spanning a large range of cosmic time needs to account for a redshift evolution.

\begin{figure}[htbp]
\begin{center}
\includegraphics[width=0.325\columnwidth]{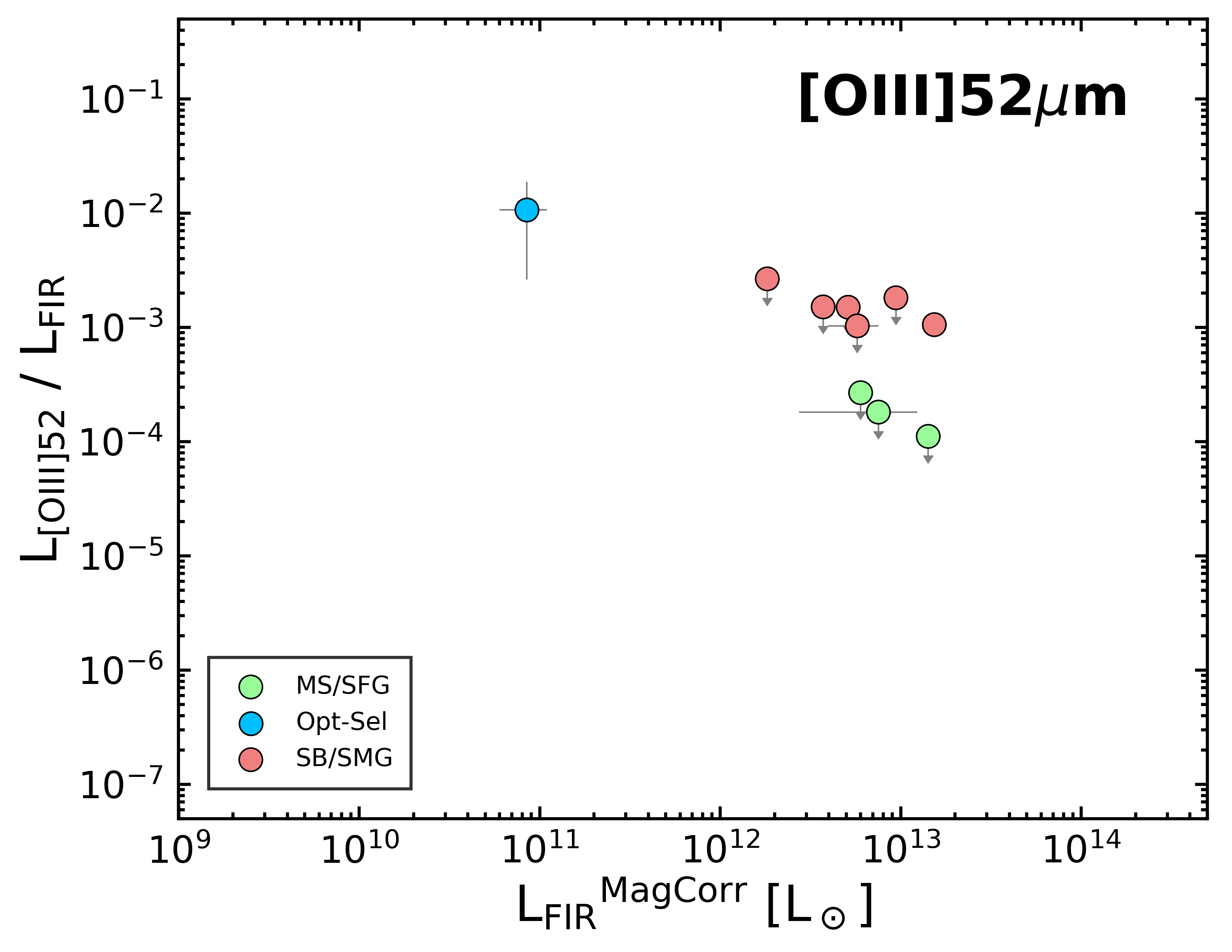}
\includegraphics[width=0.325\columnwidth]{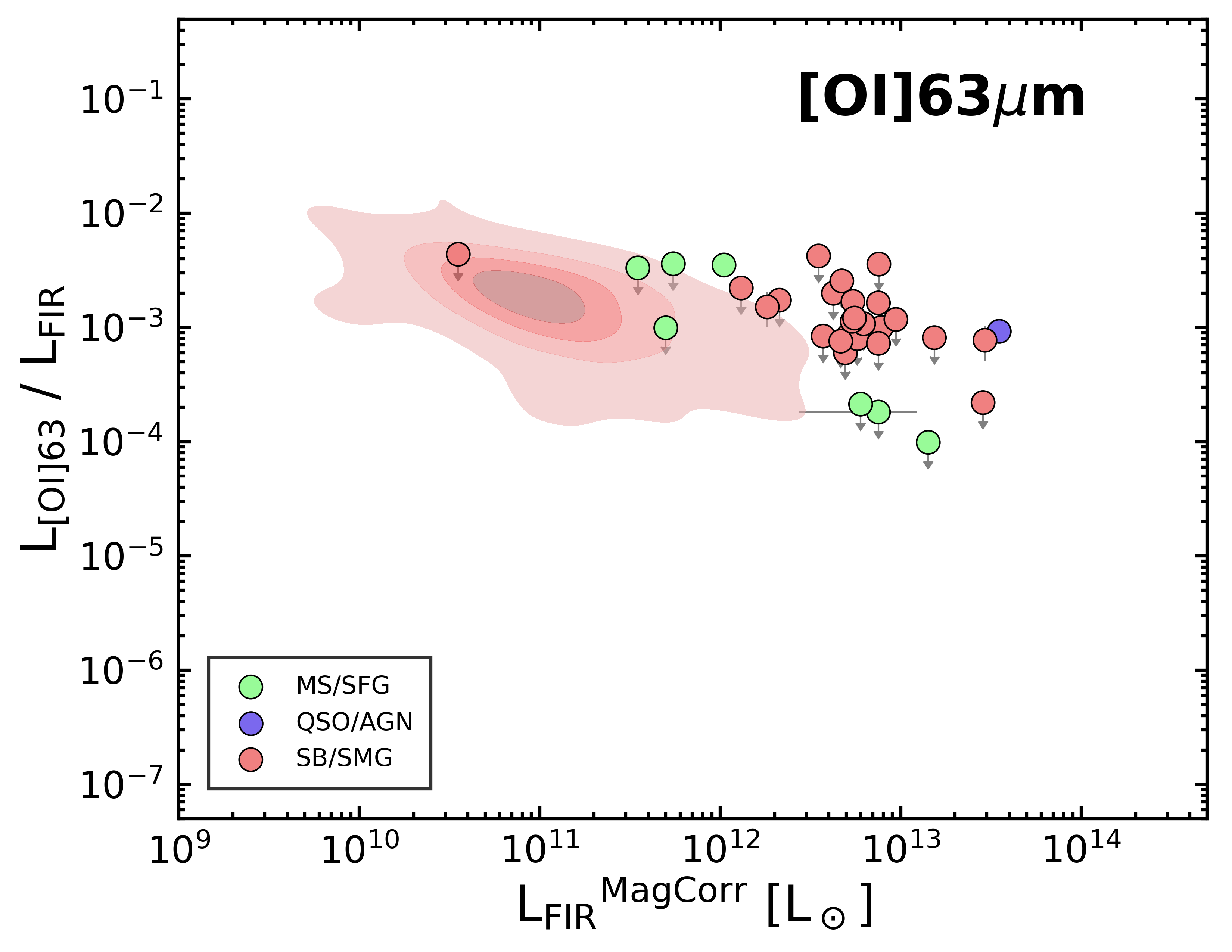}
\includegraphics[width=0.325\columnwidth]{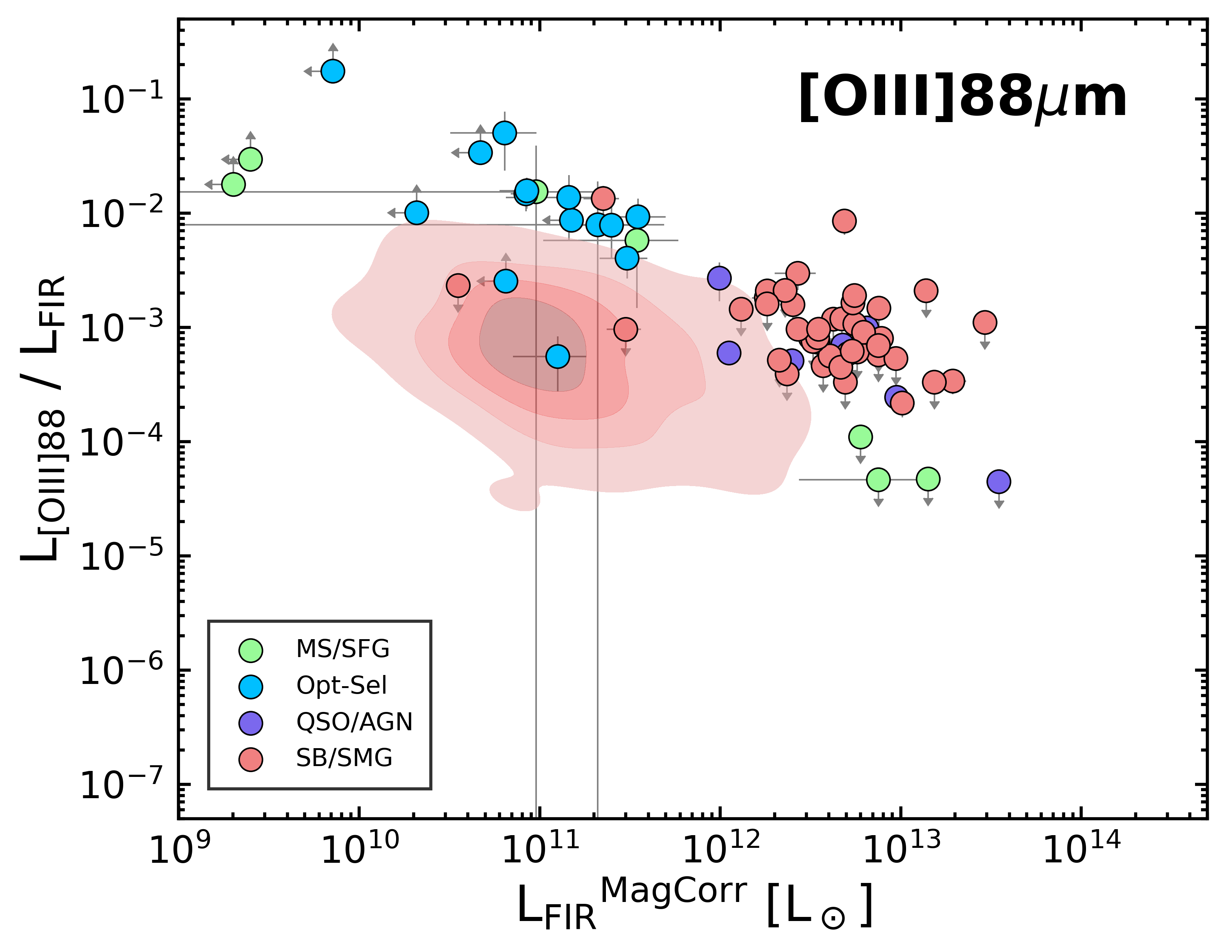}
\includegraphics[width=0.325\columnwidth]{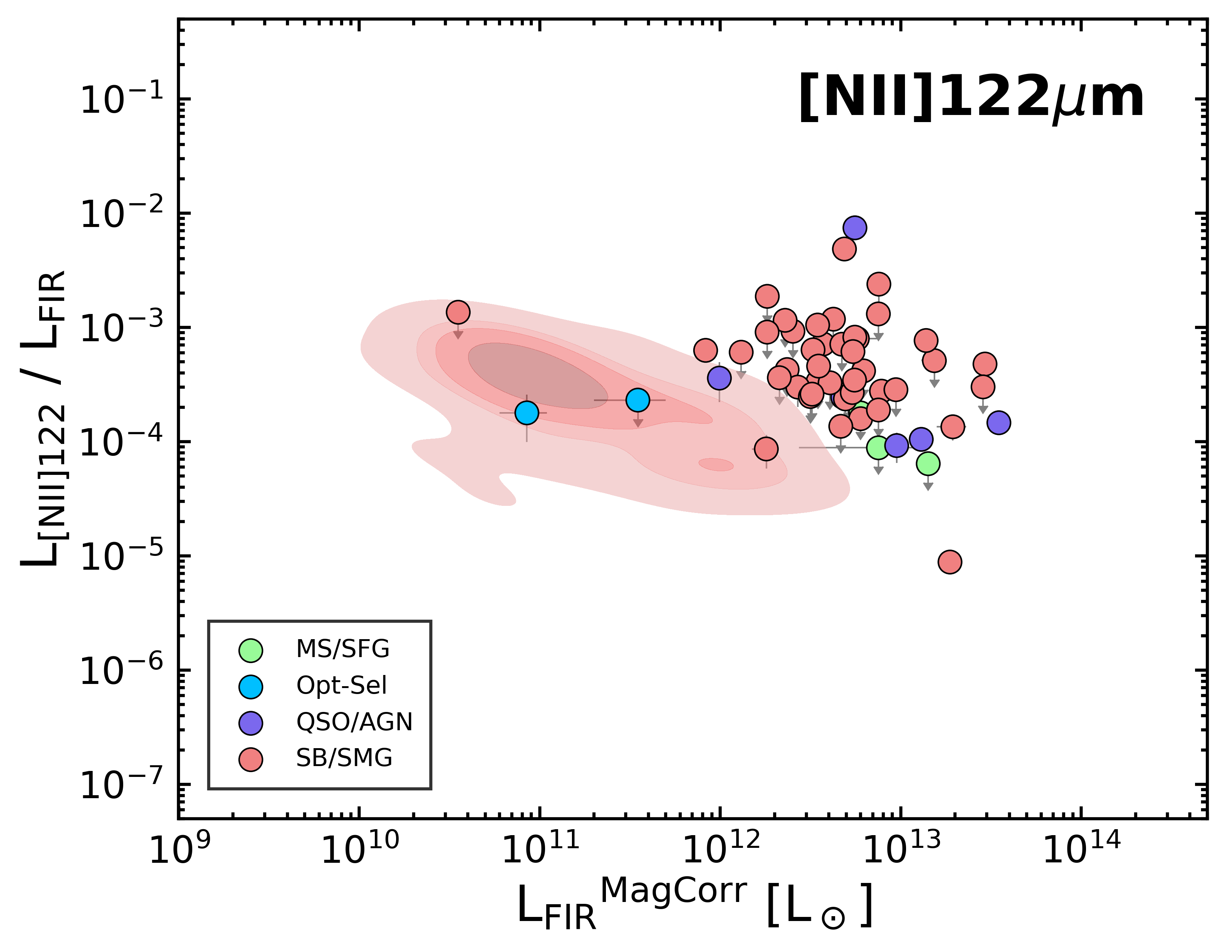}
\includegraphics[width=0.325\columnwidth]{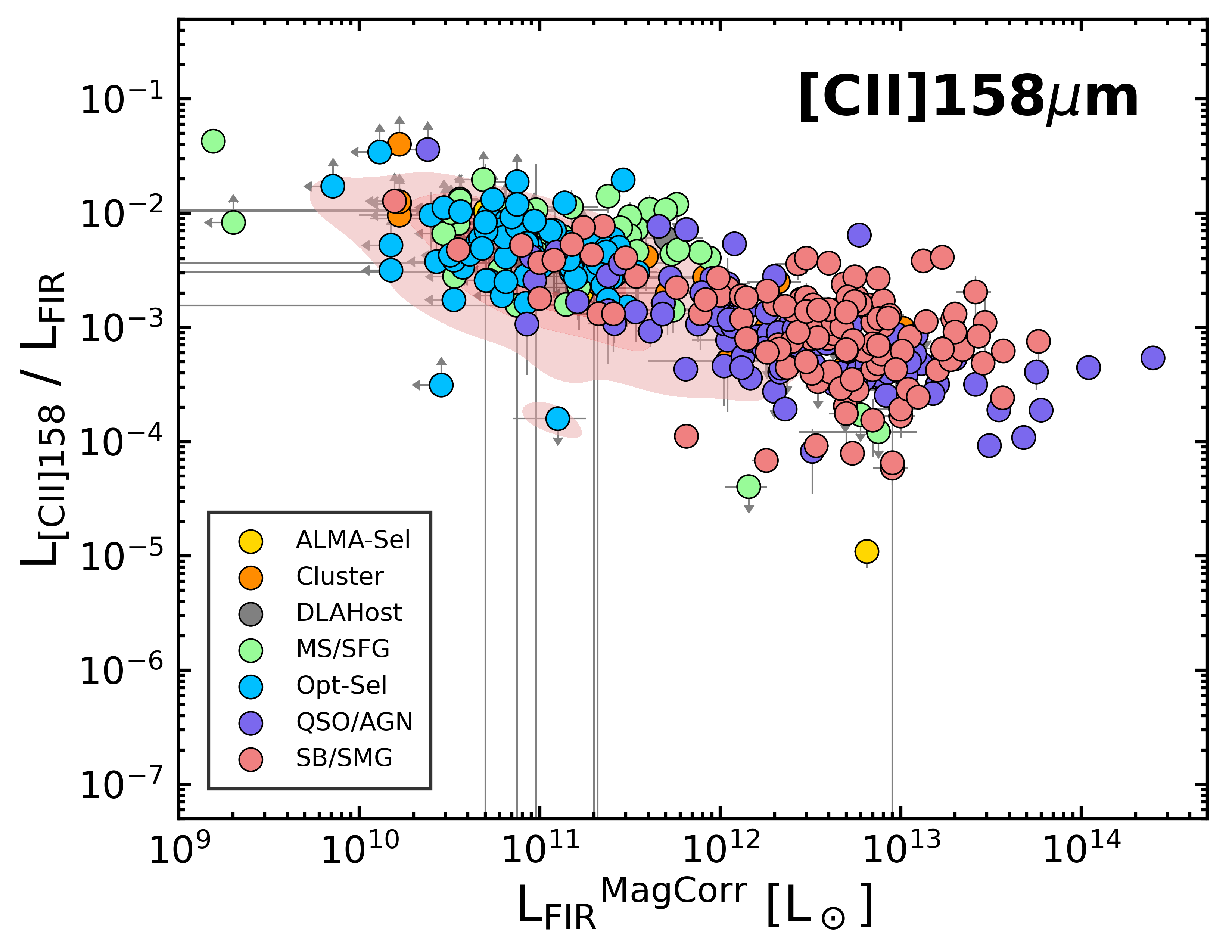}
\includegraphics[width=0.325\columnwidth]{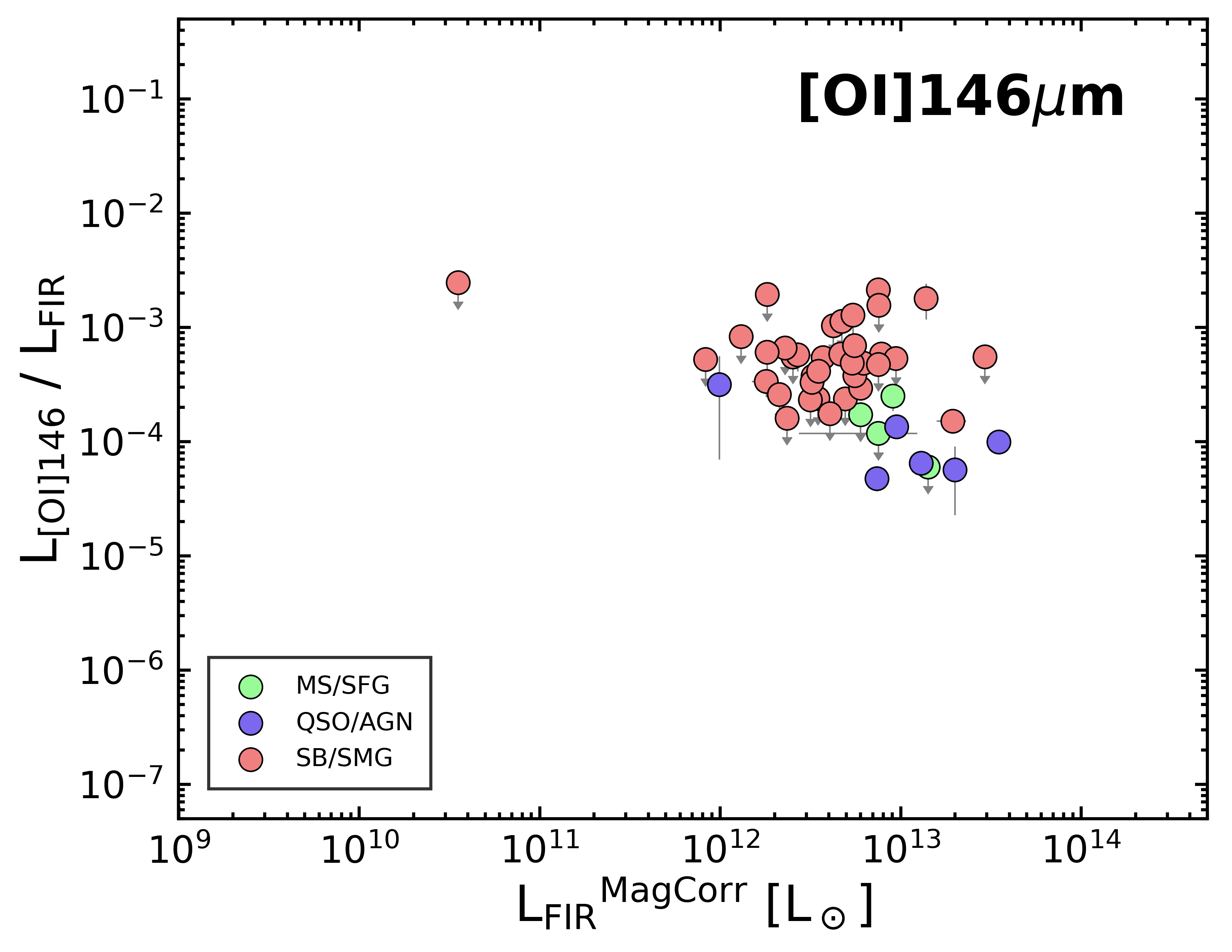}
\includegraphics[width=0.325\columnwidth]{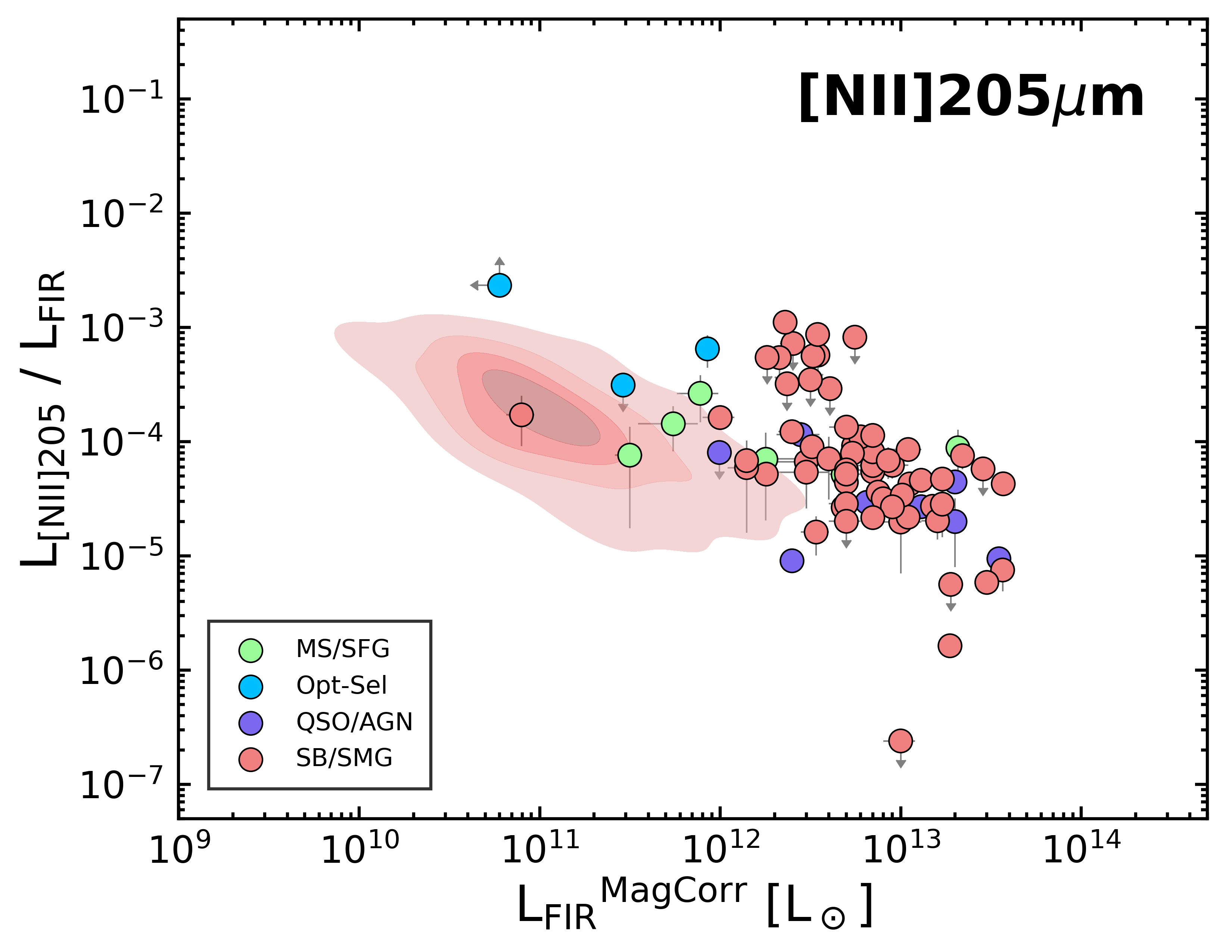}
\includegraphics[width=0.325\columnwidth]{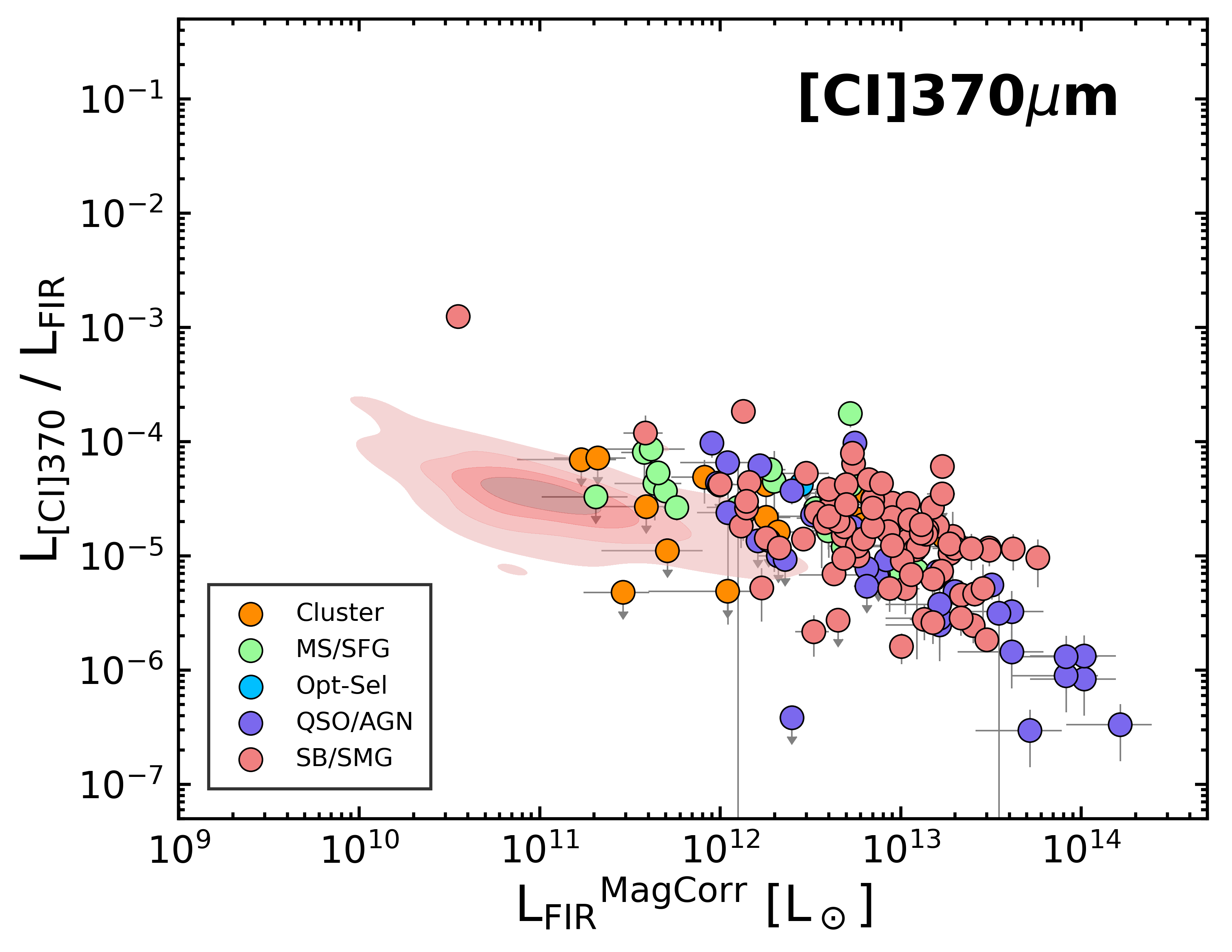}
\includegraphics[width=0.325\columnwidth]{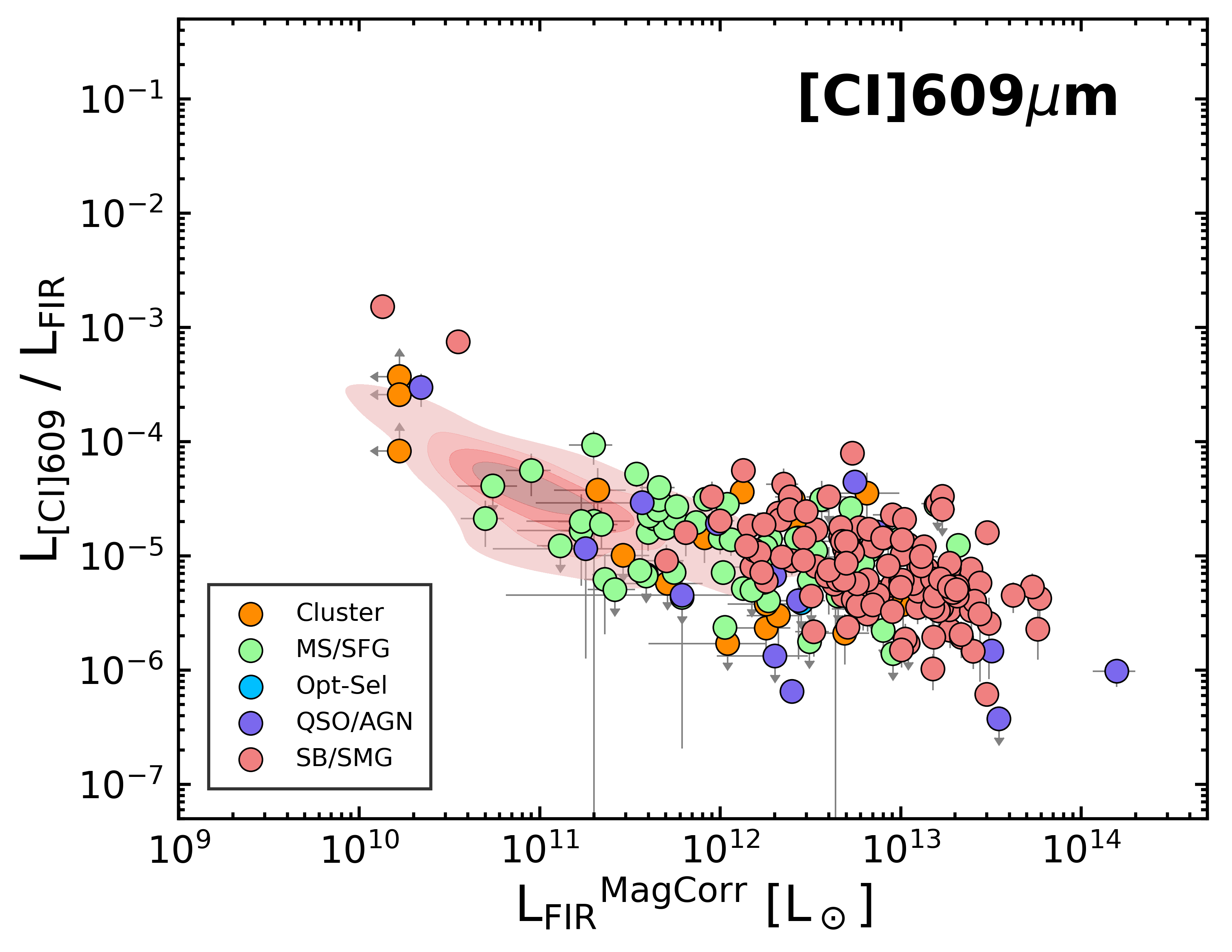}
\end{center}
\caption{Emission line to FIR luminosity ratios for the FSL catalog as a function of observed FIR luminosity (corrected for magnification) color-coded by galaxy type. The red contours represent the location in the parameter space where nearby, starbursting, dusty luminous infrared galaxies lie. In some of the panels a clear segregation among source types is visible, such as in the \Oiii{}\,88\,$\mu$m and \Cii{}\,158\,$\mu$m lines, whereby optically selected galaxies (light blue) show significantly smaller line deficits than those seen in quasars/AGN (dark blue) and starburst/SMGs (red). Normal MS/SF galaxies (green) occupy different regions of the parameter space depending on the emission line under consideration. As in Fig.~\ref{fig_deficit_z}, this figure showcase how some of the trends seen in the line deficits as a function of \iLFIR\, are not a consequence of variations of the local physical conditions of galaxies of the same population at the same redshift, but rather, in this case, due to comparing galaxies of different types (i.e., caused by selection biases).}
\label{fig_deficit_cctype}
\end{figure}

In addition to redshift, other selection biases can also complicate the interpretation or sometimes artificially create line-deficit trends. Figure~\ref{fig_deficit_cctype} shows the same line-deficit plots as in the previous figures, but color-coding the sources as a function of their type/selection. It can be easily noticed that high-$z$ sources that have FIR luminosities in the range of local, dusty galaxies (\iLFIR\,$\simeq$\,10$^{10-12}$\,\Lsun; see \Cii{}, \Oiii{} and \Ci{} panels) are mostly optically selected or main-sequence galaxies, which in nature are intrinsically different from the nearby LIRG population. Moreover, these main-sequence 'low-luminosity' high-$z$ galaxies differ from the local dusty population in different ways depending on the emission line under study. For instance, while normal, star-forming high-$z$ sources have \Ci{}\,609\,$\mu$m deficits similar to the local, dusty population, their \Cii{}\,158\,$\mu$m-to-FIR ratios are on average a factor of $\sim$\,3 larger, and the \Oiii{}\,88\,$\mu$m-to-FIR ratios a factor of $\sim$\,10 higher. Differences between the main-sequence 'low-luminosity' high-$z$ population and their dusty, high-$z$ counterparts, starbursts and SMGs, are also evident, as optically selected, high-$z$ galaxies have around one order of magnitude shallower deficits (higher line-to-FIR ratios) than SMGs in the \Cii{}\,158 and \Oiii{}\,88\,$\mu$m lines, but show similar deficits when considering the \Ci{}\,609\,$\mu$m line.

Another interesting insight to point out is the location of high redshift AGN and quasars in the \Cii{}-deficit parameter space, which is virtually indistinguishable from that of starbursts and SMGs at the same cosmic epochs (Figs.~\ref{fig_deficit_z}--\ref{fig_deficit_cctype}). This suggests that the nature of the central energy source may not play a dominant role in setting the line deficits in these two galaxy populations. Also worth noting is the group of optically selected $z\gtrsim 7$ galaxies showing a significantly large excess of \Oiii{}\,88\,$\mu$m emission with respect to \iLFIR\, when compared to nearby LIRGs and $z\sim$\,2--7 SMGs and QSOs ---much larger than those seen in other emission lines. In this case, selection effects alone are probably not the main reason why these galaxies are located in that particular region of the parameter space. Instead, physical effects such as a combination of increased ionization, lower metallicities (driving up the line-to-FIR ratios), and a decrease in obscured star-formation well beyond the peak of Cosmic Noon ($z$\,$\gtrsim$\,7) (driving down the FIR luminosities) could be the ones boosting the O$^{++}$ emission.

In summary, what this compilation of more than two decades of FSL observations indicates is that both redshift evolution and selection effects are critical in setting the line deficit in galaxies, sometimes exacerbating or sometimes compensating the effects of each other. It is important to note, however, that these take-away messages are mostly inferred from the trends observed in the \Cii{}\,158, \Oiii{}\,88 and \Ci{}\,609\,$\mu$m emission lines alone. In order to further explore the impact of redshift and selection effects in a statistical manner in other FSLs, a larger number of sources need to be targeted in the future.

A corollary of these findings is that the combined effects of redshift and selection biases can have important implications when attempting to derive universal scaling relations between IR FSLs and FIR luminosities in galaxies; which are typically used to calibrate, for instance, star formation rates or molecular gas masses. Only after accounting for those biases, the slopes of the trends shown in the panels of Figures~\ref{fig_deficit}, \ref{fig_deficit_z} and \ref{fig_deficit_cctype} will depend on the galaxies' internal physical processes alone. In addition, the fact that galaxy populations of the same type (i.e., identified with the same selection bias) but at different redshifts display roughly the same slopes while covering different \iLFIR\, dynamic ranges indicates that they are all likely governed by the same internal physical processes. However, it is important to acknowledge that these inferences are drawn from only three emission lines and a limited number of galaxies. More uniform population samples at high-$z$ are needed in order to establish whether the correlations seen up to $z$\,$\simeq$\,6 hold at earlier times.

\subsubsection{Inferring gas properties}

Most of the galaxies studied so far at high redshifts have been detected only in one line (typically, \Cii{}), but sources with multiple IR FSL observations exist thanks to fortuitous frequency combinations (see Fig.~\ref{fig_z_distr}, bottom panel).

\begin{figure}[htbp]
\begin{center}
\includegraphics[width=0.495\columnwidth]{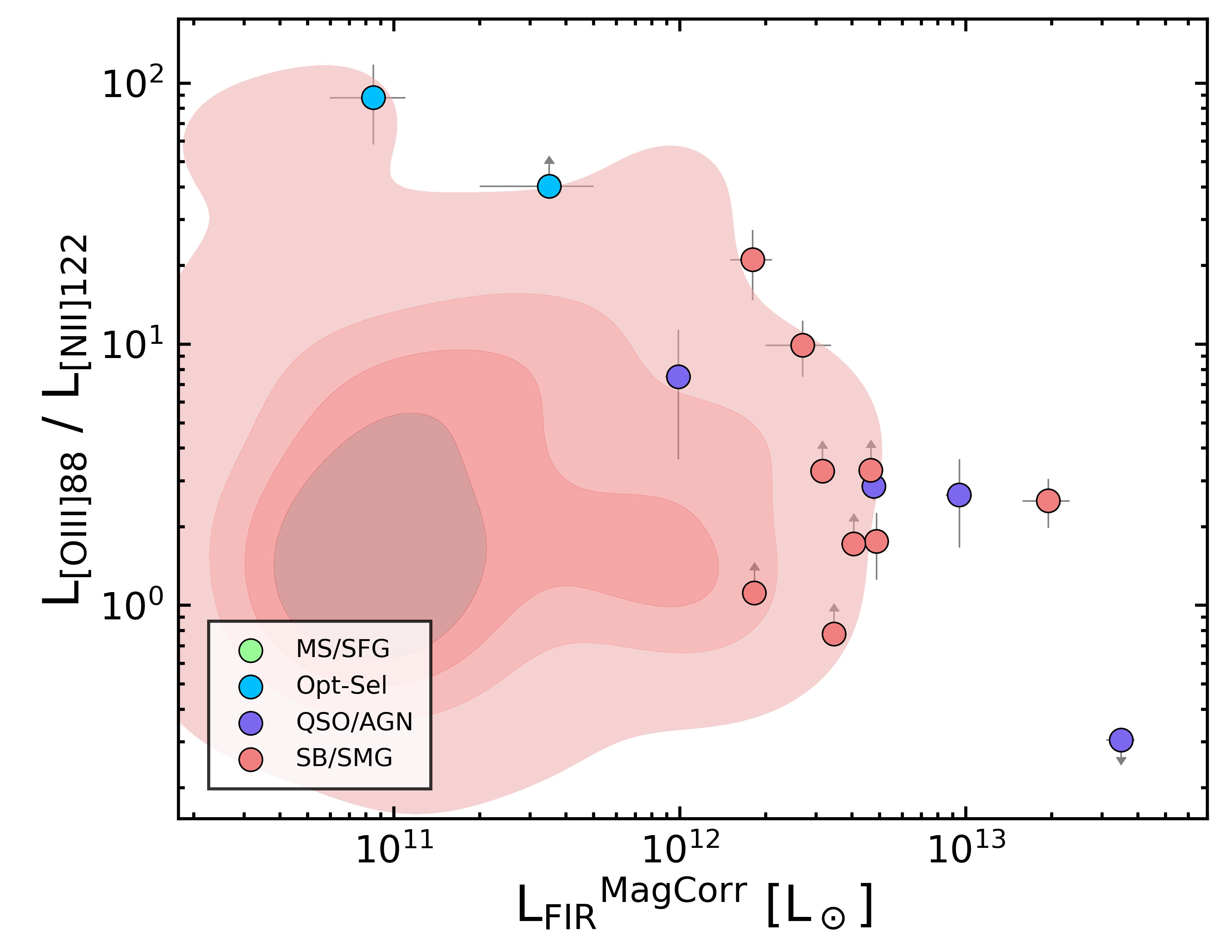}
\includegraphics[width=0.495\columnwidth]{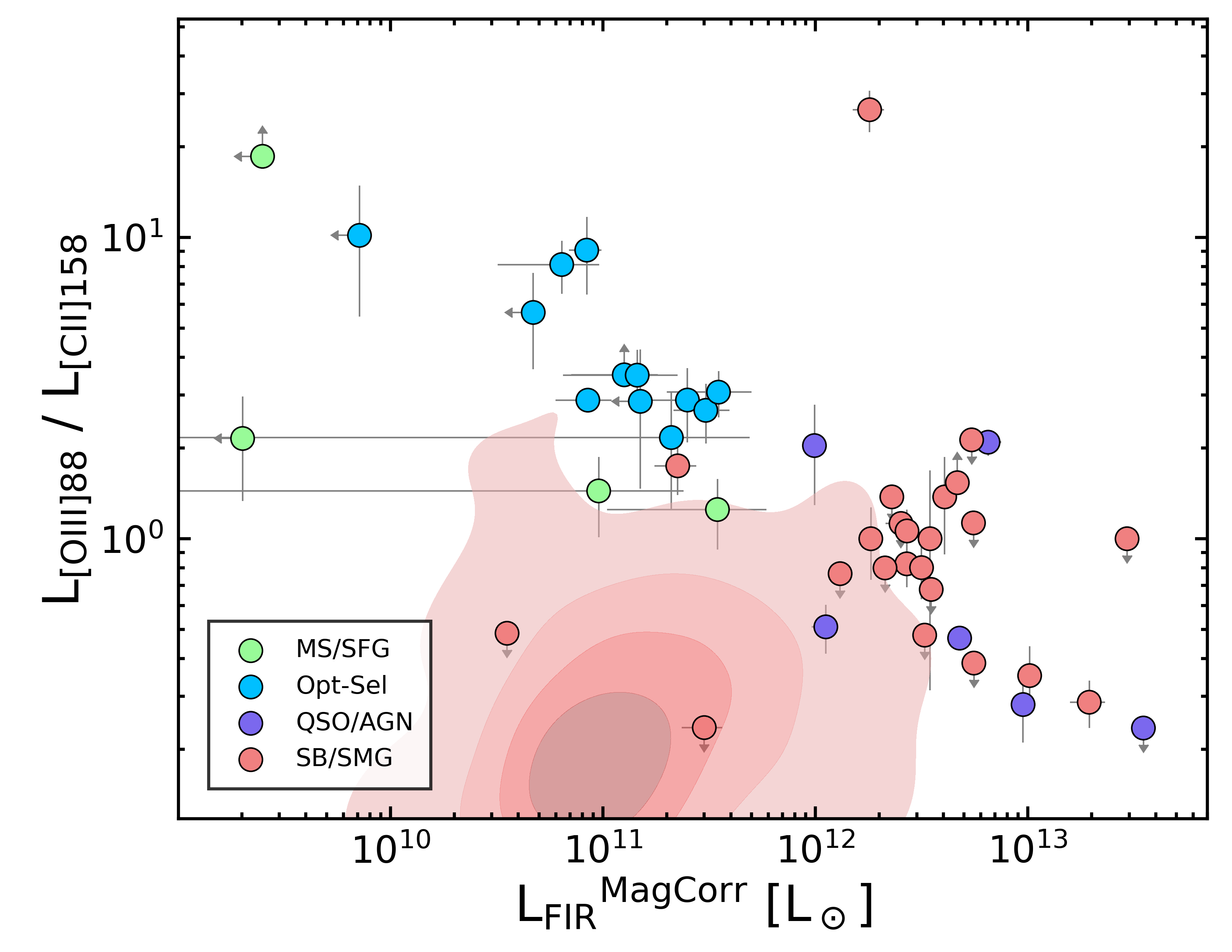}
\includegraphics[width=0.495\columnwidth]{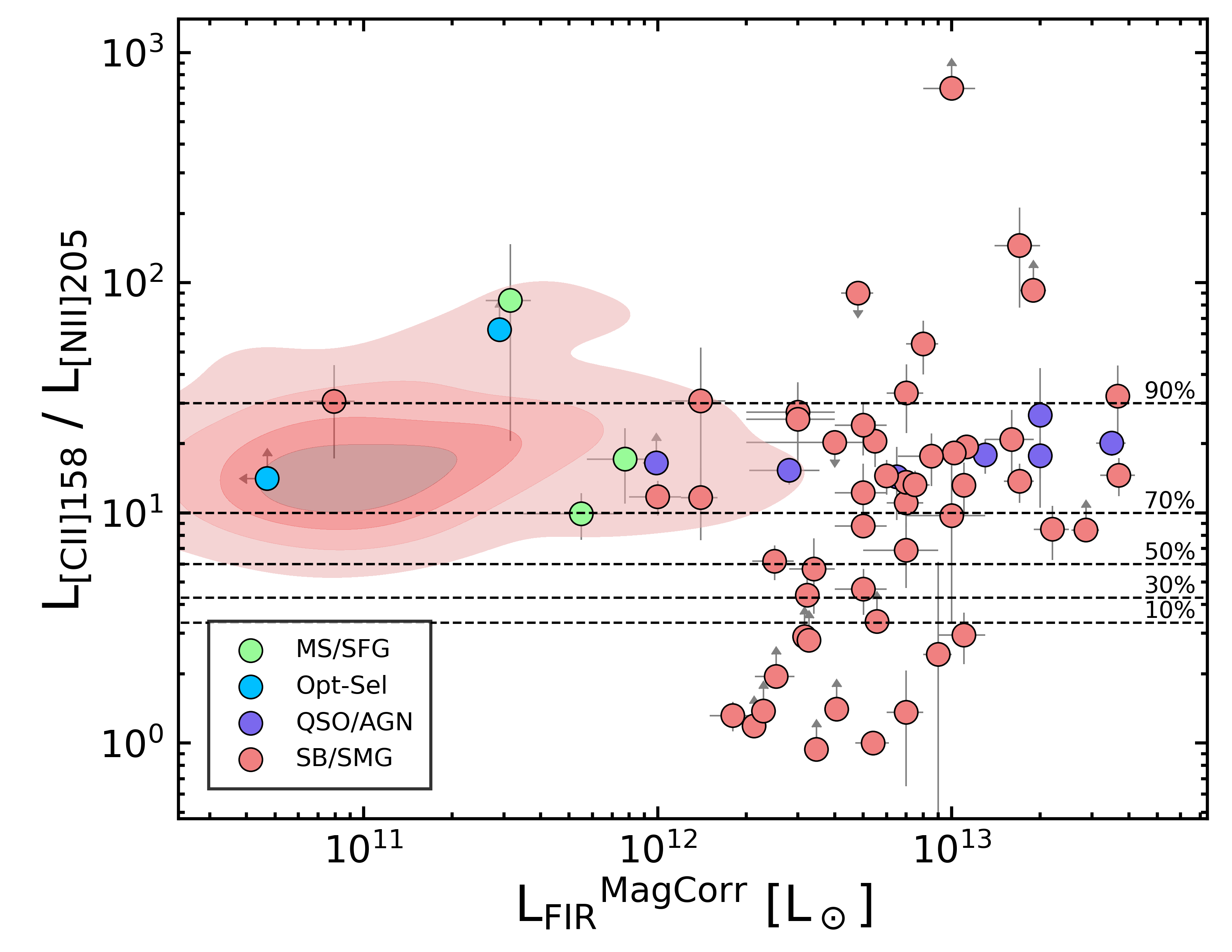}
\includegraphics[width=0.495\columnwidth]{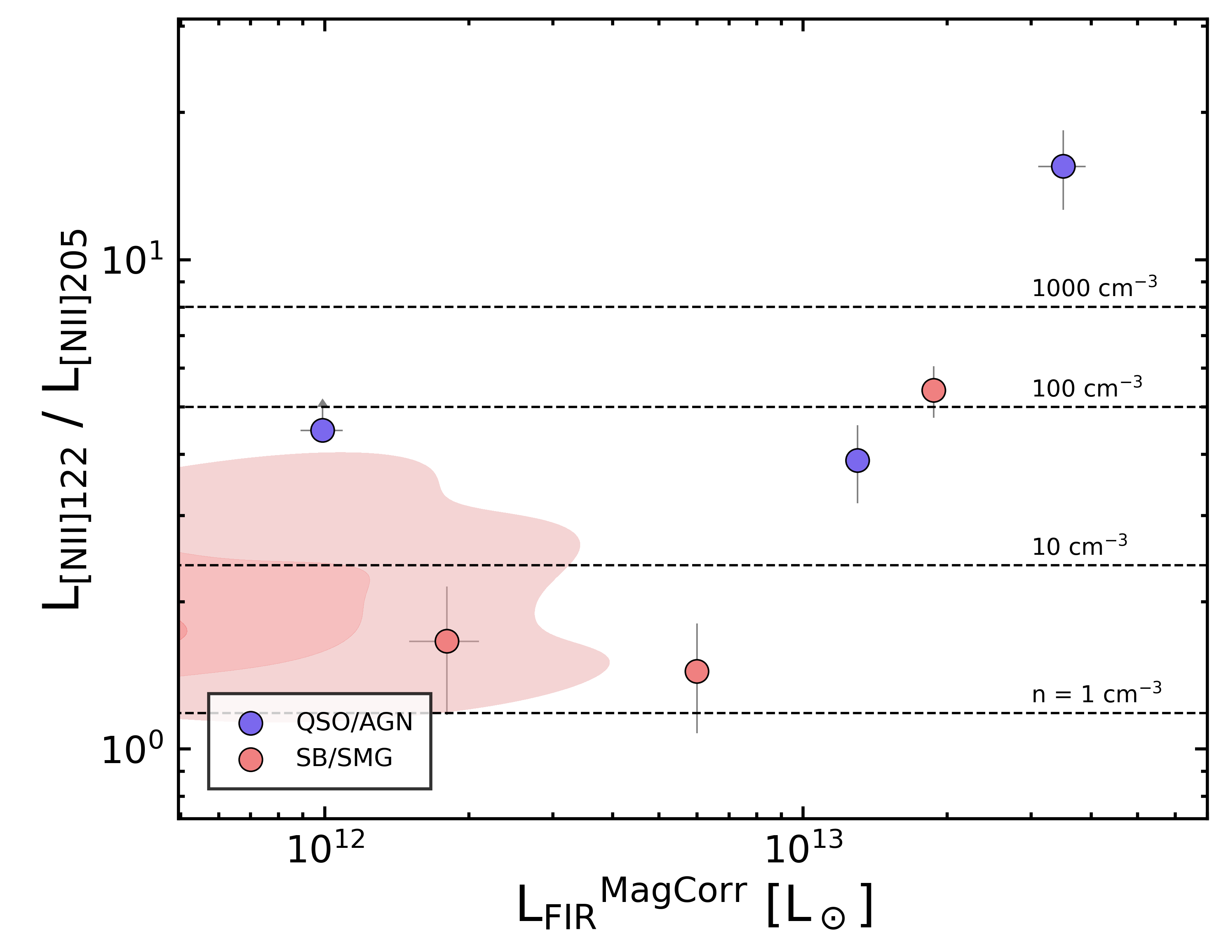}
\includegraphics[width=0.495\columnwidth]{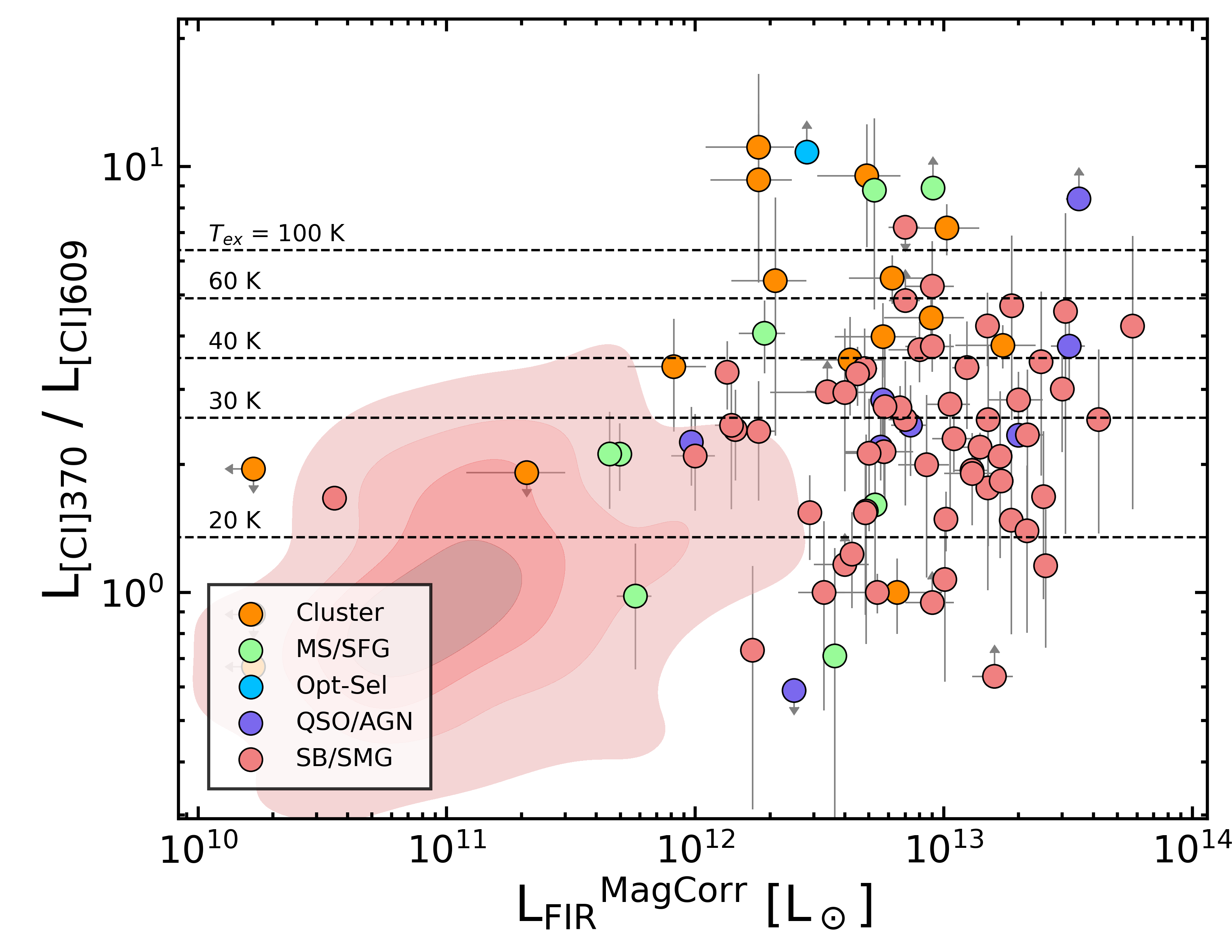}
\end{center}
\caption{Emission line ratios as a function of observed FIR luminosity (corrected for magnification) color coded as a function of galaxy types. \textit{Top-left panel}: \Oiii{}\,88\,$\mu$m/\Nii{}\,122\,$\mu$m luminosity ratio; \textit{top-right panel}: \Oiii{}\,88\,$\mu$m/\Cii{}\,158\,$\mu$m luminosity ratio; \textit{Mid-left panel}: \Cii{}\,158\,$\mu$m/\Nii{}\,205\,$\mu$m luminosity ratio; the horizontal lines show the corresponding fractions of \Cii{} arising from PDRs. \textit{Mid-right panel}: \Nii{}\,122\,$\mu$m/\Nii{}\,205\,$\mu$m luminosity ratio; the horizontal lines show the corresponding electron density. \textit{Bottom panel}: \Ci{}\,370\,$\mu$m/\Ci{}\,609\,$\mu$m luminosity ratio; the horizontal lines show the corresponding excitation temperatures. The red contours represent the location in the parameter space where nearby, starbursting, dusty infrared galaxies lie. Galaxies at $z>1$ show a much wider range of line luminosity ratios than local LIRGs in nearly all the diagnostics shown here.}
\label{fig_ratios}
\end{figure}

As described in Sect.~\ref{sec_diagnostics}, luminosity ratios between IR FSLs can be used to put constraints on the physical properties of the ISM. For instance, the \Oiii{}\,88\,$\mu$m and \Nii{}\,122\,$\mu$m emission lines have very similar critical densities, \ncrit\,$\simeq$\,500\,\ncmmm, and since they are both temperature saturated (their energy level temperatures, $E_{u}$\,$\sim$\,160--190\,K, are much lower than the typical temperature of \Hii\, regions), the ratio between them is mostly sensitive to the intensity and/or hardness of the radiation field (Fig.~\ref{fig_ionization_energy}) and to metallicity at $Z$\,$\gtrsim$\,0.1\,\Zsun (Fig.~\ref{fig_Z}). The top-left panel of Fig.~\ref{fig_ratios} shows the \Oiii{}\,88\,$\mu$m/\Nii{}\,122\,$\mu$m ratio as a function of the FIR luminosity (corrected for magnification) for the galaxies compiled in the FSL catalog, color-coded as a function of source type. If the line ratio was driven by the intensity or hardness of the radiation field, a positive correlation (or at the very least no correlation) with \iLFIR\, would be expected. However, despite the low number statistics, the data hints to an anti-correlation, with dusty starbursts showing on average significantly lower \Oiii{}\,88\,$\mu$m/\Nii{}\,122\,$\mu$m ratios at larger \iLFIR\, than optically selected galaxies, which is what would be expected if the former are more metal rich than the latter.

The top-right panel of Fig.~\ref{fig_ratios} shows an anti-correlation between the \Oiii{}\,88\,$\mu$m and \Cii{}\,158\,$\mu$m line ratio and the FIR luminosity of galaxies. However, the interpretation of this trend is not straight-forward, as these lines have very different ionization potentials (35.1\,eV vs. 12.3\,eV, respectively) and critical densities (\ncrit\,$\simeq$\,500\,\ncmmm\, vs. \ncritH\,$\simeq$\,4\,$\times$\,$10^3$\,\ncmmm, respectively). Optically selected galaxies display higher line ratios than QSOs/AGNs and SBs/SMGs, indicating that either the intensity or hardness of the radiation field is larger in optically selected sources, or that the \Oiii{}\,88\,$\mu$m line is suppressed due to high gas densities in QSOs/AGNs and SBs/SMGs (assuming most of the \Cii{} emission arises from the neutral phase). However, as mentioned above, the \Oiii{}\,88\,$\mu$m and \Nii{}\,122\,$\mu$m emission lines have similar critical densities, whilst displaying the same trend with \iLFIR, suggesting density is not the dominant process setting these trends. On the other hand, metallicity can be playing a role in the \Oiii{}\,88\,$\mu$m/\Cii{}\,158\,$\mu$m ratio as well, since there is also an evolution of the relative abundance of oxygen versus carbon as a function of absolute metallicity similar of that of oxygen versus nitrogen (Fig.~\ref{fig_Z}).

As discussed in the previous section, the \Cii{}\,158\,$\mu$m line can arise from the ionized and neutral ISM. The equations of statistical balance predict that the \Cii{}\,158\,$\mu$m/\Nii{}\,205\,$\mu$m ratio should stay roughly constant as a function of the electron density in the hot, ionized gas phase, around a value of $\simeq$\,3--5, since both lines have very similar critical densities. Any excess above that range implies that \Cii{}\,158\,$\mu$m is also contributed by emission from the warm, neutral gas phase in PDRs. The bottom-left panel of Fig.~\ref{fig_ratios} shows that high-$z$ SBs/SMGs have a large scatter in their \Cii{}\,158\,$\mu$m/\Nii{}\,205\,$\mu$m ratios, ranging from values consistent with pure contribution from ionized gas, to values in which the \Cii{} PDR fraction is $\gsim$\,95\% (\Cii{}\,158\,$\mu$m/\Nii{}\,205\,$\mu$m\,$\gsim$\,20). 

The \Nii{}\,122/205\,$\mu$m line ratio is a direct tracer of the density on the ionized gas, both in Galactic star forming regions and other galaxies. However, due to the relatively low critical densities of both lines ($\simeq$\,270 and 40\ncmmm, respectively), they mostly probe a mixture of the dense ISM (\Hii\, regions) and the diffuse warm medium, depending on the volume filling factor of each phase. The fact that densities of a few tens of \ncmmm\, are found in dense, nearby dusty galaxies, suggests that even in these environments the contribution of the diffuse warm gas to the integrated line emission is significant. The middle-right panel of Fig.~\ref{fig_ratios} shows that at high redshift there are only a handful of sources for which the two N$^+$ FIR FSLs have been secured, with only one source, a dusty QSO, reaching values of the ratio $\gsim$\,10 compatible with the high density limit \citep{fernandez-aranda24}. Future observations of larger samples of galaxies will be critical to investigate the density and filling factors of the different phases of the ionized gas at cosmic noon and beyond.

The bottom panel of Fig.~\ref{fig_ratios} presents the \Ci{}\,370/609\,$\mu$m emission line ratio as a function of the \iLFIR. This ratio can also be used as a tracer of gas density, in this case of the neutral ISM \citep[e.g.,][]{papadopoulos22}. Most sources with observations in both lines are starbursting galaxies or SMGs, and they cover a range of values spanning an order of magnitude, from $\simeq$\,1 to 10, roughly probing the entire range of densities traced by the line ratio, from the lower to the upper density limits ($\sim$\,1--10$^4$\,\ncmmm; see Fig.~\ref{fig_density_diagnostics}). Compared to nearby dusty galaxies, high-$z$ SMGs seem to display on average higher gas densities. Other galaxy populations, however, have not been targeted as intensively, but also seem to show large line ratio variations.

\begin{figure}[htbp]
\begin{center}
\includegraphics[width=0.495\columnwidth]{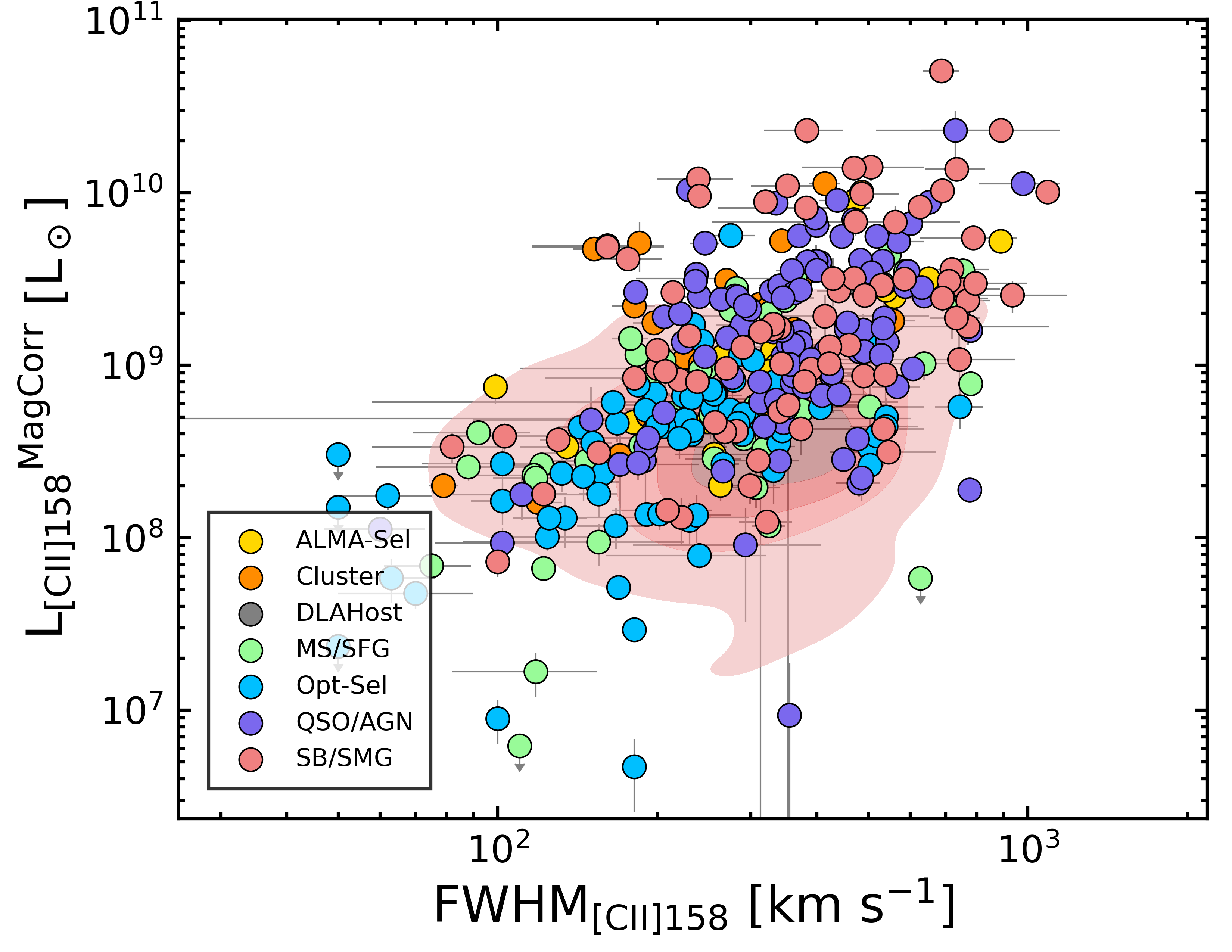}
\includegraphics[width=0.495\columnwidth]{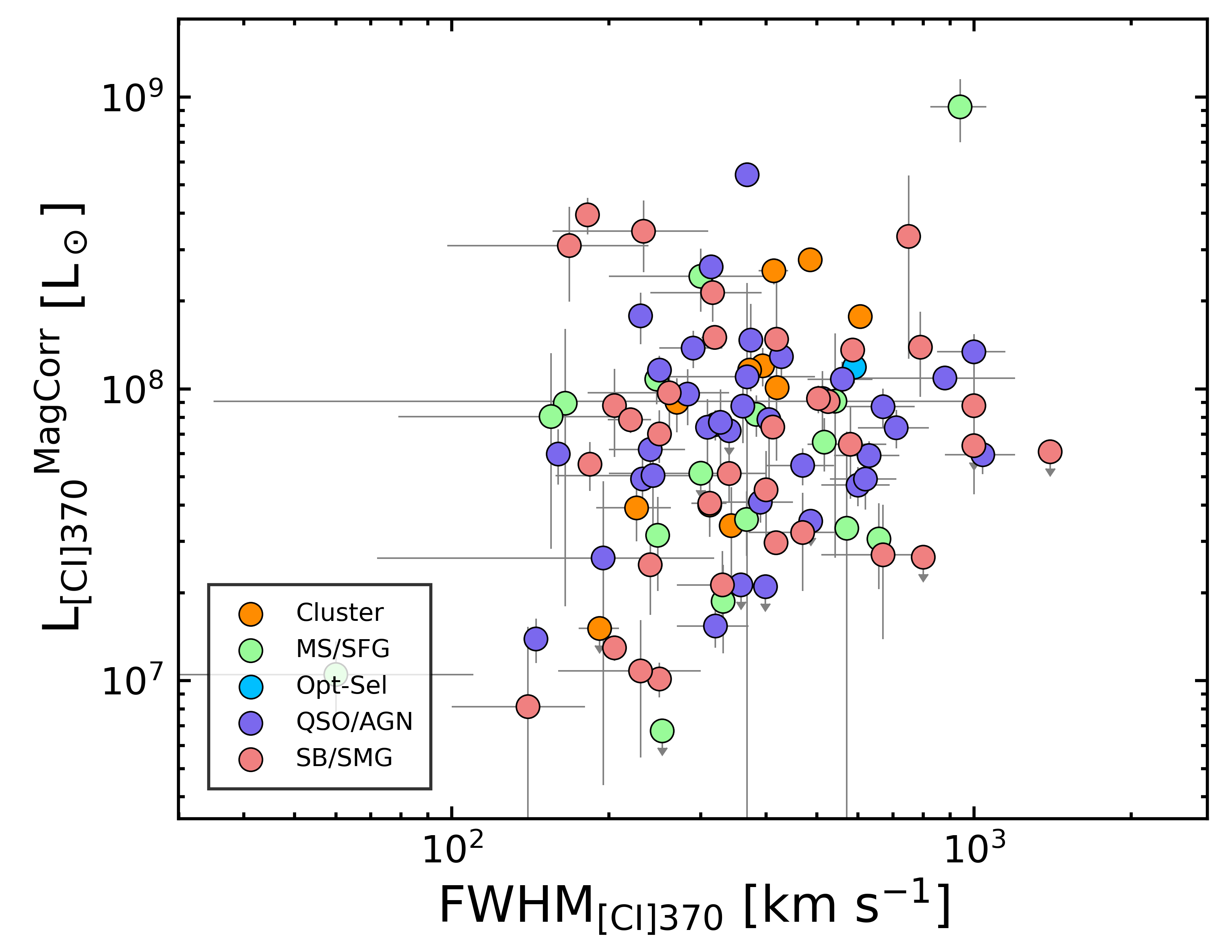}
\includegraphics[width=0.495\columnwidth]{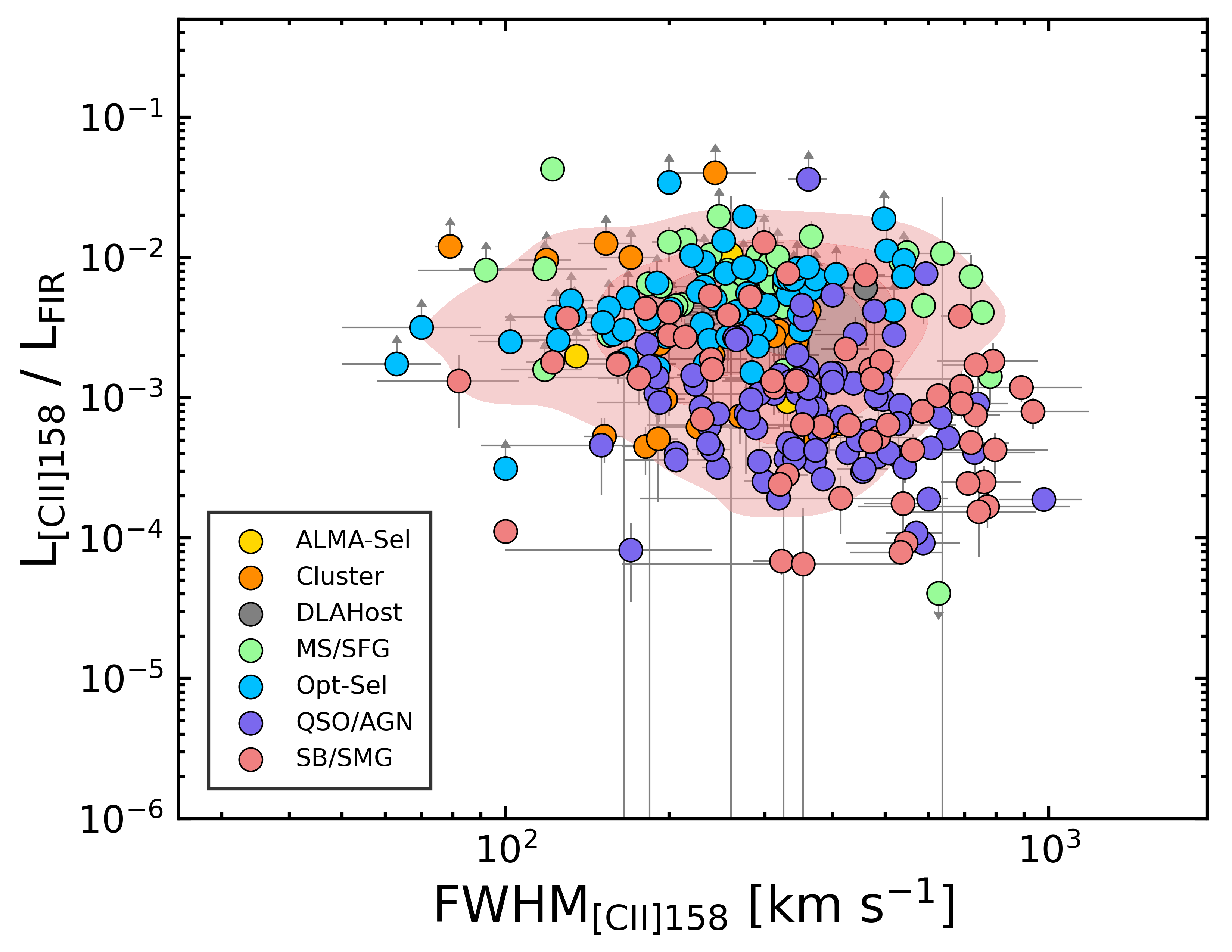}
\end{center}
\caption{Rest-frame line luminosities (corrected for magnification) as a function of the FWHM of the line, color coded as a function of galaxy type. The red contours represent the location in the parameter space where nearby, starbursting, dusty luminous infrared galaxies lie. There is a correlation between the \Cii{} line luminosity and its FWHM. Main-sequence and optically selected galaxies seem to have on average lower line luminosities and velocity dispersion than SMGs and QSOs. Such trend is not seen in the \Ci{}\,307\,$\mu$m line.}
\label{fig_fwhm}
\end{figure}

Fig.~\ref{fig_fwhm} presents the \Cii{}\,158\,$\mu$m luminosity (corrected for magnification) as a function of the full-width at half-maximum (FWHM) of the line. There is a broad, yet clear trend for more luminous sources to exhibit larger line widths. Interestingly, there is a segregation in the area of the parameter space where galaxies of different types are located, with optically selected and main-sequence galaxies showing the smallest line luminosities and FWHMs, and quasars, starbursts, and SMGs having the largest luminosities, up to $L_{\rm [CII]}$\,$\gtrsim$\,10$^{10}$\,\Lsun\, and FWHMs, up to $\sim$\,1000\,\kms. Despite the overall trend is comprised by all the different galaxy types, each one roughly follows the same trend by itself. From the well-established Tully-Fisher \citep{tully77} and Faber-Jackson \citep{Faber76} relations we would expect a broad correlation between the rotation or the velocity dispersion of galaxies, respectively, as measured from the line width of emission lines tracing the bulk of the baryonic matter, and their luminosity. However, these empirical relations predict dependencies of the luminosity with $v_{rot}$ or $\sigma$ (depending on the interpretation of the line width) that scale as $L\,\propto\,\sigma^{3-4}$. In Fig.~\ref{fig_fwhm} we see that the luminosity of the \Cii{} line only increases by about two orders of magnitude per order of magnitude increase of the line's FWHM.

A mechanism different from PDRs that can also increase the luminosity of the \Cii{} line together with its FWHM, is shock-driven turbulence. Whether it is caused by galaxy--galaxy interactions that can promote cloud-cloud collisions and/or structural heating in dense environments, like in the nearby merger system of Stephan's Quintet, stellar-driven outflows associated with supernova feedback, or jetted, mechanically driven outflows from radio-loud AGN like in NGC~4258 \citep{appleton13, appleton18}, the \Cii{} luminosity could carry an additional contribution from turbulence arising from any of these mechanisms, which would also increase the line's FWHM. In addition to increased turbulence, galactic outflows --whether they have a stellar or AGN origin-- can push gas away from the galaxy in the form of massive winds moving at velocities of many hundreds to thousands of km/s, also potentially increasing both the total line flux and its averaged FWHM.

Interestingly, the same trend is not followed by the \Ci{}\,370\,$\mu$m line (Fig.~\ref{fig_fwhm}, top-right panel), which should also be sensitive to the increase in gas kinetic temperature due to turbulent heating. However, \Cii{} can be boosted not only due to the increase in \iTkin\, in the post-shocked medium, but also because of a reduction of its optical depth due to line broadening, as the shocked gas becomes more turbulent. While \Cii{} can be close to be optically thick in dense PDRs (see Fig.~\ref{fig_optical_depth}), it can decrease its opacity (and therefore increase its net luminosity per gas mass) if there is sufficient broadening of the line wings ($\sigma_v$ in Eq.~\eqref{eq_tau_line}) where photons can escape the region ---an effect that is less pronounced in the neutral carbon FSLs due to their significantly lower optical thickness.

The contribution of turbulence to the \Cii{} line is commonly associated in the nearby Universe to galaxies showing low $L_{\rm FIR}$ and low deficit values (with \Cii{}-to-FIR ratios larger than $10^{-2}$) compared to what PDR models can reproduce. In contrast, the bottom panel of Fig.~\ref{fig_fwhm} shows that sources with the highest line widths tend to display the smallest \Cii{}/$L_{\rm FIR}$, ratios. However, a counter example to the Stephan's Quintet or NGC~4258 is NGC~6240, which hosts wide spread, low velocity shocks that dominate the observed CO line luminosity. The FWHM of the \Cii{} line of NGC~6240 is $\sim$\,550\,\kms\, and the galaxy has a \Cii{}/$L_{\rm FIR}$\,$\simeq$\,$4\,\times\,10^{-3}$ ---a regime close to the high redshift SMGs and starbursts showing large FWHMs and \Cii{} deficits.

\begin{figure}[ht]
\begin{center}
\includegraphics[width=0.495\columnwidth]{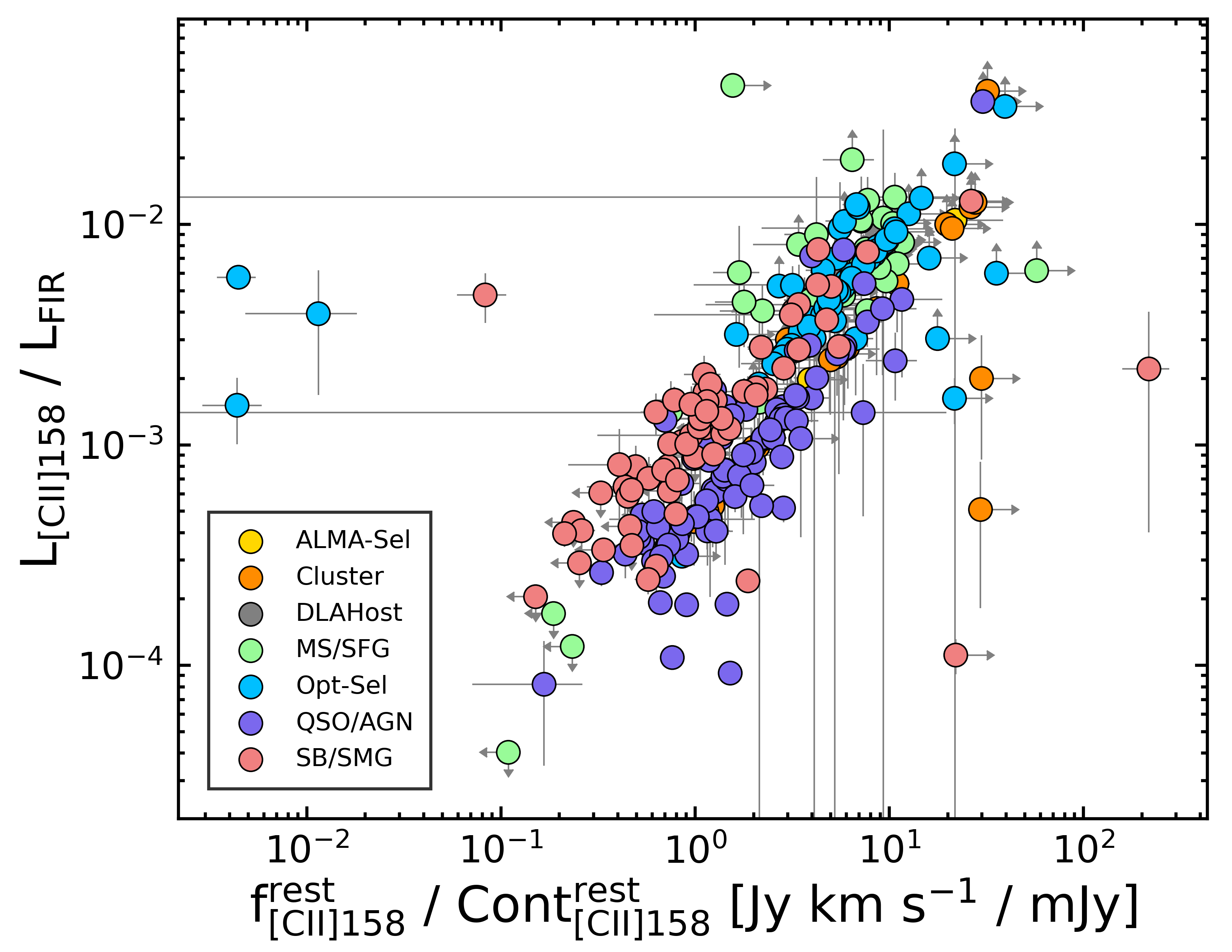}
\includegraphics[width=0.495\columnwidth]{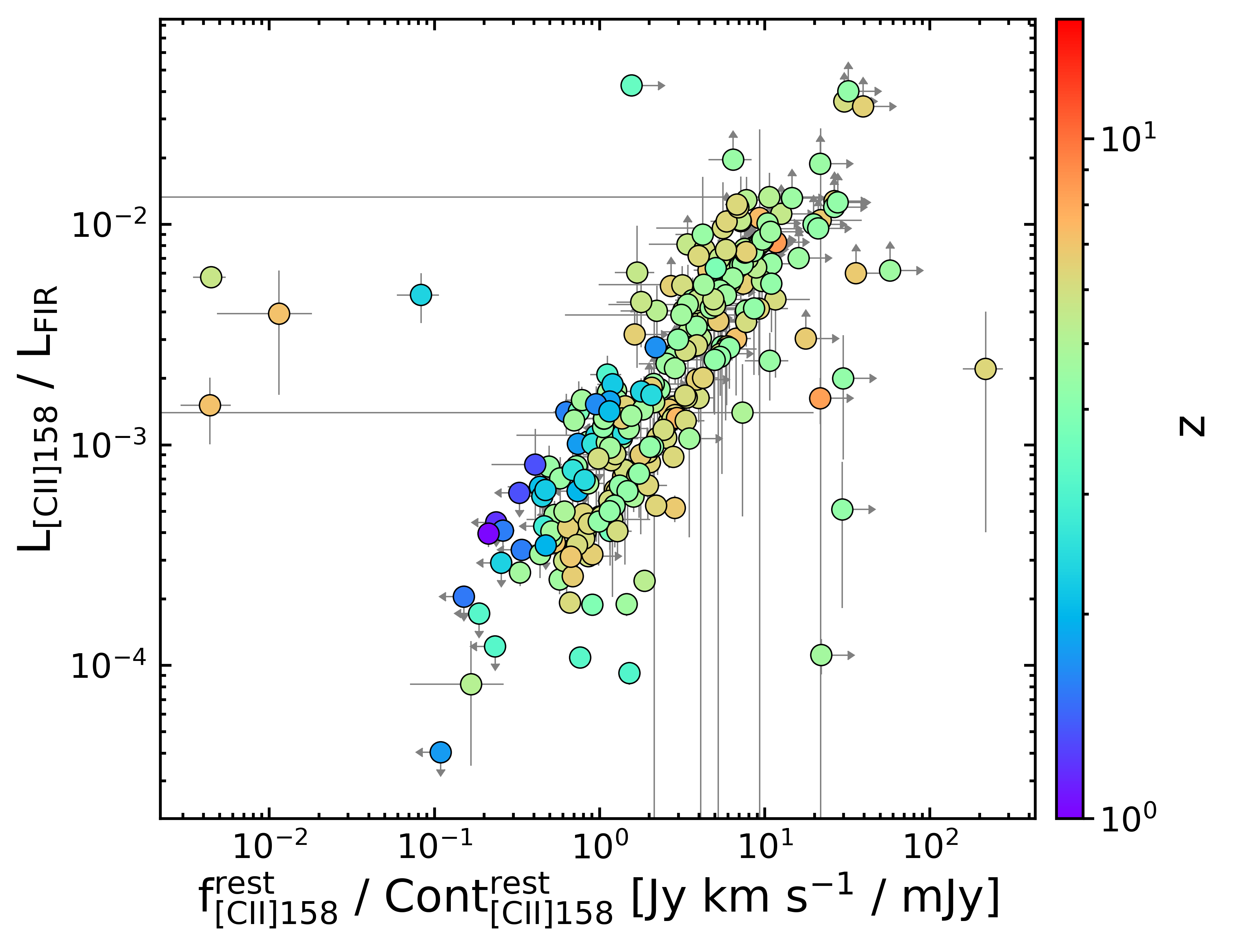}
\end{center}
\caption{\Cii{}\,158\,$\mu$m line deficit for the FSL catalog as a function of the line equivalent width, color-coded as a function galaxy type (left panel) and redshift (right panel). There is a segregation in the \Cii{}-to-FIR ratios for a given equivalent width, between starbursts/SMGs/main-sequence galaxies and QSOs/AGNs due to the common choice in the literature to pick higher dust temperature values for the latter population.}
\label{fig_eqw}
\end{figure}

Figure~\ref{fig_eqw} displays the \Cii{}\,158\,$\mu$m line deficit as a function of the line equivalent width. The left panel shows there is a segregation between galaxy types. For a given \Cii{} equivalent width, starbursts and SMGs show on average smaller line deficits by a factor of two than QSOs/AGNs. This behavior is most likely caused by the limited multi-band photometric information available for many of the sources, which forces researchers to rely on assumptions, namely to set a fixed --rather than derived-- dust temperature: the common choice in the literature is to select $T_{\rm dust}\,=\,47$\,K for high redshift quasars, and $T_{\rm dust}\,=\,$30--35\,K for starbursts, SMGs and optically selected galaxies instead. Note that this assumption seems to be independent of the redshift, as show in the right panel of the figure.

\subsubsection{Spectral line profiles}\label{sec_specline}

Similarly to other line diagnostics of the ISM, the spectral profile of FSLs can unveil precious information on the kinematics of the gas. In the previous section, we have already shown how the line width can be related to ISM properties. Because FSLs are typically optically-thin, their spectra show the luminosity-weighted line-of-sight velocity of the ISM. Galaxies showing a flat rotation curve, similar to local spiral galaxies, may display a ``double horned'' line profile, provided that the line emission extends for a sufficiently large area where the rotation velocity is constant \citep[e.g.,][]{valentino20}. Face-on disks on the other hand have relatively narrow, Gaussian-like spectral profiles due to the vertical turbulence of the gas in the disk \citep[e.g.,][]{valentino20,boogaard20}. Galaxies undergoing mergers and close interactions with gas-rich satellites will likely show irregular spectral profiles \citep[e.g.,][]{diaz-santos21}. \citet{deblok14} analyzed the dynamical information that can be extracted from the spectral analysis of spatially-unresolved observations of distant galaxies. 

A complementary, but related reason for studying the spectral profile of FSLs is the search for outflows. Quasar feedback may heat and displace significant fractions of the host galaxy ISM \citep[e.g.,][]{heckman90,veilleux20}. Feedback from AGN is a common mechanism for galaxy evolution models to explain the discrepancy between the observed bright end of the mass function of galaxies and the intrinsic halo mass function predicted by cosmological simulations \citep[see][for a review]{somerville15}. In the spectral line profile, feedback may be apparent in the form of a broad ($\gsim 1000$ \kms) component associated with outflowing gas, in addition to the `narrow' one ($\lsim 300$ \kms) tracing the galaxy potential well. This broad component might also appear as high-velocity shoulders or wings in the line spectral profile (see, e.g., \citealt{bischetti19, stanley19, ginolfi20a, herrera-camus21, khusanova22}). A robust characterization of this component is challenging: 1) if these broad components encompass most of the observed bandwidth \citep[as in the original \Cii{} observations of the archetypal $z=6.42$ quasar J1148+5251 in][]{cicone15}, the continuum flux density may be misjudged, especially in regimes of low-S/N \citep[as demonstrated in][]{meyer22}; 2) the spectral line profile might be intrinsically complex (i.e., not described by a simple gaussian curve), hence determining whether a faint, broad residual is present becomes model-dependent; 3) since these components typically carry only a small fraction of the total line flux, and they are spread over several spectral channels, the S/N ratio of the features per resolution elements is often very limited; 4) the presence of satellite or companion galaxies may mimic the presence of asymmetric outflows, especially in stacking analyses \citep[e.g., see discussion in][]{bischetti19}.

\subsubsection{Spatially-resolved studies of the gas within galaxies}\label{sec_morphology}

More than a decade ago, using cycle 0 ALMA data, \citet{debreuck14} reported the detection of spatially resolved \Cii{} observations in a starburst galaxy hosting an AGN at $z$\,=\,4.8, showing that the gas kinematics were dominated by disk rotation but with a high degree of turbulence. Following on those steps and pushing forward towards higher redshifts, a few years later \citet{smit18} reported the detection of \Cii{} over multiple resolution elements in two galaxies at $z\approx 6.8$, pointing that their study ``opens up opportunities for high angular-resolution \Cii{} dynamics in galaxies less than one billion years after the Big Bang.''. Indeed, the recent progress achieved in terms of imaging and sensitivity by submm-wavelength interferometers enables detailed maps of the FSLs in high–redshift galaxies \citep{lelli21,rizzo22}. At $z>1$, linear scales of 1\,kpc (comparable to the sizes of galaxies) correspond to $0.1'' - 0.2''$. These angular scales are within reach for submm-wavelength interferometers such as ALMA and NOEMA.
However, in the push for higher and higher angular resolution, one should account for the rapidly deteriorating surface brightness sensitivity of the observations. In an ideal situation, for a (locally) flat surface brightness, the same integration time under the same weather conditions would yield a signal-to-noise ratio, S/N, per resolution element that is smaller by a factor that is about the square of the beam area ratios (i.e., 1/4 if the beam is twice as small in linear size). To first order, S/N $\propto \sqrt{t}$, where $t$ is the integration time. Thus, if we halve the linear size of the beam (i.e., we double the angular resolution), we need $\sim16\times$ the integration time to achieve the same S/N per resolution element. In addition, observations at relatively coarse angular resolutions, secured with the arrays in compact configurations, are easier to calibrate and more stable, thanks to the shorter baselines. Finally, because each baseline is sensitive only to certain angular scales, as we push for higher angular resolution, we may loose the information from the more extended and diffused components. All these arguments make observations at high angular resolution challenging, and relatively expensive to secure even for ALMA. In order to mitigate these restrictions, it is often mandatory to compromise between angular resolution and surface brightness sensitivity, based on the scientific goals.

{\bf Morphology and spatial extent.} The most basic measurement enabled by high-angular resolution observations is the size of the emitting region. 
\citet{walter09nature} reported one of the first spatially-resolved maps of \Cii{} beyond the local Universe. They found that the dust and gas emission in the host galaxy of a $z\approx 6.4$ quasar are extremely compact ($\sim$750\,pc in radius). Similar compact emission appears to be common in the high-$z$ quasar population \citep{wang13, trakhtenbrot17, venemans17b, wang19, venemans20, yue21, shao22, walter22}. The few high-$z$ quasar host galaxies showing \Cii{} emission on scales larger than a few kpc are in interaction with a companion galaxy \citep[e.g.,][]{diaz-santos16,decarli19, banados19, venemans19, neeleman21}. The \Cii{} emission in sub-mm galaxies can be similarly compact, with sizes of 1--3 kpc \citep[e.g.,][]{riechers13, neri14, oteo16, ansarinejad22}, but with occasional exceptions extended over several kpc \citep[see, e.g.,][]{riechers14, rybak19,venkateshwaran24}. Optically-selected galaxies tend to show even smaller sizes \citep[albeit with exceptions, as shown in][see Fig.~\ref{fig_literature}b]{herrera-camus25}. Stacking analyses have reported typical sizes of $\sim2.5$ and $2.2$\,kpc at $z\sim 4.5$ and $5.5$, respectively \citep{fudamoto22,fujimoto22}, with some works providing sizes on individual objects \citep{ikeda25}.

In all the cases, \Cii{} appears more extended than the underlying dust continuum emission. This happens in part because of the ubiquitous nature of the \Cii{} emission, which traces gas along the molecular gas clouds as well as the neutral and mildly ionized ISM, resulting in an intrinsically smoother morphology; and in part because of the dust continuum observability. As discussed in Sect.~\ref{sec_deficit}, the \Cii{}/IR luminosity ratio is suppressed in regions of higher surface density \citep{diaz-santos14}, due to a combination of various astrophysical processes (temperature gradients, photoelectric efficiency, etc) that enhance proportionally more IR luminosities than \Cii{} emission in compact star-forming regions. This leads to a steeper surface brightness radial profile in the dust emission than what is observed for \Cii{} \citep[see, e.g.,][]{rybak19,venemans19,herrera-camus21}. In turn, \citet{ikeda25} found no strong \Cii{}/IR gradients in normal star-forming galaxies at $z\sim 5.5$.

Mapping the internal structure of galaxies (in order to identify spiral arms, bars, gas clumps, etc) requires superior resolution, in the $\sim$100 pc regime ($\sim0.02''$). ALMA has mapped such structures in high-$z$ galaxies via their dust continuum \citep[e.g.,][]{hodge19, rujopakarn19}; the same experiment is much more time consuming with line emission, and has now been performed in a handful of quasar host galaxies at $z>6$ \citep{venemans19, neeleman23,  meyer23, meyer24}. It appears that morphological features appear smoother when observed in their \Cii{} emission rather than in their dust continuum, due to the same arguments discussed for the radial \Cii{}/IR dependence \citep[see, e.g.,][]{venemans19,meyer24}.

\textbf{Dynamics.} A spatial resolution element of $\sim 1$ kpc ($\sim 0.1''-0.2''$), i.e., a few times smaller than the linear size of high-$z$ galaxies, may allow us to test whether the gas dynamics is dominated by rotation, dispersion, or is irregular \citep[see, e.g.,][]{neeleman21,wang24highres,venkateshwaran24}. Such observations usually struggle to constrain the inclination angle of high-$z$ galaxies and to disentangle intrinsic line broadening from beam smearing effects; but they deliver reliable order-of-magnitude estimates. For more detailed dynamical models, higher spatial resolution (a few 100\,pc or better) is required. This can be achieved either by pushing ALMA's imaging capabilities \citep[as in the observation of a warped disk in][see Fig.~\ref{fig_literature}c]{neeleman23}, or via the natural magnification offered by gravitational lensing \citep[e.g.,][]{litke19,rizzo21,fujimoto21,tamura23}. It should be emphasized that these dynamical masses gauge the total (baryon + dark matter) mass \textit{within the \Cii{}-traced emitting region}, which is a small fraction of the total dark matter halo.

Among the most interesting results from these dynamical mass measurements in high-$z$ galaxies is the discovery of disk galaxies in the young universe \citep[e.g.,][]{smit18,neeleman20,jones21,neeleman21,herreracamus22,posses23,neeleman23,romanoliveira23,rowland24,romanoliveira24,venkateshwaran24,umehata25}. Although some of these disks are not nearly as ``thin'' as the ones observed in the local universe, due to a combination of turbulence, gravitational interactions, and local instabilities, the presence of disks at high redshift questions an early idea, according to which disk galaxies might be too fragile for the early, merger-rich cosmic times when these galaxies are observed. 

Another key result from dynamical studies using FSLs is the estimate of virial masses in quasar host galaxies at cosmic dawn. Once compared to the local scaling relations, the black hole masses are comparable \citep[e.g.,][]{izumi19,izumi21} or more often higher \citep[e.g.,][]{walter09nature,shao17,wang19,pensabene20,neeleman21,shao22,neeleman23,meyer23} than what one would expect from the local black hole --- host galaxy relations \citep[see review in][]{kormendy13}. This finding may have an intrinsic nature, if black holes build up their mass faster than the host \citep{volonteri12}; or be the result of selection effects, given that black hole masses are feasible only in luminous quasars, which tend to have large black hole masses \citep[e.g.,][]{lauer07,reines15,shankar19}.

In the last few years, as astronomers used ALMA in its most extended configurations to achieve the best possible angular resolution,
we approach the possibility of mapping and reconstructing the gas dynamics within the sphere of influence of black holes at high redshift. The sphere of influence is defined as the region where the black hole (and not the diffused mass from the host galaxy) dictates the dynamics. For an isothermal sphere, with a line Full Width at Half Maximum FWH\,=\,300\,\kms{}, the radius of the sphere of influence is $r_{\rm si}=G\,M_{\rm BH}\,\sigma^{-2}$ (for FWHM=$2\,\sqrt{2\,\ln{2}}\,\sigma$ $\approx2.35\,\sigma$), corresponding to 262 pc for $M_{\rm BH}=10^9$\,\Msun{}. Observations of \Cii{} in $z>6$ quasars have already reached these scales. However, the velocity and velocity dispersion maps secured so far do not show the dynamical signature of the black holes. This is because the high concentration of gas in the central hundred parsecs of the hosts, inferred from either the dust continuum or the \Cii{} luminosity, are comparable to or even larger than the mass of the black holes \citep{walter22,neeleman23,meyer23}, hence further reducing the actual value of $r_{\rm si}$. 

\textbf{Extended emission and outflows.} An important application of spatially-resolved maps of FSL emission is the study of gas in the circum-galactic medium (CGM) of galaxies. Galaxy formation models unanimously surmise a suite of feedback processes (radiation, winds, jets, etc) arising from both star formation and nuclear activity \citep[see reviews in][]{somerville15,veilleux20} that regulate the baryonic growth of galaxies and can contaminate the CGM of metal-rich gas. Models suggest that outflows can be widespread phenomena in high-$z$ massive galaxies \citep{pizzati23} and that up to 25\% of extended emission arises from the diffuse CGM \citep{dicesare24}.

Characterizing this diffuse gaseous component would shed light on the ways feedback operates and on its impact on galaxy evolution. However, surface brightness scales as (1+$z$)$^{-4}$, hence detecting these extended features is challenging at high redshift. Furthermore, because interferometers act as spatial scale filters, a combination of array configurations is often required to isolate the CGM scales ($\gsim$10 kpc, or a few arcsec) from the central galaxy. This however leads to irregular beam sizes, and imaging issues. For instance, the standard `cleaning' approach replaces the observed beam with a 2D gaussian curve, which is often not a good approximation in the case of multi-configuration observations. As a result, low-S/N features are incorrectly treated, leading to residuals that can mimic the presence of extended emission \citep[see][for a detailed discussion of these issues]{novak20}. A further complication in the study of purely diffused emission is the presence of faint satellite galaxies (see next subsection), which might be responsible for some of the emission, especially in the low-S/N regime or in stacking analyses. Direct analysis of visibility data can overcome some of these problems (\citet{novak20,akins22,ikeda25}, but limitations in the modeling remain.

Extended \Cii{} halos have been reported in main sequence massive galaxies \citep[e.g.,][]{fujimoto19,fujimoto20,ginolfi20,herrera-camus21,lambert23,ikeda25} as well as quasar host galaxies \citep{bischetti24}. Outflows have been invoked to interpret plums \citep{solimano24}, although tidal features and satellites are also common interpretations \citep{diaz-santos16,diaz-santos18,decarli19,ginolfi20,posses24}. The \Oiii{}/\Cii{} ratio in the diffuse component may shed light on its origin: The presence of a cold, neutral gas phase, perhaps interspersed with photon-dominated regions, would imply relatively low \Oiii{}/\Cii{} ratio. 
Prominent outflows should shine bright in their \Oiii{} emission. High–frequency ALMA and JWST NIRSpec IFU observations will thus allow to decipher the nature of the extended [C ii] halos observed so far.

\textbf{Offsets among different FSL maps.} ALMA opened a window to detect and image at $<1''$ resolution FSLs at high frequency ($\nu_{\rm obs}>400$\,GHz). At $z\gsim 5$, this led to the detection of \Oiii{}\,88\,$\mu$m in galaxies that had already been detected in \Cii{}. Surprisingly, the two lines frequently showed small spatial offsets (up to a few kpc; see, e.g., \citealt{carniani17,carniani18,carniani20,ren23,fujimoto24}, but also a counter example in \citealt{wong22}). Similar offsets appear between the \Cii{} emission, the Ly$\alpha$ emission, and the UV broad-band continuum emission. These discrepancies raised questions on the intrinsic geometry of these early galaxies: Is \Cii{} tracing the bulk of the cold ISM, whereas the \Oiii{} emission maps outflows? Is the stellar component primarily located at the position of the UV-bright knots, or are these only luminous, young star clusters intersperse in a mostly dust-obscured galaxy \citep[see, e.g.,][]{hodge15}? 

Observations with JWST can now unveil the morphology of the starlight emission in high-redshift galaxies 
\citep[e.g.,][]{hodge24,herard-demanche24}. The coming years will undoubtedly see rapid progress in understanding the interplay of intrinsic morphology, obscuration, and multi-phase line emission.

\subsubsection{Searching for faint companions around known sources} 

Interferometric observations of \Cii{} known in high-$z$ galaxies led to the serendipitous discovery of \Cii{}-emitting galaxies in their immediate neighborhood (at projected separations of $\lsim$100 kpc). Thanks to the superb imaging capabilities of ALMA, \citet{carilli13_br1202} showed that the IR luminous galaxy BR1202--0725 at $z=4.7$ is actually not only composed of a sub-mm galaxy and an optical quasar \citep[as first revealed in the dust continuum by][]{omont96}, but also a number of companion, lower mass sources. This reinforced results on the multiplicity in the dust continuum of IR-luminous galaxies previously observed with single-dish telescopes \citep{hodge13}. 

ALMA observations of \Cii{} emission in massive sources at $z>4$ enabled the characterization of their close environment \citep[e.g.,][]{trakhtenbrot17, diaz-santos18, nguyen20, fudamoto21, fuentealbafuentes24}. These studies revealed that metal-enriched galaxies populate the environment of massive galaxies, and reveal a tumultuous growth via mergers at early cosmic times. The detection of \Cii{} emitters in the close environment of $z>6$ quasars \citep{decarli17} was the first, long sought-after spectroscopic confirmation that quasars at these early cosmic time inhabit overdense environment \citep[see also][]{willott17,neeleman19}. 

Expanding these searches beyond individual pointings is observationally costly. For instance, a 7-point hexagonal mosaic will only expand the radius by a factor $\sim 2$ compared to an individual pointing. But because of the rapid drop in the correlation function \citep[see, e.g.,][]{decarli17,meyer22}, the yield of new detection per pointing steeply decreases at increasing radii.

\subsubsection{Line intensity mapping} 

Line intensity mapping is an observational technique that aims at detecting the aggregated line signal from an unresolved population of sources. Fluctuations of the line signal both along the line of sight and in the sky plane enable to reconstruct the power spectrum of the global line emission \citep{chang10,kovetz17}. The technique has the potential of inferring, simultaneously, the total luminosity of the line and the clustering of the line-emitting galaxies, even without reaching the angular resolution and sensitivity required to identify individual galaxies. The observed power spectrum $P_k(z)$ corresponding to a scale wavenumber $k$ and redshift $z$ is:
\begin{equation}\label{eq_LIM_powerspec}
P_k(z) = \langle I(z)\rangle^2 b^2(z) P_m(k,z) + P_{\rm shot}(z)
\end{equation}
where $P_{\rm shot}(z)\propto \int_0^\infty L^2 \Phi(L,z)\,dL$ is the shot noise power spectrum, $\langle I(z)\rangle \propto \int_0^\infty L \Phi(L,z)\,dL$ is the average line intensity (weighted over the line luminosity function), $P_m(k,z)$ is the power spectrum of the dark matter halos, and $b(z)$ is the redshift-dependent bias of the galaxy population responsible for the line emission.

Line intensity mapping promises to reveal the power spectrum of fine-structure line emission at high redshift, in the mm and sub-mm bands, similar to the success achieved with H{\sc i}\,21\,cm investigations \citep[e.g.,][]{chang10}. An extensive effort has been devoted in forward modeling the strength and detectability of the FSL intensity mapping signal in high-redshift galaxies \citep[e.g.,][]{sun19, yang21, yang22, murmu23, sun23, fronenberg24}. The expected signal however is faint even for \Cii{}, especially compared to the foreground of molecular gas emission. Molecular deep field observations demonstrated that \Cii{} only accounts for less than a percent of the total line flux measured at 3\,mm and 1.2\,mm; the remainder is associated with carbon monoxide (CO) lines at cosmic noon \citep{decarli20}, which dominate the shot noise of the power spectrum \citep{uzgil19}. Removing the foreground CO signal requires an excellent characterization of the CO luminosity function, which would require a transformational leap in survey speed of available facilities. Cross-correlating signal from different lines could help in isolating the FSL intensity mapping signal from the foreground \citep[see predictions in, e.g.,][]{schaan21,padmanabhan22}, provided that both observational and methodological limitations (beam size, depth, etc) are controlled for.

\section{Conclusions}\label{sec_conclusions}

The rapid progress in the study of FSL emission at high redshift over the last decade, especially in the far-IR domain, has opened a new window to fully characterize of the ISM in early galaxies. Some of the most recent and on-going observational campaigns with ALMA and other sub-mm/radio interferometers have revealed both common features across sources at different redshifts but also cases in which the ISM properties of galaxies belonging to the same population can largely vary across a wide range of the parameter space. We have seen that general trends exist, e.g., in `line deficits' or in \Cii{}--SFR relations, but the location or slope of these trends may be driven by selection effects such as the particular nature of the studied sources: main-sequence vs. starburst galaxies, or star-forming sources vs. AGN/quasars. However, once these potential biases are controlled for, the evolution of the trends seen for a specific galaxy population as a function of redshift are actually dictated by the physical processes occurring within them.

Ultimately, the final goal of this collective endeavor is to understand how some of the most fundamental properties of galaxies, such as the average density and metallicity of the gas they harbor, as well as the average intensity of the interstellar radiation field produced by their dominant energy sources, evolve across cosmic time. But in order to circumvent the uncertainties and degeneracies carried by observables probing more than one physical process, energy source, or gas property, it is imperative that a wealth of multi-phase gas tracers are assembled for each of the galaxies contributing to the cause, both in terms of emission line observations and dust continuum. And while this may seem like a monumental, nearly out of reach enterprise, it is nonetheless very much akin to the decades-long effort that has been dedicated to pinning down the evolution of star formation, stellar mass, and gas content of galaxies across cosmic time. Tens of thousands of hours of observing time across multiple state-of-the-art, ground- and space-based telescopes have been poured into achieving these goals, which in many critical ways have revolutionized our understanding and transformed our views on how galaxies have formed and evolved since cosmic dawn.

The available facilities operating at mm and sub-mm wavelengths will certainly continue expanding on the existing samples. The synergy between ALMA, NOEMA, JVLA and JWST has only started to deliver its fruits: In particular, by unlocking the discovery of galaxies beyond the parameter space studied so far (in terms of cosmic time, mass, metal content, ionization conditions, mass of black holes powering AGN, etc) and by complementing observations of the cold ISM in the rest-frame FIR of high-$z$ galaxies with tracers of rest-frame optical bands.

In the near future, the power of this complementarity will be fully unleashed, including the potential synergies with the Probe Far-Infrared Mission for Astrophysics (PRIMA) space telescope. PRIMA holds the promise of accessing the so critical $\sim$25 to $\sim$250\,$\mu$m region of the electro-magnetic spectrum, a decade in wavelength that is strikingly not covered by any telescope facility today (or for any foreseeable future as of the writing of this review). The FIRESS instrument may deliver a spectral resolving power of up to R\,$\sim$\,4400 with which to observe critical FSLs that JWST is currently targeting in the mid-IR in nearby galaxies up to cosmic noon. Despite the relatively modest size of its primary mirror (1.8\,m), PRIMA will mark an orders-of-magnitude improvement over Spitzer and Herschel in sensitivity and mapping speed, thanks to the combination of active cooling and state of the art detectors in space.
PRIMA's scientific goals largely align with those current (sub-)mm facilities have in vision, including the study of the co-evolution of galaxies and their supermassive black holes across cosmic time and measuring the formation and buildup of heavy elements and interstellar dust in galaxies from cosmic dawn to today.

On a longer term, however, it is clear that a major development in the capabilities of existing (sub-)mm facilities will be essential to take the next leap in the study of IR FSLs at high redshift --an example of which is the planned upgrade of the next-generation VLA, the ngVLA. For ALMA, the achievement of higher angular resolution and superior imaging synthesis would undoubtedly be beneficial for morphological and kinematic studies, but the crucial requirement to fully unleash another era of transformational discoveries would be to step up its line sensitivity. A gain of a factor at least $\sim$\,3 in collective area, achieved by increasing the number of antennas, would push survey speed and boost the detectability of individual lines and dust continua in galaxies at the highest redshifts, and even enable molecular line absorption studies, all while simultaneously improving the image fidelity. Enhanced line sensitivity, and a more efficient use of excellent weather conditions at the unique ALMA site to push for the highest frequency observations, will enable studies of lines with rest-frame wavelengths $\lambda_\text{rest}\lsim$\,50\,$\mu$m at $z >6$, opening the ultimate door to measurements of gas metallicity, density, and ionization in galaxies born within the first Gyr of cosmic history.


\subsection*{Supplementary information}

The compilation of all the IR FSL observations of individual galaxies reported in the literature by the end of December 2024, accompanied by software to parse the catalog, select subsets of sources based on specific parameters, and reproduce all the plots in this review, is available at the following GitHub repository: \url{https://github.com/tdiazsantos/fslcat}. In addition, a python notebook is included in the repository where all the set up necessary to solve the equations of statistical equilibrium to find the population levels is laid out.

\subsection*{Acknowledgments}

We thank Chris Carilli and the anonymous referee for their constructive review of the manuscript. We thank Lee Armus, Axel Wei\ss{}, and Pierre Cox for their thorough reading of the manuscript, which led to significant improvements. We are grateful to Leindert Boogaard, Carlos de Breuck, Carl Ferkinhoff, Paul Goldsmith, Ryota Ikeda, Hanae Inami, Kotaro Kohno, Livia Vallini, Fabian Walter for the insightful discussions we had with them during the preparation of the manuscript.
R.D.~acknowledges support from the INAF GO 2022 grant ``The birth of the giants: JWST sheds light on the build-up of quasars at cosmic dawn'', INAF Minigrant 2024 `The interstellar medium at high redshift'', and by the PRIN MUR ``2022935STW'', RFF M4.C2.1.1, CUP J53D23001570006 and C53D23000950006. T.D.-S. acknowledges support by the Hellenic Foundation for Research and Innovation (HFRI) under the ``2nd Call for HFRI Research Projects to support Faculty Members \& Researchers'' (Project Number: 3382).

\end{document}